\documentclass[aps,a4paper,showkeys,nofootinbib,longbibliography,twocolumn]{revtex4-2} 
\usepackage{braket}
\usepackage[utf8]{inputenc}
\usepackage{filecontents}
\usepackage{natbib} 
\usepackage{amsmath,amssymb,bm,amsthm}
\usepackage{subfigure}
\usepackage{graphicx} 
\usepackage{dcolumn} 
\usepackage{bm} 
\usepackage[mathlines]{lineno} 
\usepackage{booktabs}
\usepackage{color}
\newcounter{one}
\usepackage{url}

\usepackage{textcase}

\usepackage{colortbl}
\usepackage{tabularx}
\usepackage{verbatim}
\usepackage{multirow}

\usepackage[T1]{fontenc} 
\usepackage{lmodern}
\usepackage{bbm}
\usepackage[utf8]{inputenc}
\usepackage{amsfonts}
\usepackage{array}

\usepackage[ruled,lined]{algorithm2e}

\usepackage{bbm}
\usepackage{enumitem}
\usepackage{umoline} 
\usepackage[usenames,svgnames]{xcolor}
\usepackage{natbib}
\usepackage[hyperindex,breaklinks]{hyperref} 
\hypersetup{
     colorlinks=true,   
     linkcolor=Navy,      
     citecolor=Navy,       
     filecolor=Navy, 
     urlcolor=Navy,  
    runcolor=cyan,
 } 
\usepackage{graphicx}
\usepackage{amsfonts}
\usepackage{amssymb}
\usepackage{amsmath}
\usepackage{bbm}
\usepackage{enumitem}
\usepackage{xcolor}
\usepackage{dsfont}

\newcommand{\tr}[0]{ {\rm tr}}

\newcommand{\ceil}[1]{\left \lceil #1  \right \rceil}
\newcommand{\half}[1]{{ \rm h}}
\newcommand{\Oorderof}{\mathcal{O}}
\newcommand{\orderof}[1]{\Oorderof(#1)} 

\newcommand{\for}[0]{\quad \textrm{for} \quad}

\newcommand{\co}{{\rm c}}

\usepackage{yhmath}

\newcommand{\poly}{{\rm poly}}

\newcommand{\ad}{{\rm ad}}

\newcommand{\Or}{\quad {\rm or} \quad}
\newcommand{\AND}{\quad {\rm and} \quad}

\def\beq{\begin{equation}}
	\def\eeq{\end{equation}}
\def\nbeq{\begin{equation*}}
	\def\neeq{\end{equation*}} 
\def\<{\langle}
\def\>{\rangle}

\def\tr{{\rm tr}}

\newcommand{\polylog}{{\rm polylog}}

\newcommand{\gs}{\Omega}

\newcommand{\mR}{{\mathcal{R}}}
\newcommand{\mE}{{\mathcal{E}}}
\newcommand{\mN}{{\mathcal{N}}}

\newcommand{\Gs}{\Omega}

\def\tr{{\rm tr}}

\newtheorem{theorem}{Theorem}
\newtheorem{subtheorem}{Subtheorem}
\newtheorem{lemma}{Lemma}
\newtheorem{corol}[lemma]{Corollary}
 
\newtheorem{definition}{Definition}   
\newtheorem{prop}[subtheorem]{Proposition}

\newcommand{\br}[1]{\left( #1 \right)}
\newcommand{\brr}[1]{\left[ #1 \right]}
\newcommand{\brrr}[1]{\left\{ #1 \right\}}
\newcommand{\norm}[1]{\left \|  #1 \right \|}

\newcommand{\abs}[1]{\left| #1 \right|}

\makeatletter

\newcommand{\supponecolumnfootnotes}{
	\onecolumn@grid@setup
	\let\set@footnotewidth\set@footnotewidth@one
	\let\compose@footnotes\compose@footnotes@one
}

\newcommand{\startsupplementarycontents}{
	\let\mainaddcontentsline\addcontentsline
	\renewcommand{\addcontentsline}[3]{
		\def\tempa{##1}
		\def\tempb{toc}
		\ifx\tempa\tempb
		\mainaddcontentsline{stoc}{##2}{##3}
		\else
		\mainaddcontentsline{##1}{##2}{##3}
		\fi
	}
}

\newcommand{\supplementarytableofcontents}{
	\begingroup
	\section*{\tocname}
	\@starttoc{stoc}
	\endgroup
}

\makeatother

\usepackage{mathtools}
\def\multiset#1#2{\ensuremath{\left(\kern-.3em\left(\genfrac{}{}{0pt}{}{#1}{#2}\right)\kern-.3em\right)}}

\renewcommand\thefootnote{*\arabic{footnote}} 

\begin{document}

\title{Quantum complexity and generalized area law in fully connected models}

\author{Donghoon Kim$^1$}
\email{donghoon.kim@riken.jp}

\author{Tomotaka Kuwahara$^{1,2,3}$}
\email{tomotaka.kuwahara@riken.jp}

\affiliation{$^{1}$
Analytical quantum complexity RIKEN Hakubi Research Team, RIKEN Center for Quantum Computing (RQC), Wako, Saitama 351-0198, Japan
}

\affiliation{$^{2}$
	RIKEN Cluster for Pioneering Research (CPR), Wako, Saitama 351-0198, Japan
}

\affiliation{$^{3}$
PRESTO, Japan Science and Technology (JST), Kawaguchi, Saitama 332-0012, Japan}

\begin{abstract}

The area law for entanglement entropy captures a fundamental constraint on the complexity of quantum many-body ground states and enables their efficient description.
While the area law is rigorously established in one dimension, its status in higher-dimensional local systems remains unresolved, and it does not hold in general for geometrically non-local systems.
Here, we establish a generalized area law for gapped ground states of fully connected Hamiltonians.
We show that the bipartite entanglement entropy grows at most logarithmically with system size despite the absence of geometric locality. 
In the ground state, each site is only weakly entangled with the rest, while configurations with extensive local fluctuations are strongly suppressed, effectively restricting the accessible Hilbert space.
As a consequence, the ground state admits a matrix product state approximation with polynomial bond dimension with respect to the system size at fixed accuracy.
In the permutation-invariant setting, we further prove a constant entanglement bound and demonstrate that the gapped ground state can be computed in time polylogarithmic in the system size.
These results show that low entanglement complexity need not rely on geometric locality, broadening the conceptual and computational scope of area laws.

\end{abstract}

\maketitle

Understanding the ground states of local Hamiltonians and developing efficient representations for them are central challenges in quantum many-body physics and quantum information theory. Quantum entanglement lies at the core of this framework, serving as a key indicator of the intrinsic complexity of quantum states~\cite{vidal2003entanglement,calabrese2004entanglement,amico2008entanglement,horodecki2009quantum,laflorencie2016quantum}.  In many-body systems, low entanglement often means that the effective Hilbert space needed to describe a quantum state is far smaller than the whole, exponentially large space. This insight has underpinned the success of tensor network methods, offering a practical and efficient approach to representing quantum states and thereby advancing our understanding~\cite{RevModPhys.77.259,schollwock2011density,verstraete2023density}.

A central principle of this analysis is the entanglement area law, which posits that for the ground state $\ket{\gs}$ of a gapped local Hamiltonian, the entanglement entropy between two subsystems $A$ and $B$ scales with the boundary separating them, rather than with the size of the entire system~\cite{RevModPhys.82.277}.  The entanglement entropy is defined as the von Neumann entropy of the reduced density matrix $\rho_{A} = \tr_{B} \br{\ket{\gs} \bra{\gs}}$, given by $S_{A,B}(\ket{\gs}) = - \tr(\rho_A \log \rho_A)$. Significant progress has been made in understanding this law, particularly for one-dimensional (1D) systems. A major breakthrough was Hastings' proof of the area law for 1D gapped systems, utilizing the Lieb-Robinson bound~\cite{Hastings_2007}. This result was further refined by Arad and collaborators~\cite{PhysRevB.85.195145,arad2013area} and extended by Brand\~ao and Horodecki, who proved the area law under the assumption of exponentially decaying correlations~\cite{brandao2013area,Brandao2015}. 
Within this progression, the Approximate Ground State Projection (AGSP) formalism emerged as a leading method, enabling the development of a polynomial-time algorithm for finding ground states~\cite{landau2015polynomial,Arad2017}.
Of the various dimensions, 1D systems stand out as the most well-understood example, highlighting the robustness of this theoretical and computational framework.

Extending the area law beyond 1D is crucial for understanding the complexity of a broader class of quantum states, yet it remains a significant challenge. Despite progress---such as extensions to tree graphs in dimensions below two~\cite{abrahamsen2019polynomial}, systems with long-range interactions~\cite{Kuwahara2020arealaw,kim2024boson}, and advances in thermal equilibrium states~\cite{PhysRevLett.100.070502,PhysRevX.11.011047}---the field is still largely underexplored. A recent advancement demonstrates the area law for two-dimensional frustration-free systems, assuming a local spectral gap~\cite{10.1145/3519935.3519962}, but these findings remain limited compared to the substantial achievements in 1D systems.

\begin{figure}[ttt]
	\centering
	\includegraphics[width=0.45\textwidth]{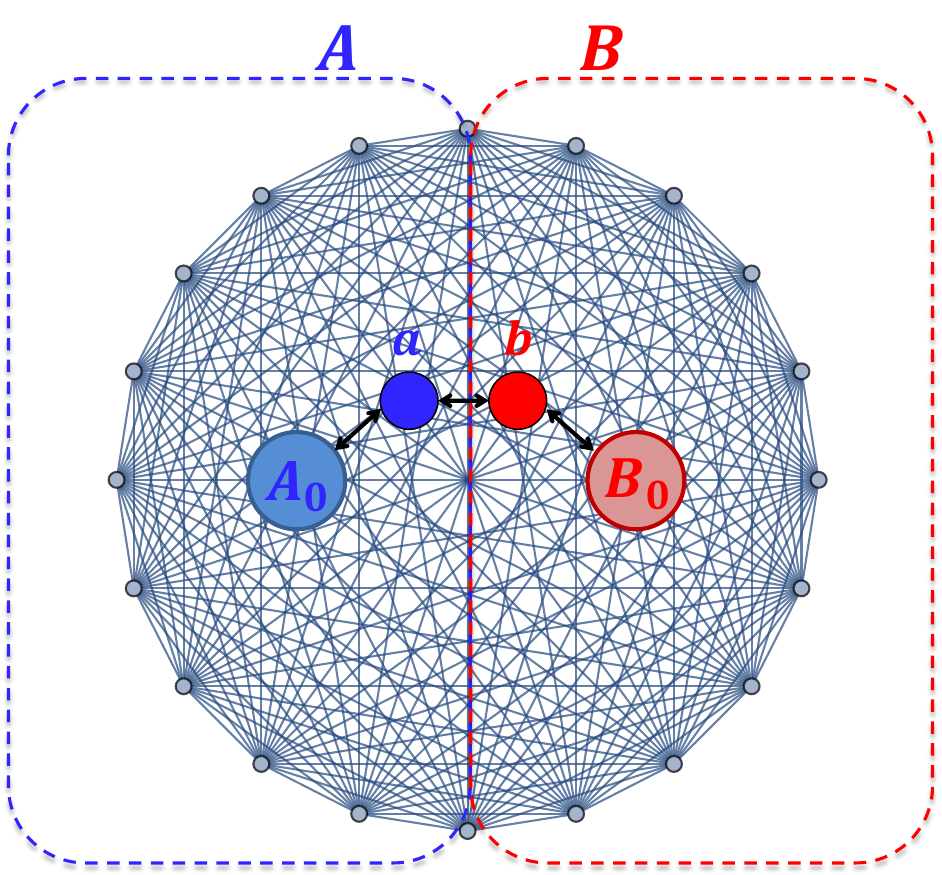}
	\caption{
		Schematic representation of the generalized area law on a fully connected system.
		The system is partitioned into two subsystems, $A$ and $B$. 
		Even though the interaction graph is fully connected, the ground state exhibits an emergent local product structure under the bounded local energy scale and spectral gap assumptions.
		As a result, the entanglement across the boundary is effectively governed by a limited set of degrees of freedom (denoted by $a$ and $b$ in the figure).
		This reflects the generalized area law, which states that the entanglement entropy between $A$ and $B$ is upper-bounded by a constant (up to logarithmic corrections).
	}
	\label{fig:generalized_area_law}
\end{figure}

A particularly intriguing extension involves fully connected systems, or all-to-all interacting systems, where each component interacts with every other component. 
In these geometrically non-local systems, spatial dimensions lose relevance, representing an extreme case in higher-dimensional physics.
Such systems are crucial in condensed matter physics, statistical mechanics, high-energy physics, quantum chemistry, etc.
Key examples include the Sherrington-Kirkpatrick (SK) model for spin glasses~\cite{Sherrington1975}, the Lipkin-Meshkov-Glick (LMG) model for nuclear structure~\cite{Lipkin1965,PhysRevA.71.064101}, lattice gauge theories reformulated as spin systems for quantum simulation~\cite{Kogut1975}, and the Sachdev-Ye-Kitaev (SYK) model, which is central to black hole physics~\cite{Sachdev1993,Kitaev2015}.
Recent advances in quantum simulators, such as trapped-ion systems~\cite{Wineland2008,Monroe2013,Blatt2012,doi:10.1126/sciadv.1500838,Puri2017,doi:10.1126/sciadv.1602273} and superconducting spin qubits~\cite{pita2025blueprint}, have also enabled the implementation of all-to-all interactions, highlighting their potential for solving complex problems via quantum algorithms.

However, the highly nonlocal nature of fully connected systems complicates the analysis of ground-state entanglement. Even among paradigmatic fully connected spin models, the entanglement structure can vary widely. For example, in the (integrable) LMG model, the bipartite entanglement entropy is known to scale at most logarithmically with system size~\cite{PhysRevA.71.064101}.
Moreover, recent numerical studies suggest that density-matrix renormalization group algorithms can perform well for certain fully connected models under suitable conditions~\cite{tindall2022quantum}, hinting at a reduced effective complexity in some regimes.
On the other hand, the SYK models typically exhibit a volume law of entanglement for generic eigenstates~\cite{PhysRevB.94.035135,PhysRevB.97.245126}. 
Even for the ground state, numerical evidence indicates the volume law throughout the parameter regimes explored for the transverse-field quantum SK spin glass~\cite{PhysRevLett.129.220401}. This diversity indicates that all-to-all connectivity alone does not determine entanglement complexity; additional structural conditions must be taken into account. Consequently, determining when an area law holds and establishing rigorous bounds on entanglement in these systems remain essential yet challenging problems.

In this work, we establish a generalized area law for the ground states of gapped Hamiltonians defined on fully connected systems.
We prove that the entanglement entropy across any bipartition grows at most logarithmically with the system size, and that the ground state admits an efficient Matrix Product State (MPS) representation with polynomial bond dimension for any fixed approximation error.
In the permutation-invariant setting, we further obtain a constant entanglement bound and a polylogarithmic-time classical algorithm for approximating the ground-state energy and constructing a corresponding approximate ground-state vector.
In other words, although any bipartition on a fully connected graph has an extensive geometric ``boundary,'' the entanglement across the cut is effectively controlled by a limited set of boundary-like degrees of freedom, yielding at most logarithmic growth.
This does not conflict with the counterexample to the generalized area law constructed in Ref.~\cite{aharonov2014local}.
That construction uses bounded-strength local terms but allows the number of interactions incident on a single site, and consequently the total incident interaction strength, to grow with the system size.
By contrast, the all-to-all interactions considered here are uniformly diluted via a Kac-type normalization, so that the total interaction strength jointly involving any fixed pair of sites decreases with the system size, while the total local energy scale remains bounded.
This uniform dilution is a key structural input in our proof and excludes the mechanism underlying the counterexample.
In the context of the long-standing quest to resolve the area law conjecture beyond one dimension, our results establish a provable area-law–type bound in an extreme infinite-dimensional (mean-field) setting despite the absence of geometric locality.

Our setup departs in essential respects from prior approaches to area laws. Traditional techniques---including the AGSP formalism~\cite{PhysRevB.85.195145,arad2013area,Kuwahara2020arealaw,10.1145/3519935.3519962}, arguments based on assumptions about the density of eigenstates~\cite{hastings2007entropy,PhysRevA.80.052104,brandao2015entanglement}, and quasi-adiabatic continuation~\cite{PhysRevB.72.045141,PhysRevLett.111.170501}---do not directly apply to fully connected systems. In particular, AGSP-based proofs exploit geometric separation from the entanglement cut, truncating Hamiltonian terms supported far from the cut to control the growth of Schmidt rank. Such strategies break down in fully connected settings, where every site interacts with every other site, and the Schmidt rank of the AGSP cannot be efficiently controlled.

Our approach is based on a quantitative trade-off inequality~\cite{Kuwahara_2016_asymptotic,Kuwahara_2017}: a nonvanishing spectral gap implies suppressed fluctuations of $k$-local observables. The only structural assumption we impose is a bounded energy scale per site (cf.~Eq.~\eqref{bar_g_extensive}), which is ensured, for instance, by $(1/n)$-scaled pairwise couplings (cf.~Eq.~\eqref{ext_cond2}). We then relate closeness to a product structure to a robustness property of the ground state under suitable local unitary transformations, and combine this robustness with the trade-off inequality. We iterate this robustness–trade-off step in a bootstrap manner. This yields weak, site-wise entanglement and ultimately a logarithmic upper bound on the ground-state entanglement. Finally, we refine the analysis by proving a concentration bound. Specifically, the ground state has negligible weight on configurations with large collective deviations from mean-field–like behavior. This concentration enables a dimension-reduction step and leads to an efficient approximation of the ground state within an effectively low-dimensional subspace.

\section*{Results}
\subsection*{Setup}
We consider a quantum system composed of $n$ qudits, each with a local dimension $d$. 
Let $\Lambda$ denote the set of all sites, with $|\Lambda| = n$. 
The system is described by a general $k$-local Hamiltonian of the form 
\begin{linenomath}
	\begin{equation} 
		H = \sum_{Z \subset \Lambda : |Z| \leq k} h_{Z}, \label{k_local}
	\end{equation}
\end{linenomath}
where each term $h_{Z}$ acts nontrivially on at most $k$ sites. 
We assume that $H$ has a nondegenerate ground state $\ket{\gs}$ and a constant spectral gap $\Delta$ which does not depend on the system size $n$.
Our primary focus is on the entanglement entropy $S_{A,B}(\ket{\gs})$ of the ground state $\ket{\gs}$.

We assume that the local interaction strengths satisfy
\begin{linenomath}
	\begin{equation} 
		\norm{h_{\{i\}}} \le g_0, \qquad \forall i \in \Lambda, \label{ext_cond1}
	\end{equation}
\end{linenomath}
\begin{linenomath}
	\begin{equation} 
		\sum_{Z:Z \ni i,i'} \norm{h_Z} \le \frac{g_1}{n}, \qquad \forall i, i' \in \Lambda, \label{ext_cond2}
	\end{equation}
\end{linenomath}
where $i\neq i'$. 
This condition ensures that the energy associated with a single site (i.e., the local energy) remains finitely bounded.
Indeed, for every $i \in \Lambda$, writing $i^{\co} := \Lambda \setminus \{i\}$ for the complement of $i$, we have
\begin{linenomath}
	\begin{equation} 
		\label{bar_g_extensive}
		\sum_{Z:Z \ni i} \norm{h_Z} \le  g_0+g_1 := \bar g  .   
	\end{equation}
\end{linenomath}

As simple examples covered by the above framework, consider first a non-uniform Lipkin--Meshkov--Glick (LMG) model, defined as follows:
\begin{linenomath}
	\begin{equation}
		H = \sum_{i<i'}\frac{1}{n} (J_{i,i'}^x \sigma_i^x \sigma_{i'}^x +  J_{i,i'}^y \sigma_i^y \sigma_{i'}^y)  ,
	\end{equation}
\end{linenomath}
where each of $J_{i,i'}^x$ and $J_{i,i'}^y$ is bounded by an $\orderof{1}$ constant. 
More generally, any all-to-all $k$-local Hamiltonian with $n^{-k+1}$ Kac scaling, 
\begin{linenomath}
	\begin{equation}
		H =  \sum_{1 \leq i_1 <i_2 \cdots <i_k \leq n}  \frac{1}{n^{k-1}} h_{i_1,i_2,\ldots, i_k}, 
	\end{equation}
\end{linenomath}
also fits into the present setting, where $\norm{ h_{i_1,i_2,\ldots, i_k}}=\orderof{1}$. 
The above class of models encompasses a wide range of physically important systems with fully connected (infinite-range) interactions and a Kac-type normalization. 
We emphasize that our assumptions still allow random couplings as long as the Kac scaling is $n^{-k+1}$, such as $\{\pm1\}$ couplings with a biased distribution.

As a final remark, the condition~\eqref{bar_g_extensive}  is stronger than mere extensivity of the total energy, since extensivity may still arise from cancellations among interaction terms. 
In the present work, the bounded local energy scale is the essential structural assumption. It guarantees that a single-site perturbation changes the energy only by 
$\orderof{1}$, which is a key input to our argument.
Models in which this local energy scale grows with system size---such as certain disordered all-to-all models---fall outside the present framework.

\subsection*{Logarithmic Bound on Bipartite Entanglement}

A key structural feature of the ground state is that each site is only weakly entangled with the rest of the system.
Consider a single site $i \in \Lambda$ and its complement $i^{\co} := \Lambda \setminus \{i\}$.
For the bipartition $i | i^{\co}$, let $\ket{\gs} = \sum_{s = 0}^{d-1} \lambda_{s} \ket{s}_{i} \ket{\phi_{s}}_{i^{\co}}$ be the Schmidt decomposition with $\lambda_{0} \geq \lambda_{1} \geq \cdots \geq \lambda_{d-1} \geq 0$.
We show (see Methods for a proof sketch) that the leading Schmidt weight is close to unity, $\lambda_{0}^{2} = 1 - \mathcal{O}((\bar{g} / \Delta)^{3}n^{-1})$; equivalently, the Schmidt weight outside the leading component satisfies
\begin{linenomath}
	\begin{equation}
		\sum_{s = 1}^{d-1} \lambda_{s}^{2} = \mathcal{O}\br{\br{\frac{\bar{g}}{\Delta}}^{3} \frac{1}{n}}, \label{Eq_Schmidt}
	\end{equation}
\end{linenomath}
where $\Delta$ is the spectral gap and $\bar{g} = g_{0} + g_{1}$.
Consequently, each single-site reduced state is nearly pure.

Throughout this paper, we refer to the dominant local Schmidt vector $\ket{0}_i$ for the bipartition $i \mid i^{\mathrm c}$ as the mean-field state at site $i$, since the exact ground state is locally concentrated on this preferred local state.

The local product structure implied by Eq.~\eqref{Eq_Schmidt} directly enforces a logarithmic upper bound on the bipartite entanglement entropy.
For a pure state $\ket{\gs}$ and a bipartition $\Lambda = A \cup B$, let $\rho_{A} = \tr_{B}(\ket{\gs}\bra{\gs})$ be the reduced state on $A$. 
The entanglement entropy between $A$ and $B$ is then given by $S(\rho_{A})$, where $S(\rho) = - \tr(\rho \log \rho)$ is the von Neumann entropy; we write this as $S_{A,B}(\ket{\gs})$.
By Eq.~\eqref{Eq_Schmidt}, each single-site reduced state $\rho_{i}$ has a dominant eigenvalue, and the total weight of the remaining spectrum satisfies $r_{i} := \sum_{s \geq 1} \lambda_{s}^{2} = \mathcal{O}((\bar{g} / \Delta)^{3} n^{-1})$.
Since the local Hilbert space dimension is $d$, the von Neumann entropy of such a nearly pure state is bounded by a constant multiple of $r_{i} \log(1 / r_{i})$.
Combining this with the above estimate on $r_{i}$, we obtain
\begin{linenomath}
	\begin{equation}
		S(\rho_{i}) \leq C \br{\frac{\bar{g}}{\Delta}}^{3} \frac{1}{n} \log n, \label{Eq_single_site_entropy}
	\end{equation}
\end{linenomath}
for some positive constant $C$ depending only on $k$ and $d$.
For an arbitrary bipartition $\Lambda = A \cup B$, subadditivity of the von Neumann entropy gives $S(\rho_{A}) \leq \sum_{i \in A} S(\rho_{i})$.
Using Eq.~\eqref{Eq_single_site_entropy} and the trivial bound $|A| \leq n$, this yields 
\begin{linenomath}
	\begin{equation}
		S_{A,B}(\ket{\gs}) \leq C \br{\frac{\bar{g}}{\Delta}}^{3} \log n , \label{Eq_bipartite_entropy}
	\end{equation}
\end{linenomath} 
so that, for a nonvanishing spectral gap $\Delta = \Omega(1)$, the entanglement entropy grows at most logarithmically with system size.

Geometric locality is therefore not essential for suppressing extensive entanglement.
Under the present assumptions, even fully connected systems with all-to-all interactions and no geometric structure exhibit at most logarithmic growth of ground-state entanglement entropy, thereby excluding volume-law scaling in this setting.

The logarithmic form of Eq.~\eqref{Eq_bipartite_entropy} corresponds to the nonvanishing-gap regime $\Delta=\Omega(1)$, which is our standing assumption. The finite-size estimate itself retains its explicit dependence on $\Delta$ and holds provided that $n\geq n_0(\Delta)=\mathcal{O}(\Delta^{-3})$. Hence, when $\Delta=\Delta_n\gtrsim n^{-a}$ with $0\leq a<1/3$, Eq.~\eqref{Eq_bipartite_entropy} yields $S_{A,B}(|\Omega\rangle)=\mathcal{O}(n^{3a}\log n)$, which is no longer logarithmic but remains subvolume.

The same emergent local product structure also constrains connected correlations.
More precisely, for observables $O_{X}$ and $O_{Y}$ supported on disjoint regions $X$ and $Y$, respectively, with $\norm{O_{X}}, \norm{O_{Y}} \leq 1$, Eq.~\eqref{Eq_Schmidt} implies that the connected correlation $\abs{\bra{\gs} O_{X} O_{Y} \ket{\gs} - \bra{\gs} O_{X} \ket{\gs} \bra{\gs} O_{Y} \ket{\gs}}$ is upper-bounded by $\mathcal{O}\br{(\bar{g} / \Delta)^{3/2} \sqrt{\min\{|X|,|Y|\} / n}}$.
For a nonvanishing gap, this bound therefore tends to zero as $n \rightarrow \infty$ whenever at least one of the two supports has size $o(n)$.
This suppression is qualitatively different from the clustering results of Hastings and Koma~\cite{ref:Hastings2006-ExpDec}.
Their bounds are controlled by the spatial separation between the observable supports, under suitable spatial-decay assumptions on the interactions.
On a fully connected graph, however, any two nonempty disjoint regions have graph distance one, so there is no nontrivial large-distance regime.
The present bound is instead controlled by the sizes of the supports relative to the total system size and follows from the emergent local product structure rather than from geometric separation.

\subsection*{Efficient MPS Approximation}

Bounds on the entanglement entropy alone are not sufficient to quantitatively characterize the complexity of the ground state.
The bound obtained in Eq.~\eqref{Eq_bipartite_entropy} relies on Eq.~\eqref{Eq_Schmidt}, which ensures that each site deviates from the mean-field state with probability at most $\orderof{1/n}$.
However, a stronger statement can be made. Rather than bounding the deviation probability at each site individually, we analyze the probability distribution of the total number of sites that deviate from the mean-field state and establish a concentration bound for this quantity.
This concentration result provides a more refined characterization of the fluctuations around the mean-field configuration. Consequently, it offers a sharper and more informative measure of the ground-state complexity than what can be inferred from entanglement entropy bounds alone.

For each site $i$, let $P_{i} = \ket{0} \bra{0}_{i}$ denote the projector onto the dominant local Schmidt component in the decomposition across $i | i^{\co}$, and define $Q_{i} = 1 - P_{i}$, which projects onto the orthogonal subspace to this leading local component.
We refer to such departures from the dominant configuration as local deviations.
The operator $M = \sum_{i \in \Lambda} Q_{i}$ then counts the total number of such local deviations.
We prove that there exist constants $C_{1},C_{2} = \mathcal{O}(1)$, depending only on $k$, $d$, $\bar{g}$, and $\Delta$, such that
\begin{linenomath} 
	\begin{equation}
		\norm{\Pi_{>z} \ket{\gs}} \leq C_{1} e^{-C_{2} \sqrt{z}}, \label{Eq_local_excitation}
	\end{equation}
\end{linenomath}
where $\Pi_{> z}$ projects onto the sector with more than $z$ local deviations (see Supplementary Theorem 2).
Thus, the weight of sectors with many local deviations decays in a stretched-exponential manner in $z$.

As a consequence, for any target accuracy $\epsilon > 0$, choosing $z = \mathcal{O}(\log^{2}(1 / \epsilon))$ ensures that $\norm{\Pi_{>z} \ket{\gs}} \leq \epsilon$.
The subspace spanned by configurations with at most $z$ local deviations has dimension bounded by $\sum_{m \leq z} \binom{n}{m} (d-1)^{m}$, which scales as $n^{\mathcal{O}(z)}$.
After arranging the sites in an arbitrary one-dimensional order, this bound controls the Schmidt rank across every cut.
Therefore, there exists a matrix product state $\ket{\gs_{D}}$ supported on this subspace, with bond dimension
\begin{linenomath} 
	\begin{equation}
		D = n^{\mathcal{O}(\log^{2}(1 / \epsilon))}, \label{Eq_MPS_bond}
	\end{equation}
\end{linenomath}
such that $\norm{\ket{\gs} - \ket{\gs_{D}}} \leq \epsilon$.
For any constant precision $\epsilon$, the bond dimension scales only polynomially with the system size $n$.

\subsection*{Permutation Invariant Systems}

As a more practically relevant class of systems, we consider permutation-invariant systems.
In this case, the Hamiltonian is invariant under arbitrary relabelings of the sites. 
Therefore, any nondegenerate ground state must be fully symmetric under permutations for $n>d$. 
This symmetry restricts the ground state to the symmetric subspace of dimension 
$\binom{n + d - 1}{d - 1} \propto n^{d-1}$ for fixed local dimension $d$. 
Consequently, by restricting $H$ to this symmetric sector and diagonalizing the resulting matrix, the ground state can be numerically obtained to arbitrary precision in polynomial time in $n$ 
(e.g., using standard dense-matrix algorithms). 
Here, the entanglement entropy of the general (including gapless) ground state is $\mathcal{O}(\log n)$~\cite{PhysRevA.71.064101}.

Building on the concentration bound established in Eq.~\eqref{Eq_local_excitation}, we improve this trivial upper bound and prove that the entanglement entropy of the gapped ground state is $\orderof{1}$. 
Furthermore, the time complexity of obtaining a $1/\poly(n)$-accurate ground-state energy, together with a corresponding approximate ground-state vector, can be reduced from $\mathrm{poly}(n)$ to $\mathrm{polylog}(n)$.

First, projecting onto configurations with at most $z$ local deviations produces a symmetric state whose Schmidt rank across any bipartition is at most $\binom{z + d - 1}{d-1}$, and whose entropy is therefore logarithmic in $z$.
Using the decay $\norm{\Pi_{>z} \ket{\gs}} \leq C_{1} e^{-C_{2} \sqrt{z}}$ to control the Schmidt tail then yields
\begin{linenomath} 
	\begin{equation}
		S_{A,B}(\ket{\gs}) \leq \mathfrak{C}, \label{Eq_constant_bound}
	\end{equation}
\end{linenomath}
where $\mathfrak{C}$ depends only on $d,k,\bar{g}$ and $\Delta$ (through $C_{1},C_{2}$), and is independent of $n$ and of the choice of bipartition (see Supplementary Theorem 3).
This constant entanglement behavior is consistent with numerical observations~\cite{PhysRevA.71.064101} for the gapped Lipkin--Meshkov--Glick model (see also Supplementary Notes 6), where the entanglement entropy saturates to an $\mathcal{O}(1)$ value as the system size increases.

Second, even for permutation-invariant systems, computing the ground state 
involves the nontrivial task of identifying the appropriate mean-field state. 
To address this issue, we treat the mean-field parameters as variational variables 
and expand the ground-state energy in generalized Dicke states 
up to excitation number $z$, which can be chosen as 
$z=\mathcal{O}(\log^2(1/\epsilon))$. 
We show that this procedure reduces the problem to optimizing 
a polynomial of degree at most $\mathcal{O}(z^{d-1})$. 
As a consequence, an $\epsilon$-approximation of the ground energy
can be obtained with computational cost as follows (see the Methods section): 
\begin{linenomath} 
	\begin{equation}
		\orderof{ \mathrm{polylog}(1/\epsilon) \log(n/\epsilon)  }
	\end{equation}
\end{linenomath}
In particular, for $\epsilon=1/\poly(n)$, we obtain a poly-logarithmic time complexity with respect to the system size $n$.

\section*{Discussion}

We compare our findings with previous studies, highlighting both consistencies and discrepancies. In earlier work~\cite{brandao2013product,Kuwahara_2017}, the approximation of ground energy using product states was investigated. More recently, the partition function of the 2-local Hamiltonian on a dense graph was studied~\cite{bravyi2022quantum}, where it was shown that the ground energy can be computed classically in polynomial time if an error of $\Omega(n)$ is allowed. Our work advances these results by focusing on the ground state itself and approximating it closely as a low-entanglement state, thereby revealing its detailed structure to arbitrary $1/\poly(n)$ precision.

Another relevant comparison is with dynamical studies of Kac-scaled all-to-all systems.
For the class of time-dependent open systems with infinite-range interactions considered in Ref.~\cite{carollo2024applicability}, and for initial states whose macroscopic fluctuation error vanishes as $n \rightarrow \infty$, the expectation values of macroscopic observables converge to the solution of the corresponding mean-field equations at every fixed finite time.
This dynamical result and ours are complementary: the former concerns the thermodynamic-limit evolution of macroscopic observables, whereas our static result uses a spectral gap to constrain local quantum fluctuations and ground-state entanglement.

Beyond the mean-field evolution of macroscopic observables, all-to-all systems also exhibit distinctive constraints on information propagation and entanglement generation.
Under their finite-energy-per-site norm assumption and product-state scrambling criterion, Ref.~\cite{lashkari2013towards} used Lieb--Robinson-type bounds to control a mean-field approximation and obtained a logarithmic lower bound on the scrambling time of dense two-body systems.
For spin-$1/2$ two-body models, stronger bounds have been obtained for infinite-temperature operator-growth diagnostics; in particular, the Kac-normalized case gives a power-law lower bound~\cite{yin2020bound}.
Separately, programmable time-dependent all-to-all Hamiltonians with order-one pairwise interaction strengths can approximately prepare globally entangled states such as GHZ and W states in time $n^{-1 + \delta}$, with inverse-polynomial error, where $\delta > 0$ is an arbitrarily small fixed constant; after an overall $1/n$ rescaling, the same protocol takes time $n^{\delta}$~\cite{yin2025fast}.
These dynamical results are compatible with our static ground-state bounds: the ability of all-to-all dynamics to generate global entanglement rapidly does not, by itself, imply large bipartite entanglement in a nondegenerate gapped ground state.

Our results can also be assessed from the perspective of the Local Hamiltonian Problem (LHP). The LHP asks whether the ground-state energy density lies within a specified energy window. 
Our findings imply that, for any constant precision $\epsilon$, the LHP for gapped fully connected models lies in NP. This follows from the fact that a polynomial-bond-dimension MPS provides a classical witness whose energy density can be verified in polynomial time.
For inverse-polynomial precision, i.e., $\epsilon = 1/\poly(n)$, our bound gives a quasi-polynomial bond dimension, $D=n^{\orderof{\log^2 n}}$, so NP-membership does not follow from the present argument. Clarifying the precise complexity at inverse-polynomial precision remains an interesting open problem.

Regarding the scope of our results, our framework assumes a uniformly bounded local energy scale per site, as in~\eqref{bar_g_extensive}. This assumption is stronger than mere extensivity of the total energy and is essential to our proof, since it ensures that a single-site perturbation changes the energy only by $\orderof{1}$. Disordered all-to-all models such as the SK and SYK models typically fall outside this regime, because their local energy scale grows with system size. In this sense, they belong to a different class from the one treated here. Consistent with this distinction, available numerical evidence for the transverse-field quantum SK model suggests volume-law-like ground-state entanglement even away from criticality~\cite{PhysRevLett.129.220401}. Whether such behavior persists for the exact ground state in the thermodynamic limit remains an interesting open question.

While our logarithmic entanglement bound does not require permutation invariance, existing studies of ground-state entanglement in gapped all-to-all systems have largely focused on permutation-invariant collective models, which fall under our stronger constant bound.
It is therefore important to identify physically motivated non-permutation-invariant models for which the general logarithmic bound provides a genuinely new constraint.
One possible example is the finite-pattern transverse-field quantum Hopfield model: generic quenched patterns break full permutation invariance, while the pairwise interactions retain the required $1 / n$ scaling, and large-system spectral analyses identify a gapped high-field paramagnetic phase away from criticality~\cite{nishimori1996quantum,xie2024quantum}.
Nevertheless, this model still retains a residual collective structure.
Identifying broader classes of non-permutation-invariant gapped all-to-all models and determining whether the logarithmic entanglement is asymptotically tight remain interesting open problems.

In closing, while the present work focuses on fully connected models, extending our framework to strongly long-range interacting systems with Kac normalization constitutes an intriguing direction for future research~\cite{RevModPhys.95.035002}.
In particular, systems with an explicit spatial geometry (e.g., one-dimensional models) provide a natural testing ground.
Indeed, our analysis suggests that the same approach may also yield subvolume-law entanglement bounds in gapped phases under suitable assumptions.
However, it remains an open question whether, and in what sense, Kac normalization restores an area-law scaling in finite-dimensional settings.
Developing tools that combine the well-established methods for finite-dimensional local systems~\cite{PhysRevB.85.195145,arad2013area,Kuwahara2020arealaw,10.1145/3519935.3519962} with the techniques introduced here may offer a promising route toward a systematic understanding of these strongly interacting long-range systems, which remain a largely unexplored frontier.

\section*{methods}

\subsection*{Robustness and Emergent Local Product Structure}
In this section, we establish a robustness property of the ground state under local unitary transformations and derive a corresponding constraint on the single-site Schmidt spectrum.
We sketch the main ideas of the proof, while the full operator-norm bookkeeping and explicit constructions are provided in Supplementary Notes~2 and~3.

Fix a single site $i \in \Lambda$ and let $\rho_{i} = \tr_{i^{\co}}(\ket{\gs} \bra{\gs})$ be the reduced density matrix on site $i$, where $i^\co$ is the complement of $i$ (i.e., $i^\co = \Lambda\setminus \{i\}$).
Consider any unitary $u_{i}$ supported on site $i$ such that $u_{i} \rho_{i}u_{i}^{\dagger} = \rho_{i}$.
We show below that such a transformation can change the ground-state energy only by $\mathcal{O}(1/n)$:
\begin{linenomath}
	\begin{equation}
		\begin{aligned}
			&\abs{\bra{\gs} u_{i}^{\dagger} H u_{i} \ket{\gs} - \bra{\gs} H \ket{\gs}} \leq \delta_{\mathrm{Rob}} ,  \\
			&\delta_{\mathrm{Rob}}:=\br{\frac{12 \gamma g_{1} d^{2}}{\Delta}}^{2} \frac{2 \bar{g}}{n}, \label{Eq_robustness}
		\end{aligned}
	\end{equation}
\end{linenomath}
where $\Delta$ is the spectral gap, $\gamma = 6k^{2}$ and $\bar{g} = g_{0} + g_{1}$ (see Supplementary Theorem 1).
This robustness bound has a direct structural implication for the bipartition $i | i^{\co}$.
Let $\ket{\gs} = \sum_{s = 0}^{d-1} \lambda_{s} \ket{s}_{i} \otimes \ket{\phi_{s}}_{i^{\co}}$ be the Schmidt decomposition across $i | i^{\co}$, with $\lambda_{0} \geq \lambda_{1} \geq \cdots \geq \lambda_{d-1} \geq 0$.
The Schmidt coefficients then obey
\begin{linenomath}
	\begin{equation}
		\lambda_{0}^{2} \geq 1 - \frac{2\delta_{\mathrm{Rob}}}{\Delta}, \qquad \sum_{s = 1}^{d-1} \lambda_{s}^{2} \leq  \frac{2\delta_{\mathrm{Rob}}}{\Delta}, \label{Eq_Schmidt_ineq}
	\end{equation}
\end{linenomath}
which is given by Supplementary Proposition~1. 
Consequently, each site is only weakly entangled with its complement.
For a constant spectral gap, the Schmidt weight outside the leading component therefore scales as $\mathcal{O}(1/n)$.

\subsection*{Proof Sketch of Inequality~\eqref{Eq_robustness}}

We now sketch the derivation of the robustness bound~\eqref{Eq_robustness}.
We isolate all interaction terms that act nontrivially on both $i$ and its complement $i^{\co}$:
\begin{linenomath} 
	\begin{equation}
		H_{i,i^{\co}} := \sum_{Z: Z \cap \{i\} \neq \emptyset, Z \cap i^{\co} \neq \emptyset} h_{Z}. \label{M_Hiico_def}
	\end{equation}
\end{linenomath}
We expand $H_{i,i^{\co}}$ in an on-site Hermitian operator basis $\{\mathcal{X}_{i,v}\}_{v = 1}^{d^{2}}$ with $\norm{\mathcal{X}_{i,v}} = 1$ (see Supplementary Note 3 B for an explicit construction):
\begin{linenomath}
	\begin{equation}
		H_{i,i^{\co}} =  \frac{1}{n} \sum_{v=1}^{d^2}  \mathcal{X}_{i,v} \otimes H_{i^{\co},v},  \label{M_Hiico_decomp}
	\end{equation}
\end{linenomath}
where each $H_{i^{\co},v} := \sum_{Z_{1} \subset i^{\co} : |Z_{1}| \leq k-1} h_{v,Z_{1}}$ is $(k-1)$-local on $i^{\co}$.
Using the interaction-strength assumption~\eqref{ext_cond2}, each $H_{i^{\co},v}$ inherits a bound similar to~\eqref{bar_g_extensive}:
\begin{linenomath}
	\begin{equation}
		\sum_{Z_1: Z_1\ni i'} \norm{h_{v,Z_1}} \leq g_1, \label{M_g_1_extensive}
	\end{equation}
\end{linenomath}
for any $v$ and $i' \in i^\co$. 
The decomposition~\eqref{M_Hiico_decomp} is not unique (it depends on the chosen basis), but any such choice suffices for the arguments below.

Let $u_{i}$ be any unitary operator acting only on site $i$ such that it preserves the single-site reduced state, $u_{i} \rho_{i}u_{i}^{\dagger} = \rho_{i}$ for $\rho_{i} := \tr_{i^{\co}}(\ket{\gs} \bra{\gs})$.
Since $u_{i}$ acts trivially on $i^{\co}$, only the coupling part $H_{i,i^{\co}}$ can contribute to the energy change:
\begin{linenomath}
	\begin{equation}
		\bra{\gs} u_{i}^{\dagger} H u_{i} - H \ket{\gs} = \bra{\gs} u_{i}^{\dagger} H_{i,i^{\co}} u_{i} - H_{i,i^{\co}} \ket{\gs}. \label{M_reduce_to_Hiico}
	\end{equation}
\end{linenomath}
We now ``center'' the on-site operators with respect to the dominant local Schmidt component $\ket{0}_{i}$ (defined from the Schmidt decomposition across $i|i^{\co}$), and center the complementary operators with respect to the global ground $\ket{\gs}$:
\begin{linenomath}
	\begin{equation}
		\begin{aligned}
			\widetilde{\mathcal{X}}_{i,v} &:= \mathcal{X}_{i,v} - \bra{0} \mathcal{X}_{i,v} \ket{0}_{i}, \\
			\widetilde{H}_{i^{\co},v} &:= H_{i^{\co},v} - \bra{\gs} H_{i^{\co},v} \ket{\gs}. \label{M_centerings}
		\end{aligned}
	\end{equation}
\end{linenomath}

The invariance condition $u_{i} \rho_{i} u_{i}^{\dagger} = \rho_{i}$ implies that for any operator $\Phi_{i}$ supported on $i$, $\bra{\gs} (u_{i}^{\dagger} \Phi_{i} u_{i} - \Phi_{i}) \ket{\gs} = 0$, and similarly for any $\Phi_{i^{\co}}$ supported on $i^{\co}$.
Applying this observation term-by-term to Eq.~\eqref{M_Hiico_decomp} shows that all purely on-site or purely-$i^{\co}$ expectation shifts cancel, and the energy response can be written in terms of the centered operators:
\begin{linenomath}
	\begin{equation}
		\begin{aligned}
			\bra{\gs} u_{i}^{\dagger} H u_{i} - H \ket{\gs} &= \frac{1}{n} \sum_{v = 1}^{d^{2}} \Big(\bra{\gs} u_{i}^{\dagger} (\widetilde{\mathcal{X}}_{i,v} \otimes \widetilde{H}_{i^{\co},v}) u_{i} \ket{\gs} \\
			&\quad- \bra{\gs} \widetilde{\mathcal{X}}_{i,v} \otimes \widetilde{H}_{i^{\co},v} \ket{\gs} \Big). \label{M_energy_centered}
		\end{aligned}
	\end{equation}
\end{linenomath}
Thus, the problem reduces to bounding a sum of centered ground-state correlations.

The operator $\widetilde{H}_{i^{\co},v}$ has zero ground-state expectation by definition, and its action on $\ket{\gs}$ is governed by the variance
\begin{linenomath}
	\begin{equation}
		\begin{aligned}
			\norm{\widetilde{H}_{i^{\co},v} \ket{\gs}}^{2} = \bra{\gs} \widetilde{H}_{i^{\co},v}^{2} \ket{\gs} =: \mathrm{Var}_{\gs} (H_{i^{\co},v}).
		\end{aligned}
	\end{equation}
\end{linenomath}
To bound this variance, we use a standard gap--variance trade-off~\cite{Kuwahara_2016_asymptotic,Kuwahara_2017}: for any $k$-local $1$-extensive observable $A$ supported on a set of size $\abs{L}$ (i.e., $A = \sum_{Z: |Z| \leq k} a_{Z}$ with $\sum_{Z: Z \ni i} \norm{a_{Z}} \leq 1$),  one has $$\mathrm{Var}_{\gs}(A) \cdot \Delta \lesssim \bar{g}|L|$$ up to a $k$-dependent constant.
Applying this relation to $A = H_{i^{\co},v} / g_{1}$ (which is $1$-extensive by Eq.~\eqref{M_g_1_extensive}) and using $\abs{L} \sim n$ yields the fluctuation bound
\begin{linenomath}
	\begin{equation}
		\norm{\widetilde{H}_{i^{\co},v} \ket{\gs}} \lesssim g_{1} \sqrt{\frac{\bar{g}n}{\Delta}}. \label{M_fluctuation_bound}
	\end{equation}
\end{linenomath}
Intuitively, the spectral gap suppresses low-energy fluctuations, forcing any extensive observable to have limited variance in the ground state.

Introduce the rank-one projector $P_{i} := \ket{0} \bra{0}_{i}$ and its complement $Q_{i} = 1 - P_{i}$.
Because $\widetilde{\mathcal{X}}_{i,v}$ is centered in the state $\ket{0}_{i}$, it satisfies $P_{i} \widetilde{\mathcal{X}}_{i,v} P_{i}=0$, and hence only terms involving at least one factor of $Q_{i}$ survive:
\begin{linenomath}
	\begin{equation}
		\widetilde{\mathcal{X}}_{i,v} = P_{i} \widetilde{\mathcal{X}}_{i,v} Q_{i} + Q_{i} \widetilde{\mathcal{X}}_{i,v} P_{i} + Q_{i} \widetilde{\mathcal{X}}_{i,v} Q_{i}
	\end{equation}
\end{linenomath}
Consequently, each centered correlation in Eq.~\eqref{M_energy_centered} can be bounded by a product of two factors; a fluctuation norm on $i^{\co}$ and a deviation amplitude on $i$,
\begin{linenomath}
	\begin{equation}
		\begin{aligned}
			\abs{\bra{\gs} \widetilde{\mathcal{X}}_{i,v} \otimes \widetilde{H}_{i^{\co},v} \ket{\gs}} &\lesssim \norm{\widetilde{H}_{i^{\co},v} \ket{\gs}} \cdot \norm{\widetilde{\mathcal{X}}_{i,v} Q_{i} \ket{\gs}}. \label{M_factorization_bound}
		\end{aligned}
	\end{equation}
\end{linenomath}

Combining Eq.~\eqref{M_energy_centered} and Eq.~\eqref{M_factorization_bound} with the fluctuation bound~\eqref{M_fluctuation_bound} and the inequality $\norm{\widetilde{\mathcal{X}}_{i,v} Q_{i} \ket{\gs}} \leq 2 \norm{Q_{i}\ket{\gs}}$, we obtain the following general control of the energy response in terms of $\norm{Q_{i} \ket{\gs}}$:
\begin{linenomath}
	\begin{equation}
		\abs{\bra{\gs} u_{i}^{\dagger} H u_{i} - H \ket{\gs}} \lesssim d^{2} g_{1} \sqrt{\frac{\bar{g}}{\Delta n}} \norm{Q_{i} \ket{\gs}}. \label{M_general_bound_delta_Q}
	\end{equation}
\end{linenomath}
One may use the trivial bound $\norm{Q_{i} \ket{\gs}} \leq 1$ in Eq.~\eqref{M_general_bound_delta_Q}, which immediately yields a nontrivial $n^{-1/2}$-type bound:
\begin{linenomath}
	\begin{equation}
		\abs{\bra{\gs} u_{i}^{\dagger} H u_{i} - H \ket{\gs}} \leq \delta_{\mathrm{Rob}}^{(0)} , \label{M_trivial_bound_delta_Q}
	\end{equation}
\end{linenomath}
where $\delta_{\mathrm{Rob}}^{(0)} \propto d^{2} g_{1} \sqrt{\frac{\bar{g}}{\Delta n}}$.
However, Eq.~\eqref{M_general_bound_delta_Q} suggests a sharper route: if we can show that $\norm{Q_{i} \ket{\gs}}$ is itself small, then the energy response is smaller by the same factor.

The key idea here is to employ a bootstrap-type methodology. 
First, we bound $\norm{Q_{i} \ket{\gs}}$ using the already obtained 
robustness parameter $\delta_{\rm Rob}^{(0)}$. 
Next, using this bound on $\norm{Q_{i} \ket{\gs}}$, 
we update the robustness parameter to $\delta_{\rm Rob}^{(1)}$. 
We then use the improved robustness parameter to derive a sharper bound 
on $\norm{Q_{i} \ket{\gs}}$. 
By iterating this procedure and taking the limit, we obtain the desired bound.

Here, we employ the inequality that connects $\delta_{\rm Rob}^{(0)}$ and $\norm{Q_{i} \ket{\gs}}^{2}$ in the following way:
\begin{linenomath}
	\begin{equation}
		\norm{Q_{i} \ket{\gs}}^{2} \leq \frac{2\delta_{\mathrm{Rob}}^{(0)}}{\Delta}, \label{M_Q_bound_from_delta0}
	\end{equation}
\end{linenomath}
which is equivalent to inequality~\eqref{Eq_Schmidt_ineq}.
We recall that $\ket{0}_{i}$ is a dominant Schmidt vector in the Schmidt decomposition of $\ket{\gs} = \sum_{s} \lambda_{s} \ket{s}_{i} \ket{\phi_{s}}_{i^{\co}}$ across $i | i^{\co}$ and $\norm{Q_{i} \ket{\gs}}^{2} = \sum_{s \geq 1} \lambda_{s}^{2}$.

We now return to the general bound~\eqref{M_general_bound_delta_Q}.
Inserting Eq.~\eqref{M_Q_bound_from_delta0} into Eq.~\eqref{M_general_bound_delta_Q}, we obtain the improved estimate
\begin{linenomath}
	\begin{equation}
		\abs{\bra{\gs} u_{i}^{\dagger} H u_{i} - H \ket{\gs}} \lesssim d^{2} g_{1} \sqrt{\frac{\bar{g}}{\Delta n}} \sqrt{\frac{\delta_{\mathrm{Rob}}^{(0)}}{\Delta}} =: \delta_{\mathrm{Rob}}^{(1)}. \label{M_delta1}
	\end{equation}
\end{linenomath}
The procedure can now be iterated.
Starting from the initial bound $\delta_{\mathrm{Rob}}^{(0)}$ in Eq.~\eqref{M_trivial_bound_delta_Q}, we define recursively
\begin{linenomath}
	\begin{equation}
		\delta_{\mathrm{Rob}}^{(s+1)} \propto d^{2} g_{1} \sqrt{\frac{\bar{g}}{\Delta n}} \sqrt{\frac{\delta_{\mathrm{Rob}}^{(s)}}{2 \Delta}}. \label{M_bootstrap_recursion}
	\end{equation}
\end{linenomath}
This bootstrapping relation has a nontrivial fixed point of order
\begin{linenomath}
	\begin{equation}
		\delta_{\mathrm{Rob}} \sim \br{\frac{d^{2} g_{1}}{\Delta}}^{2} \frac{\bar{g}}{n}. \label{M_bootstrap_fixed}
	\end{equation}
\end{linenomath}
up to $k$-dependent numerical constants.
Hence the spectral gap converts the initial $n^{-1/2}$-type control into the sharper robustness scaling $\delta_{\mathrm{Rob}} = \mathcal{O}(1/n)$.
It is worth noting that this bootstrapping yields an improvement only when the refined bound satisfies $\delta_{\mathrm{Rob}} \lesssim \delta_{\mathrm{Rob}}^{(0)}$.
In terms of the spectral gap, this condition is equivalent to $\Delta \gtrsim n^{-1/3}$.

\subsection*{Concentration of Local Deviations}

We sketch the main ideas behind the stretched-exponential concentration bound~\eqref{Eq_local_excitation}.
Complete derivations with operator-norm bookkeeping, explicit constants, and auxiliary lemmas are given in the Supplementary Note 4.

Fix the dominant local Schmidt vector $\ket{0}_{i}$ for each site $i$ and define $P_{i} := \ket{0} \bra{0}_{i}$ and $Q_{i} := 1 - P_{i}$.
We consider the local deviation number operator
\begin{linenomath}
	\begin{equation}
		M := \sum_{i \in \Lambda} Q_{i},
	\end{equation}
\end{linenomath}
whose eigenvalue equals the number of sites deviating from the dominant local component.
Let $\Pi_{> z}$ denote the projector onto the $M$-eigenspaces with eigenvalue strictly larger than $z$.
Our goal is to prove the inequality of
\begin{linenomath}
	\begin{equation}
		\norm{\Pi_{> z} \ket{\gs}} \leq C_{1} e^{-C_{2} \sqrt{z}}  \label{M_goal_tail}
	\end{equation}
\end{linenomath}
with $C_1$ and $C_2$ depending only on $k,d,\bar{g},\Delta$.

The argument has three components:
(i) a constant-mass statement showing that $\ket{\gs}$ has non-negligible weight on a fixed low-deviation sector [see inequality~\eqref{M_low_sector_mass}];
(ii) a linear bound on the projected Hamiltonian $\Pi_{\leq z} (H - E^{\ast}) \Pi_{\leq z}$ [see inequality~\eqref{M_linear_proj_bound}], which is the main technical step, where $E^{\ast} := \bra{0} H \ket{0}_{\Lambda}$ is the energy of the reference product state $\ket{0}_{\Lambda} := \bigotimes_{i \in \Lambda} \ket{0}_{i}$;
and (iii) an energy-filtering step exploiting the spectral gap to convert the linear bound into the stretched-exponential tail~\eqref{M_goal_tail}.

By construction, $\bra{\gs} M \ket{\gs} = \sum_{i} \bra{\gs} Q_{i} \ket{\gs}$.
Using the robustness-to-Schmidt-tail implication established in~\eqref{Eq_Schmidt_ineq}, each $\bra{\gs} Q_{i} \ket{\gs} = \norm{Q_{i} \ket{\gs}}^{2}$ is $\mathcal{O}(1 / n)$, so that
\begin{linenomath}
	\begin{equation}
		\bra{\gs} M \ket{\gs} = \mathcal{O}(1), \label{M_mean_flip_O1}
	\end{equation}
\end{linenomath}
with an $n$-independent constant determined by $k,d,\bar{g},\Delta$.
Markov’s inequality then gives
\begin{linenomath}
	\begin{equation}
		z \norm{\Pi_{> z} \ket{\gs}}^{2} \leq \bra{\gs} M \ket{\gs},
	\end{equation}
\end{linenomath}
so choosing a constant $z_{0} = \mathcal{O}(1)$ yields
\begin{linenomath}
	\begin{equation}
		\norm{\Pi_{\leq z_{0}} \ket{\gs}}^{2} \geq \frac{1}{2}. \label{M_low_sector_mass}
	\end{equation}
\end{linenomath}
This ensures that projecting $\ket{\gs}$ onto the low-deviation sector of $M$ retains a constant overlap.

The key technical estimate is that, after an appropriate energy shift $E^{\ast}$,
\begin{linenomath}
	\begin{equation}
		\norm{\Pi_{\leq z} (H - E^{\ast}) \Pi_{\leq z}} \leq \eta z \qquad \forall z \geq \tilde{z}, \label{M_linear_proj_bound}
	\end{equation}
\end{linenomath}
for some $\eta = \mathcal{O}(\bar{g})$ and a constant $\tilde{z} = \mathcal{O}(1)$ depending only on $k,d,\bar{g},\Delta$.
This statement captures the idea that, restricted to states with at most $z$ local deviations from $\ket{0}_{\Lambda}$ (i.e., to the low-$M$ sector), the Hamiltonian behaves in the $M$-eigenbasis like an effective tight-binding operator whose effective bandwidth (operator norm) grows at most linearly with the cutoff $z$.

The proof of Eq.~\eqref{M_linear_proj_bound} relies on a convenient re-decomposition of $H$ into terms that are centered with respect to the reference product state $\ket{0}_{\Lambda}$.
Concretely, each original $k$-local term is expanded into a sum of operators supported on subsets of its support such that the partial expectation in $\ket{0}_{i}$ vanishes on every site in their support.
This procedure does not change $H$ as an operator, but it guarantees that every multi-site interaction term, when projected onto the low-deviation subspace, necessarily contains at least one local deviation operator $Q_i$.

We decompose the Hamiltonian as (see Supplementary Note 4 D)
\begin{linenomath}
	\begin{equation}
		H = \sum_{i \in \Lambda} \acute{H}_{i} + \acute{H}^{(\geq 2)} + E^{\ast}
	\end{equation}
\end{linenomath}
where $E^{\ast} = \bra{0} H \ket{0}_{\Lambda}$ is the energy of the reference product state, $\acute{H}_{i}$ collects the centered terms supported only on site $i$, and $\acute{H}^{(\geq 2)}$ contains the centered interaction terms supported on at least two sites.
Inside $\Pi_{\leq z}$, each centered interaction in $\acute{H}^{(\geq 2)}$ necessarily carries at least two deviation factors, so its matrix elements can be bounded by quantities such as $\sqrt{\bra{\psi} Q_{i} Q_{i'} \ket{\psi}}$ for $\Pi_{\leq z} \ket{\psi} = \ket{\psi}$.
Since $\sum_{i}Q_{i} \leq z$ holds on this subspace, one obtains the linear control (see Supplementary Proposition 3)
\begin{linenomath}
	\begin{equation}
		\norm{\Pi_{\leq z} \acute{H}^{(\geq 2)} \Pi_{\leq z}} = \mathcal{O}(z). \label{M_high_order_linear}
	\end{equation}
\end{linenomath}
The on-site part $\sum_{i} \acute{H}_{i}$ is subtler because it can act without changing the deviation number.
Here, the earlier robustness bound enters again: it implies that $\ket{0}_{i}$ is an approximate eigenvector for the effective on-site potential, so the off-diagonal block $Q_{i} \acute{H}_{i} P_{i}$ is suppressed by the small single-site deviation weight (see Supplementary Proposition 4).
When summed over sites and restricted to $\Pi_{\leq z}$, these off-diagonal contributions scale as $\mathcal{O}(\sqrt{z})$ and can be absorbed into $\eta z$ once $z \geq \tilde{z}$.
Together with the diagonal $Q_{i} \acute{H}_{i} Q_{i}$ contribution (which is directly $\mathcal{O}(z)$), this yields Eq.~\eqref{M_linear_proj_bound}.
We emphasize that Eq.~\eqref{M_linear_proj_bound} is the central technical estimate; its full proof is provided in the Supplementary Information.

We now explain why Eq.~\eqref{M_linear_proj_bound} implies the tail bound Eq.~\eqref{M_goal_tail}.
Decompose $\ket{\gs}$ into $M$-eigenstates:
\begin{linenomath}
	\begin{equation}
		\ket{\gs} = \sum_{x = 0}^{n} c_{x} \ket{\omega_{x}}, \qquad M \ket{\omega_{x}} = x \ket{\omega_{x}}. \label{M_gs_decomp_M_eigen}
	\end{equation}
\end{linenomath}
Because $H$ is $k$-local, it can change the deviation number by at most $k$,
\begin{linenomath}
	\begin{equation}
		\Pi_{= x} H \Pi_{= x'}= 0 \qquad \text{if }|x-x'| > k, \label{M_bandwidth_k}
	\end{equation}
\end{linenomath}
so $H$ has a banded structure in the $M$-eigenbasis.

Consider the truncated vector $\ket{\gs_{\leq \tilde{z} }} := \Pi_{\leq \tilde{z}} \ket{\gs}$, which has overlap at least $1 / \sqrt{2}$ as long as $\tilde{z} \ge z_0$ by Eq.~\eqref{M_low_sector_mass}.
Without loss of generality, we shift the Hamiltonian so that $H \ket{\gs} = 0$.
We construct a degree-$m$ Chebyshev-based polynomial $F_{m}$ with $F_{m}(0) = 1$ and with the following two complementary features.
First, on the energy window just above the gap, whose upper edge is determined by an $m$-dependent cutoff $E_{m}$ chosen appropriately, one can ensure
\begin{linenomath}
	\begin{equation}
		|F_{m}(x)| \leq e^{-2 m / \xi_{m}} \quad \text{for }\Delta \leq x \leq 2 E_{m} + \Delta, \label{M_Fm_window}
	\end{equation}
\end{linenomath}
where the ``filter width'' parameter $\xi_{m}$ satisfies
\begin{linenomath}
	\begin{equation}
		\xi_{m} = \sqrt{1 + \frac{2E_{m}}{\Delta}}. \label{M_xim_def}
	\end{equation}
\end{linenomath}
Second, outside the energy window ($x \ge 2E_m+\Delta$), $|F_m(x)|$ may rapidly grow in $m$ and $x$. 
To control this exterior contribution, we use an exponential high-energy tail for the truncated state $\ket{\gs_{\leq \tilde{z}}}$.
For any $\beta > 0$, one has
\begin{linenomath}
	\begin{equation}
		\norm{\Pi_{>E}^{H} \ket{\gs_{\leq \tilde{z}}}} \leq e^{-\beta E} \norm{e^{\beta H} \ket{\gs_{\leq \tilde{z}}}}, \label{M_high_energy_tail_start}
	\end{equation}
\end{linenomath}
where $\Pi_{>E}^{H}$ denotes the spectral projector of $H$ onto energies larger than $E$.
The remaining factor is bounded using the same low-deviation norm estimate~\eqref{M_linear_proj_bound}.
Indeed, because $H$ is $k$-local, each application of $H$ can increase the deviation number by at most $k$; hence the $r$-th term in the expansion of $e^{\beta H} \Pi_{\leq \tilde{z}}$ only probes sectors with at most $\tilde{z} + kr$ deviations.
On these sectors, Eq.~\eqref{M_linear_proj_bound} gives an effective norm bounded by $\eta(\tilde{z} + kr)$.
More explicitly, using $H \ket{\gs} = 0$, we write
\begin{linenomath}
	\begin{equation}
		e^{\beta  H} \ket{\gs_{\leq \tilde{z}}} = \sum_{r = 0}^{\infty} \frac{\beta^{r}}{r!} \mathrm{ad}_{H}^{r}(\Pi_{\leq \tilde{z}}) \ket{\gs}. \label{M_exp_moment_series}
	\end{equation}
\end{linenomath}
Since adding the scalar shift $E^{\ast}$ does not change commutators, we may replace $H$ by $H - E^{\ast}$ in the adjoint action.
Moreover, the $r$-fold commutator only probes the sector with at most $\tilde{z} + kr$ deviations, and therefore
\begin{linenomath}
	\begin{equation}
		\norm{\mathrm{ad}_{H - E^{\ast}}^{r}(\Pi_{\leq \tilde{z}}) \ket{\gs}} \leq \br{2 \eta (\tilde{z}+ kr)}^{r}. \label{M_ad_bound}
	\end{equation}
\end{linenomath}
Consequently,
\begin{linenomath}
	\begin{equation}
		\begin{aligned}
			\norm{e^{\beta H} \ket{\gs_{\leq \tilde{z}}}} &\leq 1 + \sum_{r =1}^{\infty} \frac{(2 \beta \eta(\tilde{z} + kr))^{r}}{r!} \\
			&\leq 1 + e^{\tilde{z} / k} \sum_{r = 1}^{\infty} (2 e \beta \eta k)^{r} \label{M_exp_moment_sum}
		\end{aligned}
	\end{equation}
\end{linenomath}
Choosing $\beta_{0} = 1 / (4e\eta k)$ makes the last geometric series convergent and gives
\begin{linenomath}
	\begin{equation}
		\begin{aligned}
			\norm{e^{\beta_{0}H} \ket{\gs_{\leq \tilde{z}}}} \leq 1 + e^{\tilde{z} / k} \leq C_{\tilde{z}} \label{M_exp_moment_bound}
		\end{aligned}
	\end{equation}
\end{linenomath}
where $C_{\tilde{z}}$ depends only on the fixed cutoff $\tilde{z}$ and microscopic constants, but not on $m$ or $n$.
Therefore, the truncated state satisfies the exponential high-energy spectral tail
\begin{linenomath}
	\begin{equation}
		\begin{aligned}
			\norm{\Pi_{>E}^{H} \ket{\gs_{\leq \tilde{z}}}} \leq C_{\tilde{z}} e^{-\beta_{0} E}. \label{M_high_energy_tail_bound}
		\end{aligned}
	\end{equation}
\end{linenomath}

The key point is then to choose the filtering window so that the exponential tail above compensates for the possible growth of $F_{m}$ outside the window.
We take the window scale $E_{m}$ proportional to $m$ (indeed $E_{m} \sim m / \beta_{0}$ with a constant $\beta_{0} \propto (\eta k)^{-1}$), so that Eq.~\eqref{M_xim_def} gives $\xi_{m} \sim \sqrt{m}$, and hence
\begin{linenomath}
	\begin{equation}
		e^{-2m / \xi_{m}} \leq e^{-C \sqrt{m}} , \label{M_stretched_form_m}
	\end{equation}
\end{linenomath}
where $C$ is a constant. 
With this choice, the Chebyshev window suppression~\eqref{M_Fm_window} and the high-energy tail control combine to show that 
\begin{linenomath}
	\begin{equation}
		\norm{F_{m}(H) \ket{\gs_{\leq \tilde{z}}} - a_{\tilde{z}} \ket{\gs}} \leq C e^{-c \sqrt{m}}, \label{M_filter_approx_gs}
	\end{equation}
\end{linenomath}
where $a_{\tilde{z}} := \braket{\gs | \gs_{\leq \tilde{z}}} = \norm{\Pi_{\leq \tilde{z}} \ket{\gs}}^{2}$. 
Since $\tilde{z} \geq z_{0}$, Eq.~\eqref{M_low_sector_mass} gives $a_{\tilde{z}} \geq 1/2$.

Independently, the banded property~\eqref{M_bandwidth_k} implies that each application of $H$ can increase the deviation number by at most $k$.
Therefore, any degree-$m$ polynomial in $H$ maps $\Pi_{\leq \tilde{z}}$ into $\Pi_{\leq \tilde{z} + km}$: 
\begin{linenomath}
	\begin{equation}
		\Pi_{> \tilde{z} + km} F_{m}(H) \Pi_{\leq \tilde{z}} = 0. \label{M_filter_support_constraint}
	\end{equation}
\end{linenomath}
Applying $\Pi_{>\tilde{z} + km}$ to Eq.~\eqref{M_filter_approx_gs} and using Eq.~\eqref{M_filter_support_constraint}, we obtain
\begin{linenomath}
	\begin{equation}
		a_{\tilde{z}} \norm{\Pi_{>\tilde{z} + km} \ket{\gs}} \leq C e^{-c \sqrt{m}}.
	\end{equation}
\end{linenomath}
Since $a_{\tilde{z}} \geq 1/2$, this yields
\begin{linenomath}
	\begin{equation}
		\norm{\Pi_{>\tilde{z} + km} \ket{\gs}} \leq C' e^{-c \sqrt{m}}.
	\end{equation}
\end{linenomath}
Setting $z = \tilde{z} + km$, i.e. $m = \Theta(z)$ because $\tilde{z} = \mathcal{O}(1)$, we arrive at inequality~\eqref{M_goal_tail}, which corresponds to Eq.~\eqref{Eq_local_excitation} in the main text.

We note that the form $e^{-C \sqrt{z}}$ in Eq.~\eqref{M_goal_tail} comes from balancing the Chebyshev filtering window with the high-energy spectral tail.
A degree-$m$ Chebyshev filter suppresses spectral components inside the chosen energy window, but the polynomial may grow outside that window.
The exterior region is controlled by the exponential high-energy tail of $\ket{\gs_{\leq \tilde{z}}}$, and this balance requires the cutoff energy to scale as $E_{m} \sim m / \beta_{0}$.
For a window of length of order $E_{m}$, the Chebyshev suppression above the gap has the form $\exp[-cm \sqrt{\Delta / E_{m}}]$.
Substituting $E_{m} \sim m / \beta_{0}$ gives $e^{-c' \sqrt{m}}$.
Finally, because $H$ is $k$-local, a degree-$m$ polynomial in $H$ can increase the deviation number by at most $km$; hence $z = \tilde{z} + km$ implies $m = \Theta(z)$.
This converts the filtering error $e^{-c' \sqrt{m}}$ into the deviation tail $e^{-C \sqrt{z}}$.

\subsection*{Efficient Simulation of Permutation Invariant Ground States}

We describe a classical algorithm that computes an $\mathcal{O}(\epsilon)$-approximate ground-state energy and, under the standing nondegeneracy and gap assumptions, a corresponding approximate ground-state vector in the permutation-invariant setting~\cite{gilmore1972geometry,harrow2013church}.
A complete description of the algorithm and the detailed complexity analysis are provided in Supplementary Note~5.

The concentration bound implies that the exact ground state is well approximated by the sector with only a small number of local deviations from its dominant local Schmidt direction.
Let $\ket{0}$ denote that dominant local Schmidt vector.
Then the weight outside the sector with more than $z$ deviations decays as $e^{- c \sqrt{z}}$, so choosing 
\begin{linenomath}
	\begin{equation}
		z=\Theta \bigl(\log^2(1/\epsilon)\bigr)
	\end{equation}
\end{linenomath}
is sufficient to achieve error $\epsilon$.

For permutation-invariant systems, this low-deviation sector lies inside the symmetric subspace and admits an occupation-number basis of dimension~\cite{gilmore1972geometry,harrow2013church,Sagan2001} 
\begin{linenomath}
	\begin{equation}
		m(z)= \binom{z+d-1}{d-1},
	\end{equation}
\end{linenomath}
which is polynomial in $z$ for fixed local dimension $d$. 
Because the optimal local direction is not known a priori, we introduce a variational family of local bases $\{\ket{e_{a}(u)}\}_{a=0}^{d-1}$ generated by a local unitary $U(u)$, and treat $\ket{e_{0}(u)}$ as a trial mean-field direction.

For each $u$, let $H^{(z)}(u)$ denote the restriction of $H$ to the symmetric sector with at most $z$ flips relative to $\ket{e_{0}(u)}^{\otimes n}$, and let $\lambda_{\min}(H^{(z)}(u))$ denote its smallest eigenvalue.
We then define
\begin{linenomath}
	\begin{equation}
		E_{z}^{\star} := \inf_{u} \lambda_{\min}(H^{(z)}(u)).
	\end{equation}
\end{linenomath}
We parametrize $U(u)$ by the real and imaginary parts of its matrix elements, collected in $x_{u} \in \mathbb{R}^{2d^{2}}$, subject to polynomial constraints $\mathcal{C}(x_{u})$ expressing unitarity (and optional gauge fixing).
Every matrix element of $H^{(z)}(u)$ in the truncated symmetric basis is then a polynomial in $x_{u}$.
Introducing an auxiliary real variable $\lambda$, we define 
\begin{linenomath}
	\begin{equation}
		P(x_{u},\lambda) = \det(H^{(z)}(u) - \lambda I).
	\end{equation}
\end{linenomath}
For any threshold $t$, the condition
\begin{linenomath}
	\begin{equation}
		\exists x_{u}, \lambda: \quad \mathcal{C}(x_{u}) \, \wedge \, \lambda \leq t \, \wedge \, P(x_{u},\lambda) = 0 \label{eq:ETR_threshold}
	\end{equation}
\end{linenomath}
is equivalent to the existence of a basis choice $u$ such that $\lambda_{\min}(H^{(z)}(u)) \leq t$.

For fixed local dimension $d$, the number of real variables in $(x_{u},\lambda)$ and the number of polynomial constraints remain constant.
Moreover, each matrix entry of $H^{(z)}(u)$ is a polynomial in $x_{u}$, and the matrix size is $m(z) = \mathrm{poly}(z)$.
Hence, the characteristic polynomial $P(x_{u},\lambda)$ has degree $\mathrm{poly}(z)$ and can be constructed symbolically in $\mathrm{poly}(m(z)) = \mathrm{poly}(z)$ time using a division-free algorithm such as Berkowitz's algorithm~\cite{berkowitz1984computing}.
Standard ETR algorithms then solve the threshold problem~\eqref{eq:ETR_threshold} in time $(sD)^{\mathcal{O}(r)} \times \mathrm{poly}(\text{input bit-length})$, where $r$ is the number of variables, $s$ the number of constraints, and $D$ the maximum polynomial degree~\cite{renegar1992computational,basu2006algorithms}. 
Since here $r,s = \mathcal{O}(1)$ and $D = \mathrm{poly}(z)$, the threshold problem is solvable in $\mathrm{poly}(z)$ time, up to polynomial factors in the input bit-length.

To approximate $E_{z}^{\star}$ to additive error $\epsilon$, we first choose coarse lower and upper bounds $\mathcal{L}$ and $\mathcal{U}$ such that $\mathcal{L} \leq E_{z}^{\star} \leq \mathcal{U}$.
For permutation-invariant $k$-local Hamiltonians, such bounds can be obtained, for example, from simple operator-norm estimates, and may be chosen so that $\mathcal{U} - \mathcal{L} = \mathcal{O}(n)$.
Binary search then requires $\mathcal{O}(\log((\mathcal{U} - \mathcal{L}) / \epsilon)) = \mathcal{O}(\log(n / \epsilon))$ threshold queries.
Let $\widehat{E}_{z}$ denote the midpoint of the final interval returned by the binary search, so that $|\widehat{E}_{z} - E_{z}^{\star}| \leq \epsilon$.
We then solve one final threshold query at $t = \widehat{E}_{z} + \epsilon$ (or equivalently retain the witness from the last feasible query), extract a feasible parameter $\widehat{x}_{u}$, let $\widehat{u}$ denote the corresponding basis choice, form the numeric matrix $H^{(z)}(\widehat{u})$, and diagonalize it to obtain the corresponding approximate ground-state vector $\ket{\varphi_{z}}$ and its energy.
Combining this with the concentration bound and the choice $z = \Theta(\log^{2}(1 / \epsilon))$, we obtain a classical algorithm for the permutation-invariant setting with total runtime $\mathrm{poly}(z) \, \log(n / \epsilon) = \mathrm{polylog}(1 / \epsilon) \, \log(n / \epsilon)$ for fixed local dimension $d$ and fixed locality $k$.
The algorithm outputs an $\mathcal{O}(\epsilon)$-accurate ground-state energy estimate and a corresponding approximate ground-state vector.

In the absence of permutation invariance, by contrast, the dominant local directions can be site dependent. The resulting variational problem then contains a number of continuous parameters that grows linearly with $n$. Under our present assumptions, it is highly nontrivial to provide a polynomial-time guarantee for this global optimization, even though the subsequent low-deviation diagonalization remains efficient once the local directions are known.

\begin{acknowledgments}
Both authors acknowledge the Hakubi projects of RIKEN. 
T. K. was supported by JST PRESTO (Grant No.
JPMJPR2116), ERATO (Grant No. JPMJER2302),
and JSPS Grants-in-Aid for Scientific Research (No.
JP23H01099, JP24H00071), Japan.
\end{acknowledgments}

\bibliography{All_to_all}

\clearpage
\newpage
\setcounter{page}{1}

\onecolumngrid
\supponecolumnfootnotes

\renewcommand\thefootnote{*\arabic{footnote}}

\setcounter{equation}{0}
\setcounter{section}{0}
\addtocounter{figure}{-1}

\renewcommand{\theequation}{S.\arabic{equation}}
\renewcommand{\figurename}{Supplementary Figure}

\renewcommand{\tablename}{Supplementary Table}
\renewcommand{\thetable}{\arabic{table}}

\renewcommand{\thesection}{Supplementary Note~\arabic{section}}

\begin{center}
{\large \bf Supplementary Information for "Quantum complexity and generalized area law in fully connected models"}\\
\vspace*{0.3cm}
Donghoon Kim$^{1}$ and Tomotaka Kuwahara$^{1,2,3}$ \\
\vspace*{0.1cm}
$^{1}${\small \it Analytical quantum complexity RIKEN Hakubi Research Team, \\ RIKEN Center for Quantum Computing (RQC), Wako, Saitama 351-0198, Japan} \\ 
$^{2}${\small \it RIKEN Cluster for Pioneering Research (CPR), Wako, Saitama 351-0198, Japan}  \\
$^{3}${\small \it PRESTO, Japan Science and Technology (JST), Kawaguchi, Saitama 332-0012, Japan}
\end{center}

\vspace*{0.3cm}

\begin{center}
\footnotesize{
This Supplementary Information contains detailed proofs and further technical analyses of the results reported in the main text.}
\end{center}

\supplementarytableofcontents
\startsupplementarycontents

\section{Setup and Assumptions}

We consider a quantum system comprising $n$ qudits and denote the set of total sites by $\Lambda$, where $|\Lambda| = n$.
Let $d$ represent the dimension of the Hilbert space associated with a single site, where we consider $d < n$.
For an arbitrary subset $X \subseteq \Lambda$, we define $|X|$ as the number of sites contained in $X$. 
We also denote the complementary subset of $X$ and $X \setminus \{i\}$ by $X^{c}$ and $X_i$, respectively.

For an arbitrary operator $O$, we define it to be $k$-local and $g$-extensive if it can be expressed as 
\begin{align}
	O= \sum_{Z: |Z|\le k} o_Z ,\qquad \max_{i\in \Lambda} \sum_{Z:Z\ni i} \norm{o_Z} \le g . 
\end{align} 
Here, each local term $o_Z$ is supported on the subset $Z\subseteq\Lambda$, meaning that it acts nontrivially only on the degrees of freedom in $Z$ and acts as the identity on the complement $Z^{c}$. Equivalently, we can write  
\begin{align}
	o_Z = o_Z \otimes \hat{1}_{Z^{c}} .
\end{align}
The norm of a $k$-local and $g$-extensive operator $O$ is bounded above by $gn$ as demonstrated below:
\begin{align}
	\norm{O} \le \sum_{Z: |Z|\le k} \norm{o_Z} \le \sum_{i\in \Lambda} \sum_{Z:Z\ni i} \norm{o_Z} \le g |\Lambda| = gn. 
\end{align}

We now consider a system governed by a general $k$-local Hamiltonian of the form
\begin{align}
	\label{def:k-local}
	H = \sum_{Z:|Z|\le k} h_Z
\end{align}
such that 
\begin{align}
	\label{g_0_extensive_2}
	& \norm{h_{\{i\}}} \le g_0 \quad \forall i \in \Lambda  ,\notag \\
	&\sum_{Z:Z \ni i,i'} \norm{h_Z} \le \frac{g_1}{n} \quad \forall i,i' \in \Lambda .
\end{align}

The above assumption ensures the $g$-extensiveness of the Hamiltonian as follows:
\begin{align}
	\label{g_ext_full_Ham}
	&\sum_{Z:Z \ni i} \norm{h_Z} \le \norm{h_{\{i\}}} + \sum_{i'\in i^\co}  \sum_{Z:Z \ni i,i'} \norm{h_Z} \le g_0+g_1 =: \bar{g}.    
\end{align}

Throughout the paper, we set the ground energy equal to zero:
\begin{align}
	H \ket{\Omega} = 0 . 
\end{align}

The subset Hamiltonian $H_L$ on the subset $L$ is defined as
\begin{align}
	\label{def:k-local_subset_L}
	H_L = \sum_{Z:Z\subseteq L} h_Z .
\end{align}
We further define $\widehat{H}_L$ as 
\begin{align}
	\label{def:k-local_subset_L_widehat}
	\widehat{H}_L = \sum_{Z:Z\cap L \neq \emptyset} h_Z  = H- H_{L^\co},
\end{align}
which includes all interaction terms that involve at least one site within the subset $L$, even if the interactions extend to sites outside of $L$.
Using the condition~\eqref{g_0_extensive_2}, we immediately obtain 
\begin{align}
	\label{upp_bound_norm_widehat_H_L}
	\norm{\widehat{H}_L} \le \sum_{i\in L} \sum_{Z: Z\ni i} \norm{h_{Z}} \le  \sum_{i\in L} \bar{g} = \bar{g} |L|. 
\end{align}

For a density matrix $\rho$, the variance of an arbitrary operator $O$ with respect to $\rho$ is given by:
\begin{align}
	\mathrm{Var}_{\rho}\br{O} = \tr\br{\rho O^{2}} - \tr\br{\rho O}^{2}.
\end{align}
The entanglement entropy of the state $\ket{\psi}$ between the bipartitions $A$ and $B$ of $\Lambda$ is defined as:
\begin{align}
	\label{Def:entanglement_entropy}
	S_{A,B}(\ket{\psi}) := - \tr\left( \rho_{A} \log \rho_{A} \right) = - \tr\left( \rho_{B} \log \rho_{B} \right),
\end{align}
where $\rho_{A} := \tr_{B}\left( \ket{\psi} \bra{\psi} \right)$ and $\rho_{B} := \tr_{A}\left( \ket{\psi} \bra{\psi} \right)$ represent the reduced density matrices on $A$ and $B$, respectively.

\section{Robustness theorem and the logarithmic entanglement}

\subsection{Preliminaries: Robustness of ground states and the largest local Schmidt coefficient}

\begin{definition}[Robustness under local unitaries]
	\label{def:robustness_local_unitary}
	
	Let $\ket{\Omega}$ be the ground state of a Hamiltonian $H$, and
	let $\rho_i = \mathrm{tr}_{\Lambda_i} \bigl( 
	\ket{\Omega}\bra{\Omega} 
	\bigr)$ be the reduced density matrix at site $i$.
	
	Consider a local unitary operator $u_i$ acting on site $i$ that leaves
	$\rho_i$ invariant:
	\begin{align}
		u_i \rho_i u_i^\dagger = \rho_i .
		\label{eq:local_invariance}
	\end{align}
	We define the \emph{robustness parameter} $\delta_{\mathrm{Rob}}$
	as the smallest non-negative constant such that
	\begin{align}
		\left|
		\bra{\Omega} 
		u_i^\dagger H u_i 
		\ket{\Omega}
		-
		\bra{\Omega} 
		H 
		\ket{\Omega}
		\right|
		\le
		\delta_{\mathrm{Rob}}
		\quad
		\text{for all $u_i$ satisfying \eqref{eq:local_invariance}}.
		\label{eq:robustness_bound}
	\end{align}
\end{definition}

Using the robustness of the ground state, we obtain the following proposition: the ground state locally has a single large Schmidt coefficient, and the remaining Schmidt coefficients are small. Physically, it means that the local state is an approximately product state.
\begin{prop}\label{prop:Largest local Schmidt coefficient}
	We define the Schmidt decomposition of the ground state $\ket{\Gs}$ between a single site $i \in \Lambda$ and the remainder $i^{c} = \Lambda \setminus \{i\}$ as
	\begin{align}
		\ket{\gs} =\sum_{s = 0}^{d-1} \lambda_s\ket{s}_i\ket{\phi_s}_{i^{c}} , \label{Schmidt_decomposition}
	\end{align}
	with descending ordered singular values $\lambda_{0} \geq \lambda_{1} \geq \lambda_{2} \geq \cdots \geq \lambda_{d-1}$. Then, we obtain 
	\begin{align}
		\label{mean/field_error_bo8nd}
		|\lambda_0|^2 \ge 1-  \frac{2\delta_{\rm Rob}}{\Delta} ,\qquad  \sum_{j = 1}^{d-1} | \lambda_j|^2 \le  \frac{2\delta_{\rm Rob}}{\Delta}.
	\end{align}
\end{prop}

\textit{Proof of Proposition~\ref{prop:Largest local Schmidt coefficient}}.
We need to consider the case of 
\begin{align}
	\label{condition_delta_rob_1/2}
	\frac{\delta_{\rm Rob}}{\Delta} < \frac{1}{2} ,
\end{align}
since inequality~\eqref{mean/field_error_bo8nd} trivially holds for $\delta_{\rm Rob}/\Delta \ge 1/2$. 

We begin by constructing the reference state as 
\begin{align}
	\ket{\widetilde{\gs}_{d_1}} := \sum_{s \le d_1} \lambda_s \ket{s}_{i} \ket{\phi_s}_{i^{c}} - \sum_{s >d_1}  \lambda_s \ket{s}_{i} \ket{\phi_s}_{i^{c}}
\end{align}
with $d_1\in[0,d]$, 
where we can construct the above state by using the unitary operator that satisfies the condition~\eqref{eq:local_invariance}. 
By denoting
\begin{align}
	\ket{\widetilde{\gs}_{d_1}} = a\ket{\gs} + b\ket{\gs_\bot},  
\end{align}
we have 
\begin{align}
	a = \braket{\Omega | \widetilde{\Omega}_{d_1}} = p_{\le d_1} - p_{>d_1} ,\quad p_{\le d_1}=  \sum_{j \le d_1} | \lambda_j|^2,\quad p_{> d_1}=  \sum_{j > d_1} | \lambda_j|^2  .
\end{align}
Hence, by letting $\bra{\gs}H\ket{\gs}=E_0=0$, we have 
\begin{align}
	|\bra{\widetilde{\gs}_{d_1}}H\ket{\widetilde{\gs}_{d_1}}| = |b|^2 \abs{ \bra{\gs_\bot}H\ket{\gs_\bot} } \ge |b|^2\Delta.
\end{align}

The energy difference between $\ket{\widetilde{\gs}_{d_1}}$ and $\ket{\gs}$ is upper-bounded by $\delta_{\rm Rob}$ [see Ineq.~\eqref{eq:robustness_bound}], 
which gives 
\begin{align}
	\abs{\bra{\widetilde{\gs}_{d_1}}H\ket{\widetilde{\gs}_{d_1}}} = \abs{\bra{\widetilde{\gs}_{d_1}}H\ket{\widetilde{\gs}_{d_1}} - E_{0}} \le \delta_{\rm Rob}.
\end{align}
We thus obtain 
\begin{align}
	\delta_{\rm Rob} \ge |b|^2\Delta \quad \Rightarrow \quad |b|^2 \le  \frac{ \delta_{\rm Rob}}{\Delta}. 
\end{align}
Using $|a|^2 =1 - |b|^{2}$, we get 
\begin{align}
	\abs{p_{\le d_1} - p_{>d_1}}^2= |a|^2 \ge 1-  \frac{ \delta_{\rm Rob}}{\Delta} .
\end{align}
When $\delta_{\rm Rob}/\Delta=0$, we trivially obtain $p_{>d_1}=0$ for any $d_1$, which ensures inequality~\eqref{mean/field_error_bo8nd} by setting $d_1=0$. 
Hence, we consider $\delta_{\rm Rob}/\Delta>0$.
By combining the above inequality with $p_{\le d_1} + p_{>d_1}=1$, we derive 
\begin{align}
	\label{two_solution_d_1}
	&p_{\le d_1} \le \frac{1-\sqrt{1-\delta_{\rm Rob}/\Delta}}{2} < \frac{ \delta_{\rm Rob}}{2\Delta}=:\bar{\delta} , \notag \\
	& \Or p_{> d_1} < \bar{\delta}   ,
\end{align}
where we use $\sqrt{1-x} \ge 1-x$ for $0<x<1$.

Now, we exclude the possibility of the first solution in~\eqref{two_solution_d_1}.
Because of 
\begin{align}
	p_{\le d_1} \ge \frac{d_1+1}{d} , 
\end{align}
the inequality $p_{\le d_1} \le \bar{\delta} $ cannot be satisfied for 
\begin{align}
	d_1+1 \ge d \bar{\delta}  , 
\end{align}
Note that $\lambda_{0} \geq \lambda_{1} \geq \lambda_{2} \geq \cdots \geq \lambda_{d-1}$ has been assumed. 
Hence, under the above condition, we have 
\begin{align}
	p_{> d_1} < \bar{\delta} \AND p_{\le d_1} > 1 - \bar{\delta} . 
\end{align}

In the next step, we derive the same inequality by replacing $d_1$ with $d_2$ ($d_2 \in [0,d_1]$), where we choose $\ket{\Omega_{\le d_2}}$ as the reference state. 
Then, we have 
\begin{align}
	\label{two_solution_d_2}
	p_{\le d_2} <\bar{\delta}  \Or p_{> d_2} < \bar{\delta}   .
\end{align}
Then, from $p_{> d_1} < \bar{\delta}$ ($p_{\le d_1} > 1-\bar{\delta}$), we obtain 
\begin{align}
	p_{\le d_2} \ge \frac{d_2+1}{d_1+1} p_{\le d_1} > \frac{d_2+1}{d_1+1} (1-\bar{\delta}) , 
\end{align}
which prohibits $p_{\le d_2} < \bar{\delta}$ as long as 
\begin{align}
	\frac{d_2+1}{d_1+1} (1-\bar{\delta}) \ge \bar{\delta} \longrightarrow d_2+1 \ge \frac{\bar{\delta}}{1-\bar{\delta}}\br{d_1+1}.
\end{align}

By repeating the same process, for a set of $\{d_j\}_j$, we set the conditions of 
\begin{align}
	d_j +1 \ge \frac{\bar{\delta}}{1-\bar{\delta}}\br{d_{j-1}+1}. 
\end{align}
Under the above condition, we also obtain $p_{> d_j} \le \bar{\delta}$ for all $j$.
By choosing each of $d_j$ as  
\begin{align}
	d_j = \ceil{ \frac{\bar{\delta}}{1-\bar{\delta}}\br{d_{j-1}+1} } -1, 
\end{align}
we can ensure $d_j$ converges to $0$ for $\bar{\delta} \le 1/3$ by a finite step. 
Therefore, under the condition $ \bar{\delta}=\delta_{\rm Rob}/(2\Delta) \le 1/3$, we have $p_{> 0} \le \bar{\delta}=\delta_{\rm Rob}/(2\Delta)$, or equivalently, 
\begin{align}
	\sum_{j = 1}^{d-1} | \lambda_j|^2 \le  \frac{\delta_{\rm Rob}}{2\Delta} \le \frac{2\delta_{\rm Rob}}{\Delta}  ,
\end{align}
which proves the main inequality in the regime $\delta_{\rm Rob}/\Delta<1/2$ by \eqref{condition_delta_rob_1/2}. 

This completes the proof. $\square$

\subsection{Robustness theorem and logarithmic entanglement bound}

Here, we prove that the gapped ground state satisfies the robustness in Def.~\ref{def:robustness_local_unitary} with $\delta_{\rm Rob} = \mathcal{O}(1/n)$:
\begin{theorem} \label{prop:Robusness for local unitary}
	When the Hamiltonian is defined on the fully connected graph, satisfying the condition~\eqref{g_0_extensive_2}, the ground state satisfies the robustness in the sense that 
	\begin{align}
		\delta_{\mathrm{Rob}} = \br{\frac{12 \gamma g_1 d^2}{\Delta}}^2 \frac{2\bar{g}}{n} , \label{delta_Rob_explicit}
	\end{align}
	where $\Delta$ is the spectral gap and $\gamma=6k^2$. 
\end{theorem}

{\bf Remark.}
More precisely, we can derive a slightly better bound as follows:
\begin{align}
	\delta_{\mathrm{Rob}} =\min \brr{ \br{\frac{12 \gamma g_1 d^2}{\Delta}}^2 \frac{2\bar{g}}{n},\ 12 \gamma g_1 d^2 \sqrt{\frac{ \bar{g}}{n \Delta}}} .
	\label{delta_Rob_explicit_2}
\end{align}
The above form is meaningful for $\Delta \lesssim 1/n^{1/3}$. 

{~} \\

As a straightforward corollary, we prove the bound on the entanglement entropy:
\begin{corol} \label{corol:log_entanglement}
	We assume that the system size $n$ is larger than $n_0$ with 
	\begin{align}
		n \ge n_0,\quad n_0:=  \br{\frac{12 \gamma g_1 d^2}{\Delta}}^2 \frac{8\bar{g}}{\Delta}  .
	\end{align}
	Then, for an arbitrary decomposition of the total system $\Lambda=A\cup B$, the entanglement entropy is upper-bounded by
	\begin{align}
		\label{corol:log_entanglement/main_ineq}
		S_{A,B}(\ket{\Omega} ) \le  \frac{|A|}{n} \br{\frac{12 \gamma g_1 d^2}{\Delta}}^2 \frac{8\bar{g}}{\Delta}  \log \brr{\frac{ d^{1/2}\Delta }{\bar{g}} \cdot \br{\frac{\Delta}{12 \gamma g_1 d^2}}^2 n} .
	\end{align}
\end{corol}

{\bf Remark.}
From the inequality~\eqref{corol:log_entanglement/main_ineq}, we obtain 
\begin{align}
	\label{corol:log_entanglement/main_ineq_apply}
	S_{A,B}(\ket{\Omega} ) \le C  \br{\frac{\bar{g}}{\Delta}}^3 \log (n) 
\end{align}
for an arbitrary $A$ ($|A| \le n$). 

{~} \\

\textit{Proof of Corollary~\ref{corol:log_entanglement}.}
For an arbitrary subset, which we denote $A=\{1,2,\ldots, n_1\}$ ($n_1\le n$) without loss of generality, we consider the von-Neumann entropy $S(\rho_{1,2,\ldots, n_1}) $, which is equal to the entanglement entropy $S_{A,B}(\ket{\Omega} ) $.
Using the subadditivity of the von-Neumann entropy, we derive  
\begin{align}
	S(\rho_{1,2,\ldots, n_1}) \le \sum_{i=1}^{n_1}S(\rho_{i}) .
\end{align}
Using the Schmidt decomposition~\eqref{Schmidt_decomposition}, we obtain 
\begin{align}
	S(\rho_{i}) &\le - \lambda_0^2 \log(\lambda_0^2) +  (1-\lambda_0^2) \log\br{ \frac{d}{1-\lambda_0^2}}   \notag \\
	&\le \tilde{p}_{\rm Rob} \log(d) +\brr{ -\br{1 - \tilde{p}_{\rm Rob} }\log (1- \tilde{p}_{\rm Rob})   - \tilde{p}_{\rm Rob} \log (\tilde{p}_{\rm Rob}) } \notag \\
	&\le \tilde{p}_{\rm Rob} \log(d) - 2 \tilde{p}_{\rm Rob} \log (\tilde{p}_{\rm Rob}) ,
\end{align}
where $ \tilde{p}_{\rm Rob}:= 2\delta_{\rm Rob}/\Delta=\br{\frac{12 \gamma g_1 d^2}{\Delta}}^2 \frac{4\bar{g}}{n \Delta}$, and the assumption $n\ge n_0$ ensures $\tilde{p}_{\rm Rob}\le 1/2$.
Then, we obtain 
\begin{align}
	S(\rho_{i}) \le \frac{1}{\Delta} \br{\frac{12 \gamma g_1 d^2}{\Delta}}^2 \frac{8\bar{g}}{n} \log \brr{d^{1/2} \cdot   \br{\frac{\Delta}{12 \gamma g_1 d^2}}^2 \frac{n\Delta}{\bar{g}}} ,
\end{align}
which yields 
\begin{align}
	S(\rho_{1,2,\ldots, n_1}) &\le \frac{n_1}{n} \br{\frac{12 \gamma g_1 d^2}{\Delta}}^2 \frac{8\bar{g}}{\Delta}  \log \brr{\frac{ d^{1/2}\Delta }{\bar{g}} \cdot \br{\frac{\Delta}{12 \gamma g_1 d^2}}^2 n} .
\end{align}
This completes the proof. $\square$

\section{Proof of Theorem~\ref{prop:Robusness for local unitary}}

\subsection{Trade-off between variance and spectral gap}

We here consider an arbitrary Hermitian operator $A_L$ in the form of 
\begin{align}
	A_L= \sum_{Z : |Z| \le k_A, Z\subset L} a_Z, \qquad  \sum_{Z : Z \ni i}\norm{a_Z} \le 1 \quad  \forall i\in L.  
	\label{supp_def:A_observable}
\end{align}
Note that $A_L$ is supported on a subset $L \subseteq \Lambda$
We then define $\mathrm{Var}_{\Gs} (A_L)$ as the variance of $A_L$ in the ground state $\ket{\Gs}$: 
\begin{align}
	\mathrm{Var}_{\Gs} (A_L) = \bra{\Gs} A_L^2 \ket{\Gs} - \bra{\Gs} A_L\ket{\Gs}^2.
	\label{supp_def:A_observable_Var}
\end{align}

We here prove the following lemma which has been obtained in Ref.~\cite[Ineq.~(8)]{Kuwahara_2017}:
\begin{lemma} \label{trade_off_variance}
	Let $A_L$ be a $k_{A}$-local and $1$-extensive operator as in Eq.~\eqref{supp_def:A_observable}. 
	For an arbitrary ground state $\ket{\Gs}$ of the $k$-local Hamiltonian $H$ with the property~\eqref{g_0_extensive_2},  
	the variance $\mathrm{Var}_{\Gs} (A_L)$ and the spectral gap $\Delta$ satisfies the following trade-off inequality: 
	\begin{align}
		\mathrm{Var}_{\Gs} (A_L) \cdot \Delta \le \gamma_{k_A}^{2} \bar{g} |L| \label{trade_off_Lemma_main} ,
	\end{align}
	where we define
	\begin{align}
		\gamma_{k_A} := \sqrt{ 18 k_A^2 k (k+k_A)}.
	\end{align}
	In particular, we simply denote $\gamma = 6k^{2}$ when $k_A=k$ by omitting the index $k_A$.
\end{lemma}
\textit{Proof of Lemma~\ref{trade_off_variance}}.
We begin with the fact that $\ket{\Gs}$ is the ground state of $H$, satisfying $H \ket{\Gs} = E_{0} \ket{\Gs}$. We utilize the following inequality:
\begin{align}
	\langle \Omega | A_{L} (H - E_{0}) A_{L} | \Omega \rangle  = \langle \Omega | A_{L} (1 - \Pi_{\Omega}) (H - E_{0}) (1 - \Pi_{\Omega}) A_{L} | \Omega \rangle &\geq \langle \Omega | A_{L} (1 - \Pi_{\Omega}) A_{L} | \Omega \rangle \cdot \Delta \nonumber \\
	&= \mathrm{Var}_{\Gs} (A_L) \cdot \Delta \label{Start_lem_trade_off_proof}
\end{align}
with $\Pi_{\Omega} := |\Omega \rangle \langle \Omega |$, where we use $H \Pi_{\Omega} = E_{0} \Pi_{\Omega}$ and $(H - E_{0}) (1 - \Pi_{\Omega}) \geq \Delta (1 - \Pi_{\Omega})$.
Then, 
\begin{align}
	\langle \Omega | A_{L} (H - E_{0}) A_{L} | \Omega \rangle = - \frac{1}{2} \langle \Omega | [A_{L},[A_{L},H - E_{0}]] | \Omega \rangle = - \frac{1}{2} \langle \Omega | [A_{L},[A_{L},H]] | \Omega \rangle.
\end{align}
The inequality~\eqref{Start_lem_trade_off_proof} reduces to
\begin{align}
	\mathrm{Var}_{\Gs} (A_L) \cdot \Delta \leq - \frac{1}{2} \langle \Omega | [A_{L},[A_{L},H]] | \Omega \rangle \leq \frac{\Vert [A_{L},[A_{L},H]] \Vert}{2}. \label{sec_lem_trade_off_proof}
\end{align}
Finally, we need to estimate the norm $\Vert [A_{L},[A_{L},H]] \Vert$. For the purpose, we utilize the statement given in Ref.~\cite[Theorem~1]{Kuwahara_2016_njp}, which gives the following upper bound
\begin{align}
	\left\Vert \brr{A_{L},\Gamma^{(q)}} \right\Vert \leq 6 k_{A} q \left\Vert \Gamma^{(q)} \right\Vert, \label{njp_Kuwahara_Theorem1}
\end{align}
where $\Gamma^{(q)}$ is an arbitrary $q$-local operator.

In order to use the bound~\eqref{njp_Kuwahara_Theorem1}, we first note that
\begin{align}
	[A_{L},H] = \brr{A_{L},\widehat{H}_{L}}, \qquad \widehat{H}_{L} := \sum_{Z : Z \cap L \neq \emptyset} h_{Z}. 
\end{align}
Then, $[A_{L},H]$ is at most $(k + k_{A})$-local, and hence by applying Eq.~\eqref{njp_Kuwahara_Theorem1} with $\Gamma^{(q)} = [A_{L},H]$, we have
\begin{align}
	\left\Vert [A_{L},[A_{L},H]] \right\Vert \leq 6k_{A} (k + k_{A}) \left\Vert \brr{A_{L},\widehat{H}_{L}} \right\Vert. \label{[A_L, [A_L, H ] ]_norm}
\end{align}
In the same way, we obtain the upper bound for $\norm{\brr{A_{L},\widehat{H}_{L}}}$ as follows
\begin{align}
	\left\Vert \brr{A_{L},\widehat{H}_{L}} \right\Vert \leq 6k_{A} k \norm{\widehat{H}_{L}}. \label{[A_L, H ]_norm}
\end{align}
By combining the inequalities~\eqref{[A_L, [A_L, H ] ]_norm} and \eqref{[A_L, H ]_norm}, we obtain
\begin{align}
	\left\Vert \brr{A_{L},[A_{L},H]} \right\Vert \leq 36 k_{A}^{2} k (k + k_{A}) \norm{\widehat{H}_{L}} \leq 36 k_{A}^{2} k (k + k_{A}) \bar{g} |L|, \label{[A_L, [A_L, H ] ]_norm_final}
\end{align}
where the second inequality results from the condition~\eqref{upp_bound_norm_widehat_H_L}. Finally, by combining the inequalities~\eqref{sec_lem_trade_off_proof} and \eqref{[A_L, [A_L, H ] ]_norm_final}, we prove the main inequality~\eqref{trade_off_Lemma_main}. This completes the proof of Lemma~\ref{trade_off_variance}. $\square$

\subsection{Decomposition of local interactions}

The proof of the robustness strongly depends on the convenient decomposition as 
\begin{align}
	\label{H_i^co_v_decomposition_assumption_1form}
	H_{i,i^\co} =\sum_{v=1}^{d^2}   \frac{1}{n} \mathcal{X}_{i,v} \otimes H_{i^\co,v} ,
\end{align}
with 
\begin{align}
	\label{H_i^co_v_decomposition}
	H_{ i^\co,v} = \sum_{Z_1 \subset i^\co: |Z_1|\le k-1} h_{v, Z_1} .
\end{align}
Here, for $Z= \{i\} \cup Z_1$, we have 
\begin{align}
	h_{\{i\} \cup Z_1} = \frac{1}{n} \sum_{v=1}^{d^2} \mathcal{X}_{i,v} \otimes h_{v, Z_1} ,
\end{align}
We can always find an appropriate set $\{\mathcal{X}_{i,v}\}_v$ such that $\norm{\mathcal{X}_{i,v}}=1$ and 
\begin{align}
	\label{mathjcal_P_i_v_prop}
	\sum_{Z_1: Z_1 \ni i'} \norm{h_{v, Z_1}} \le n \sum_{Z_1: Z_1 \ni i'} \norm{h_{\{i\} \cup Z_1}} =n \sum_{Z: Z \ni i,i'} \norm{h_Z} \le  g_1,\quad \forall v.
\end{align}

For $Z=\{i\} \cup Z_1$, by decomposing 
\begin{align}
	&h_Z = \sum_{w_1=1}^{d}  \ket{w_1} \bra{w_1}_i \otimes \bra{w_1} h_{Z} \ket{w_1}_i + 
	\frac{1}{2} \sum_{w_1< w_2}^{d}\brr{  \sigma_{i, w_1,w_2}^{x} \otimes \tr_{i} \br{ \sigma_{i, w_1,w_2}^{x}  h_{Z} } +  \sigma_{i, w_1,w_2}^{y} \otimes \tr_{i} \br{ \sigma_{i, w_1,w_2}^{y}h_{Z} }} , \notag \\
	& \sigma_{w_1,w_2}^{x} := \ket{w_1} \bra{w_2} +  \ket{w_2} \bra{w_1},\quad   \sigma_{w_1,w_2}^{y} := i \ket{w_1} \bra{w_2} - i  \ket{w_2} \bra{w_1} ,
\end{align}
each of the operators $\bra{w_1} h_{Z} \ket{w_1}_i$, $\tr_{i} \br{ \sigma_{i, w_1,w_2}^{x}  h_{Z} } /2$, and $\tr_{i} \br{ \sigma_{i, w_1,w_2}^{y}h_{Z} } / 2$ corresponds to $h_{v, Z_1}/n$ in Eq.~\eqref{H_i^co_v_decomposition} with a different $v$, where $\{\ket{v}_i\}_{v=1}^{d}$ is orthonormal base on $i$. 
Then, we prove $\norm{ \sigma_{w_1,w_2}^{x}}= \norm{\sigma_{w_1,w_2}^{y}}=1$.
Moreover, we can immediately ensure that the norms of  $\norm{\bra{w_1} h_{Z} \ket{w_1}_i}$, $\norm{\tr_{i} \br{ \sigma_{i, w_1,w_2}^{x}  h_{Z} }/2}$, and $\norm{\tr_{i} \br{ \sigma_{i, w_1,w_2}^{y}h_{Z} }/2}$ are smaller than or equal to $\norm{h_Z}= \norm{h_{\{i\} \cup Z_1}}$, which also yields $\norm{h_{v, Z_1}} \le n\norm{h_Z}$ and inequality~\eqref{mathjcal_P_i_v_prop}.

Therefore, we conclude that each of $\{H_{i^\co,v} \}_{v=1}^{d^2}$ satisfies the $g_1$ extensiveness as follows:  
\begin{align}
	\label{g_1_extensive_decomped_ham}
	H_{ i^\co,v} = \sum_{Z_1 \subset i^\co: |Z_1|\le k-1} h_{v, Z_1},\quad  \sum_{Z_1: Z_1 \ni i'}\norm{h_{v, Z_1}} \le g_1, \quad \forall v .
\end{align}

\subsection{Proof of Theorem~\ref{prop:Robusness for local unitary}}

The key idea is to utilize the decomposition as follows:
\begin{align}
	\label{start_bootstrapped_Robustness}
	\bra{\Omega} \left( u_i^\dagger H u_i - H \right) \ket{\Omega}
	&=\bra{\Omega} \left( u_i^\dagger H_{i,i^\co} u_i - H_{i,i^\co} \right) \ket{\Omega} \notag \\
	&= \frac{1}{n} \sum_{v=1}^{d^2}  \bra{\Omega} \br{ u_i^\dagger \mathcal{X}_{i,v}  \otimes H_{i^\co,v}u_i - \mathcal{X}_{i,v} \otimes H_{i^\co,v}}\ket{\Omega}  \notag \\
	&= \frac{1}{n} \sum_{v=1}^{d^2}  \bra{\Omega}u_i^\dagger   \br{ \mathcal{X}_{i,v}- \bra{0}\mathcal{X}_{i,v} \ket{0}_i} \otimes 
	\br{ H_{i^\co,v} -\bra{\Omega} H_{i^\co,v}  \ket{\Omega}} u_i  \ket{\Omega}  \notag \\
	&\quad -  \frac{1}{n} \sum_{v=1}^{d^2}  \bra{\Omega}  \br{  \mathcal{X}_{i,v}  - \bra{0}\mathcal{X}_{i,v} \ket{0}_i} \otimes 
	\br{ H_{i^\co,v} -\bra{\Omega} H_{i^\co,v} \ket{\Omega}} \ket{\Omega} ,
\end{align}
where we use the decomposition~\eqref{H_i^co_v_decomposition_assumption_1form} and the fact that the unitary $u_{i}$ does not change the local reduced density matrix for $\rho_0=\ket{\Omega}\bra{\Omega}$; that is, 
\begin{align}
	\bra{\Omega} \left( u_i^\dagger \Phi_i u_i - \Phi_i \right) \ket{\Omega}= 
	\bra{\Omega} \left( u_i^\dagger \Phi_{i^\co} u_i - \Phi_{i^\co} \right) \ket{\Omega}= 0
\end{align}
for arbitrary operators $\Phi_i $ and $\Phi_{i^\co}$ that are supported on $i$ and $i^\co$ respectively.\footnote{In this equation, $\ket{0}_i$ is chosen as the mean-field state which is constructed from the Schmidt decomposition with the largest coefficient.}

Next, we upper-bound
\begin{align}
	&\abs{ \bra{\Omega}u_i^\dagger \tilde{\mathcal{X}}_{i,v}  \otimes 
		\tilde{H}_{i^\co,v} u_i  \ket{\Omega}}   \notag \\
	&\tilde{\mathcal{X}}_{i,v}:= \mathcal{X}_{i,v}- \bra{0}\mathcal{X}_{i,v} \ket{0}_i ,\quad 
	\tilde{H}_{i^\co,v}:= H_{i^\co,v} -\bra{\Omega} H_{i^\co,v} \ket{\Omega} . 
\end{align}
We introduce the projection $P_i$ and $Q_i$ as follows:
\begin{align}
	\label{P_i_Q_i_definitiion/mean_field}
	P_i = (\ket{0} \bra{0})_i ,\quad Q_i = 1- P_i  ,
\end{align}
which are supported on the site $i$.
Then, we obtain 
\begin{align}
	\tilde{\mathcal{X}}_{i,v}=  P_i\tilde{\mathcal{X}}_{i,v} Q_i +Q_i\tilde{\mathcal{X}}_{i,v} P_i+ Q_i \tilde{\mathcal{X}}_{i,v}Q_i   .
\end{align}

Using the above decomposition, we have 
\begin{align}
	&\abs{ \bra{\Omega}u_i^\dagger \tilde{\mathcal{X}}_{i,v}  \otimes 
		\tilde{H}_{i^\co,v} u_i  \ket{\Omega}} 
	\le 3 \norm{ \tilde{H}_{i^\co,v}  \ket{\Omega}} \cdot \norm{\tilde{\mathcal{X}}_{i,v} Q_i \ket{\Omega} } ,
\end{align}
where we use 
\begin{align}
	\abs{ \bra{\Omega}u_i^\dagger P_i\tilde{\mathcal{X}}_{i,v} Q_i  \otimes 
		\tilde{H}_{i^\co,v} u_i  \ket{\Omega}} 
	&= \abs{ \bra{\Omega} \tilde{H}_{i^\co,v} u_i^\dagger P_i\tilde{\mathcal{X}}_{i,v} Q_i  \otimes 
		u_i  \ket{\Omega}} \notag \\ 
	&\le   \norm{ \tilde{H}_{i^\co,v}  \ket{\Omega}} \br{\bra{\Omega} u_i^\dagger  Q_i\tilde{\mathcal{X}}_{i,v} P_i \tilde{\mathcal{X}}_{i,v}Q_i u_i\ket{\Omega}}^{1/2}  \notag \\
	&\le  \norm{ \tilde{H}_{i^\co,v}  \ket{\Omega}} \cdot \norm{\tilde{\mathcal{X}}_{i,v} Q_i \ket{\Omega} } .
\end{align}
The same upper bound holds for $Q_i\tilde{\mathcal{X}}_{i,v} P_i$ and $Q_i \tilde{\mathcal{X}}_{i,v}Q_i$. 

Then, we consider $ \norm{ \tilde{H}_{i^\co,v}  \ket{\Omega}}$ and $\norm{\tilde{\mathcal{X}}_{i,v} Q_i \ket{\Omega} }$.
For $ \norm{ \tilde{H}_{i^\co,v}  \ket{\Omega}}$, we can write 
\begin{align}
	\norm{ \tilde{H}_{i^\co,v}  \ket{\Omega}}^2 = \bra{\Omega} \br{H_{i^\co,v} -\bra{\Omega} H_{i^\co,v} \ket{\Omega}}^2  \ket{\Omega}= \mathrm{Var}_{\Omega}(H_{i^\co,v}).
\end{align}
Hence, from the condition~\eqref{g_1_extensive_decomped_ham} ($g_1$ extensiveness of $H_{i^\co,v}$), we can utilize Lemma~\ref{trade_off_variance}  with $k_A =k$ and $A_L= H_{i^\co,v}/g_1$\footnote{We remind that the operator $A_L$ is normalized so that it is $1$-extensive.}
From inequality~\eqref{trade_off_Lemma_main} in the lemma, we can obtain $\mathrm{Var}_{\Omega}(H_{i^\co,v}/g_1) \le \gamma^2  \bar{g} n/\Delta$, where we use that the  full Hamiltonian is $\bar{g}$ extensive from \eqref{g_ext_full_Ham}.
Therefore, we arrive at the inequality of  
\begin{align}
	\norm{ \tilde{H}_{i^\co,v}  \ket{\Omega}} \le \gamma g_1 \sqrt{\frac{ \bar{g} n}{\Delta}} .
\end{align}

On the other hand, for $\norm{\tilde{\mathcal{X}}_{i,v} Q_i \ket{\Omega} }$, we trivially obtain 
\begin{align}
	\label{norm_tilde_X_Q_i_trivial_bound}
	\norm{\tilde{\mathcal{X}}_{i,v} Q_i \ket{\Omega} } \le \norm{\tilde{\mathcal{X}}_{i,v}} \le 2\norm{\mathcal{X}_{i,v}} =2 , 
\end{align}
where we use $\norm{\mathcal{X}_{i,v}} =1$. 
Thus, we upper-bound
\begin{align}
	\abs{ \bra{\Omega}u_i^\dagger \tilde{\mathcal{X}}_{i,v}  \otimes 
		\tilde{H}_{i^\co,v} u_i  \ket{\Omega}} 
	&\le 3 \norm{ \tilde{H}_{i^\co,v}  \ket{\Omega}} \cdot \norm{\tilde{\mathcal{X}}_{i,v} Q_i \ket{\Omega} }   \notag \\
	&\le 3 \cdot 2 \cdot \gamma g_1 \sqrt{\frac{ \bar{g} n}{\Delta}}  = 6 \gamma g_1\sqrt{\frac{ \bar{g} n}{\Delta}} .
\end{align}
By applying the above to Eq.~\eqref{start_bootstrapped_Robustness}, 
we obtain 
\begin{align}
	\label{bootstrapped_Robustness_0th_ver}
	\abs{ \bra{\Omega} \left( u_i^\dagger H u_i - H \right) \ket{\Omega}} 
	\le 12 \gamma g_1 d^2 \sqrt{\frac{ \bar{g}}{n \Delta}} =:\delta_{\rm Rob}^{(1)} . 
\end{align}

To improve Inequality~\eqref{bootstrapped_Robustness_0th_ver}, we rely on the bootstrapping method as follow.
We first note that Proposition~\ref{prop:Largest local Schmidt coefficient} gives
\begin{align}
	\label{Q_i_proj_norm}
	\norm{Q_i \ket{\Omega} }^2 =\sum_{j\ge 1} | \lambda_j|^2 \le \frac{2\delta_{\rm Rob}}{\Delta} , \quad \forall i\in \Lambda, 
\end{align}
where we used inequality~\eqref{mean/field_error_bo8nd}.
This improves inequality~\eqref{norm_tilde_X_Q_i_trivial_bound} to 
\begin{align}
	\label{norm_tilde_X_Q_i_trivial_bound_refine}
	\norm{\tilde{\mathcal{X}}_{i,v} Q_i \ket{\Omega} } \le \norm{\tilde{\mathcal{X}}_{i,v}} \cdot \norm{Q_i \ket{\Omega} } \le 
	2 \sqrt{\frac{2\delta_{\rm Rob}}{\Delta} } .
\end{align}
By applying it with $\delta_{\rm Rob} \le \delta_{\rm Rob}^{(1)}$, we also refine the upper bound~\eqref{bootstrapped_Robustness_0th_ver} by 
\begin{align}
	\label{bootstrapped_Robustness_1st_ver}
	\abs{ \bra{\Omega} \left( u_i^\dagger H u_i - H \right) \ket{\Omega}} 
	\le 6 \gamma g_1 d^2 \sqrt{\frac{ \bar{g}}{n \Delta}} \cdot 2\sqrt{\frac{2\delta_{\rm Rob}^{(1)}}{\Delta} }  =:\delta_{\rm Rob}^{(2)} . 
\end{align}
Thus, by defining $\tilde{C}$ as 
\begin{align}
	\tilde{C}:=  \frac{12\gamma g_1 d^2}{\Delta} \sqrt{\frac{2 \bar{g}}{n}} ,
\end{align}
we iteratively improve the inequality~\eqref{bootstrapped_Robustness_1st_ver} to 
\begin{align}
	\abs{ \bra{\Omega} \left( u_i^\dagger H u_i - H \right) \ket{\Omega}} \le \delta_{\rm Rob}^{(s+1)} := \tilde{C} \sqrt{\delta_{\rm Rob}^{(s)}}  .
\end{align}
By setting $\delta_{\rm Rob}^{(0)}=\Delta/2$, which gives $\delta_{\rm Rob}^{(1)}=\tilde{C} \sqrt{\Delta/2}$ in the same form as in \eqref{bootstrapped_Robustness_0th_ver}, we can solve the above relation as follows:
\begin{align}
	\delta_{\rm Rob}^{(1)} = \tilde{C}\sqrt{\delta_{\rm Rob}^{(0)}},\quad 
	\delta_{\rm Rob}^{(2)} = \tilde{C}\brr{  \tilde{C}\sqrt{\delta_{\rm Rob}^{(0)}}}^{1/2}, \quad 
	\delta_{\rm Rob}^{(3)} = \tilde{C} \brr{  \tilde{C}\brr{  \tilde{C}\sqrt{\delta_{\rm Rob}^{(0)}}}^{1/2}}^{1/2}  ,\ldots 
\end{align}
which eventually gives 
\begin{align}
	\label{delta_rm_rob_final_up}
	\delta_{\rm Rob}^{(\infty)} =\lim_{s_0\to\infty} \tilde{C}^{\sum_{s=0}^{s_0-1} 2^{-s}} \br{\delta_{\rm Rob}^{(0)}}^{2^{-s_0}} = \tilde{C}^2 = 
	\br{\frac{12 \gamma g_1 d^2}{\Delta}}^2 \frac{2\bar{g}}{n}   , 
\end{align}
which is the desired upper bound~\eqref{delta_Rob_explicit} for $ \delta_{\rm Rob}$.

This completes the proof. $\square$

{~} \\

As a short remark, we emphasize that the above bootstrapping method yields a better bound only for $\Delta \gtrsim 1/n^{1/3}$, beyond which the weak bound~\eqref{bootstrapped_Robustness_0th_ver}, i.e., $\delta_{\rm Rob}^{(1)} \propto (n\Delta)^{-1/2}$, gives a better upper bound than $ \delta_{\rm Rob}^{(\infty)} \propto (n\Delta^2)^{-1}$.
This point leads to the refined upper bound~\eqref{delta_Rob_explicit_2}. 

\section{Probability distribution for the number of flipped spins}

\subsection{Main statement} \label{sec:Main statement}

Here, we are interested in the probability of observing the deviated spins from the mean-field states. 
To formulate the problem, we also consider the 1-local 1-extensive operator $M$ as follows:
\begin{align}
	M= \sum_{i\in \Lambda} Q_i =  \sum_{i\in \Lambda} \brr{1- (\ket{0}\bra{0})_i} , 
\end{align}
where we use the notation~\eqref{P_i_Q_i_definitiion/mean_field}.
Then, the observable $M$ characterizes the number of spins that are deviated from the mean-field states.

In the new formalism, we aim to analyze the probability distribution as follows:
\begin{align}
	\norm{ \Pi_{> z} \ket{\Omega} }^2 ,
\end{align}
where we define $\Pi_{> z}$ as projection onto the eigenbases of $M$ such that the eigenvalues are larger than $z$.
Our task is to derive an upper bound in the form of 
\begin{align}
	\norm{ \Pi_{> z} \ket{\Omega} }^2 \le  e^{-{\rm const.} \times z^{1/2} } .
\end{align}

We prove the following theorem:
\begin{theorem} \label{Them:prob_dist_z}
	As in Corollary~\ref{corol:log_entanglement}, we assume 
	\begin{align}
		\label{assump_for_n_ge_n_0}
		n \ge n_0,\quad n_0:=  \br{\frac{12 \gamma g_1 d^2}{\Delta}}^2 \frac{8\bar{g}}{\Delta} .
	\end{align}
	We define the parameters $\eta$ and $\tilde{z}$ as follows:
	\begin{align}
		\label{Them:prob_dist_z_basic_para}
		&\eta = 2^{2k+4} \bar{g}  , \quad 
		\tilde{z}:= 2^{12} \gamma^2 d^4  \br{\bar{g}/\Delta}^{3} .
	\end{align}
	Then, for $z\ge \tilde{z}$, we upper-bound the square root of the probability $\bra{\Omega} \Pi_{> z}  \ket{\Omega}$ as follows:
	\begin{align}
		\label{Them:prob_dist_z_main_ineq}
		\norm{\Pi_{> z}  \ket{\Omega}}\le C_1 e^{-C_2 \sqrt{z}}  
	\end{align} 
	with
	\begin{align}
		\label{Them:prob_dist_z_main_ineq_parameter}
		C_1:= 2 + 16 e^{\tilde{z}/k}    , \quad C_2= \brr{ \frac{4\Delta}{(\tilde{z}+2k)(16 e \eta k+\Delta)}}^{1/2} . 
	\end{align}
\end{theorem}

{\bf Remark.}
Several key points are summarized as follows:
\begin{enumerate}
	\item{} We can approximate the ground state up to an error $\epsilon$ by considering at most $\log^2(1/\epsilon)$ deviated spins, 
	and hence the total number of parameters to describe the ground state is given by $$d^{\orderof{z}}\binom{n}{z} = e^{\log(n) \log^2(1/\epsilon)}.$$
	This also characterizes the sufficient bond dimension to represent the ground state by using the matrix product state.
	\item{} In the case of the uniform (exchange invariance) systems, we can construct the ground state by $D=\orderof{z^d}$ (see also Sec.~\ref{Sec:Uniform Hamiltonians}), and hence we only require the poly-logarithmic complexity when $z=\log^2(1/\epsilon)$. 
\end{enumerate}

\subsection{Method 1: lower bound on the probability for small $z$} \label{Lower_bound_probability}

We begin with the lower bound of  
\begin{align}
	\norm{ \Pi_{\le z} \ket{\Omega} }^2 .
\end{align}
We trivially obtain 
\begin{align}
	\bra{\Omega} M  \ket{\Omega} \ge z \norm{\Pi_{> z} \ket{\Omega}}^2 .
\end{align}
On the other hand, from inequality~\eqref{Q_i_proj_norm} and 
\begin{align}
	\bra{\Omega} M  \ket{\Omega} = \sum_{i \in \Lambda} \bra{\Omega} Q_i  \ket{\Omega} \le \frac{2n \delta_{\rm Rob}}{\Delta} \le 
	\frac{4\bar{g}}{\Delta} \br{\frac{12 \gamma g_1 d^2}{\Delta}}^2 =: z'_0 ,
\end{align}
we can derive 
\begin{align}
	z \norm{\Pi_{> z} \ket{\Omega}}^2 \le z_0' \longrightarrow \norm{\Pi_{> z} \ket{\Omega}}^2 \le \frac{z_0'}{z}.
\end{align}
Here, we note that the constant $z'_0$ does not depend on the system size. 
Therefore, by choosing $z_0=2z_0'$, we obtain 
\begin{align}
	\label{definitiion_z_0}
	&\norm{ \Pi_{> z_0} \ket{\Omega} }^2 \le \frac{1}{2}  ,\quad  \norm{ \Pi_{\le z_0} \ket{\Omega} }^2 \ge \frac{1}{2}   \notag \\
	&\for z_0=   \frac{8\bar{g}}{\Delta} \br{\frac{12 \gamma g_1 d^2}{\Delta}}^2  . 
\end{align}

\subsection{Method 2: probability distribution under a weak hopping amplitude}

We consider the tight-binding formalism as in Ref.~\cite{Kuwahara_2016_asymptotic}. 
For this purpose, we here decompose the ground state $\ket{\Omega}$ by using the eigenstates of $M$:
\begin{align}
	\ket{\Omega} = \sum_{x=0}^{n} c_x \ket{\omega_x} , \qquad  M \ket{\omega_x} = x \ket{\omega_x} ,
\end{align}  
where $\ket{\omega_x}$ represents the normalized state derived from projecting the ground state onto the eigenspace of $M$ corresponding to the eigenvalue $x$, i.e., $\ket{\omega_x} \propto \Pi_{=x} \ket{\Omega}$.
Then, we decompose the Hamiltonian $H$ by using the projection $\Pi_{=x}$ as follows:
\begin{align}
	\label{def:breve_Hami}
	H= \sum_{x,x'=0}^{n}  \Pi_{=x}  H\Pi_{=x'}  , 
\end{align}  
which is similar to the one-particle tight-binding model. 
Also, since the $h_Z$ term with $|Z| \le k$ influence less than or equal to $k$ sites, we have 
\begin{align}
	\label{Hopping_length_xx'}
	\Pi_{=x}  H\Pi_{=x'} =0  \for  |x-x'|>k . 
\end{align}  

Here, the primary task is to estimate the norm of $\Pi_{\le z}  H\Pi_{\le z}$.
In general, the following upper bound holds 
\begin{align}
	\norm{\Pi_{\le z}  H\Pi_{\le z}}= \orderof{n}  ,
\end{align}  
which is too loose for our purpose. 
In the proof of Theorem~\ref{Them:prob_dist_z}, the following estimation plays a key role: 
\begin{align}
	\label{condition_weak_hopping}
	\norm{\Pi_{\le z}  H\Pi_{\le z}} = \orderof{z},
\end{align}  
which means that the hopping amplitude increases at most linearly with $z$. 

After obtaining Eq.~\eqref{condition_weak_hopping}, we can derive the sub-exponential decay of the probability distribution, which is ensured by the following proposition:
\begin{prop} \label{prob_tight_binding_new}
	For a target $z \in \mathbb{N}$ and given $E^\ast \in \mathbb{R}$ and $\tilde{z}$, we assume the following bound for $z\ge \tilde{z}$: 
	\begin{align}
		\label{Assump_breve_H_lez}
		\norm{  \Pi_{\le z} (H -E^\ast) \Pi_{\le z}} \le \eta z  \for z\ge \tilde{z}.
	\end{align}  
	Then, the probability for the operator $M$ is upper-bounded as follows:
	\begin{align}
		\label{prob_tight_binding_new/main}
		\norm{\Pi_{> z}  \ket{\Omega}}
		&\le \frac{1 + 4 C_{\tilde{z}}}{\bra{\Omega} \Pi_{\le \tilde{z}} \ket{\Omega} }  e^{-2m /\xi_m}  \for z\ge \tilde{z} , \quad  m=\left \lfloor \frac{z-\tilde{z}}{k}  \right \rfloor , 
	\end{align}  
	with 
	\begin{align}
		\xi_m= \br{ 1+ \frac{4m}{\beta_0 \Delta} }^{1/2}, \quad 
		\beta_0= \frac{1}{4e \eta k},\quad C_{\tilde{z}} := 2e^{\tilde{z}/k}   .
	\end{align}
\end{prop}

{\bf Remark.}
From the upper bound, we can ensure that the probability of observing a large $z$ decays subexponentially $e^{-\orderof{\sqrt{z}}}$, and our task is to derive the parameter $\eta$ for an appropriate regime of $z$.

To simplify the bound~\eqref{prob_tight_binding_new/main}, we use 
\begin{align}
	m=\left \lfloor \frac{z-\tilde{z}}{k}  \right \rfloor \ge \frac{z}{\tilde{z}+2k},  
\end{align}  
where we assume $z\ge \tilde{z}+k$, which also ensure $m\ge 1$. 
We also derive
\begin{align}
	\frac{m}{\xi_m} = m^{1/2} \br{ m^{-1}+ \frac{16 e \eta k}{\Delta} }^{-1/2}
	\ge z^{1/2}\brr{ \frac{ \Delta}{(\tilde{z}+2k)(16 e \eta k+\Delta)}}^{1/2}  .
\end{align}
Then, the upper bound~\eqref{prob_tight_binding_new/main} reduces to 
\begin{align}
	\label{prob_tight_binding_new/main_simple}
	\norm{\Pi_{> z}  \ket{\Omega}}
	&\le \frac{1 + 8e^{\tilde{z}/k}   }{\bra{\Omega} \Pi_{\le \tilde{z}} \ket{\Omega} }  \exp \brrr{- z^{1/2}\brr{ \frac{4\Delta}{(\tilde{z}+2k)(16 e \eta k+\Delta)}}^{1/2}} 
\end{align}  
for $z\ge \tilde{z}+k$.
The above inequality still holds for $z\le \tilde{z}+k$ since the RHS of the inequality is larger than $1$ for $z\le \tilde{z}+k$.
Therefore, we adopt inequality~\eqref{prob_tight_binding_new/main_simple} for all $z$. 

Therefore, we can see that the main inequality~\eqref{Them:prob_dist_z_main_ineq} with \eqref{Them:prob_dist_z_main_ineq_parameter} immediately follows from inequality~\eqref{prob_tight_binding_new/main_simple} after we get the parameters $\eta$ and $\tilde{z}$, where the parameter $\tilde{z}$ is set so that the condition~\eqref{Assump_breve_H_lez} holds for $z\ge \tilde{z}$.

{~}
\\

\textit{Proof of Proposition~\ref{prob_tight_binding_new}.}
Throughout the proof, we use the notation
\begin{align}
	\ket{\Omega_{\le \tilde{z}}} := \Pi_{\le \tilde{z}} \ket{\Omega} .
\end{align}  
We employ a similar proof technique to establish local reversibility in Ref.~\cite{Kuwahara_2017}.  
First, we aim to derive the inequality of 
\begin{align}
	\label{AKL_type_upp}
	&\norm{\Pi^{H}_{> E} \ket{\Omega_{\le \tilde{z}}}} \le C_{\tilde{z}} e^{-\beta_0 E}, \notag \\
	& \beta_0= \frac{1}{4e \eta k},\quad C_{\tilde{z}} := 2e^{\tilde{z}/k}  .
\end{align}
To prove it, we calculate 
\begin{align}
	\label{target_quantity_ineq}
	\norm{\Pi^{H}_{> E}e^{-\beta H} e^{\beta H} \ket{\Omega_{\le \tilde{z}}}} \le e^{-\beta E} \norm{e^{\beta H} \ket{\Omega_{\le \tilde{z}}}} . 
\end{align}

To upper-bound the norm $\norm{e^{\beta H} \ket{\Omega_{\le \tilde{z}}}} $, we first decompose
\begin{align}
	e^{\beta H} \ket{\Omega_{\le \tilde{z}}} = e^{\beta H} \Pi_{\le \tilde{z}} \ket{\Omega}  =\sum_{m=0} \frac{\beta^m}{m!} \ad_H^m( \Pi_{\le \tilde{z}}) \ket{\Omega} 
	=\sum_{m=0} \frac{\beta^m}{m!} \ad_{H^\ast}^m( \Pi_{\le \tilde{z}}) \ket{\Omega}  ,
\end{align}
with 
\begin{align}
	H^\ast := H - E^\ast ,
\end{align}  
where we defined $\ad_H(\cdot) := [H,\cdot]$ and use $H\ket{\Omega } = 0$ and $\ad_H(\cdot) = \ad_{H^\ast} (\cdot)$. 
Then, we note that 
\begin{align}
	&\br{H^\ast}^m  \Pi_{\le \tilde{z}} = \Pi_{\le \tilde{z}+ km}  \br{H^\ast}^m  \Pi_{\le \tilde{z}} =  \br{\Pi_{\le \tilde{z}+ km}  H^\ast  \Pi_{\le \tilde{z}+ km}}^m  \Pi_{\le \tilde{z}}  ,\notag \\
	&  \Pi_{\le \tilde{z}} \br{H^\ast}^m =  \Pi_{\le \tilde{z}}    \br{\Pi_{\le \tilde{z}+ km}  H^\ast  \Pi_{\le \tilde{z}+ km}}^m ,
\end{align}  
where we  use the fact that $\br{H^\ast}^m$ is at most $(km)$ local. 
Therefore, we can also ensure that $\ad_{H^\ast}^m( \Pi_{\le \tilde{z}})$ reduces to 
\begin{align}
	\ad_{H^\ast}^m( \Pi_{\le \tilde{z}})  = \ad_{ \Pi_{\le km + \tilde{z}} H^\ast  \Pi_{\le km + \tilde{z}}  }^m( \Pi_{\le \tilde{z}}) 
	=  \ad_{H_{\le km + \tilde{z}}^\ast}^m \br{ \Pi_{\le \tilde{z}} } 
\end{align}  
by denoting $$H^\ast_{\le z} = \Pi_{\le z} H^\ast \Pi_{\le z}.$$ 
Therefore, under the condition~\eqref{Assump_breve_H_lez}, we obtain 
\begin{align}
	\norm{e^{\beta H} \ket{\Omega_{\le \tilde{z}}}} 
	&\le \sum_{m=0}^\infty \frac{\beta^m}{m!}\norm{\ad_{H_{\le km + \tilde{z}}^\ast}^m  \br{ \Pi_{\le \tilde{z}} } } \notag \\
	&\le 1 + \sum_{m=1}^\infty \frac{(2\beta)^m}{m!} \brr{ \eta (k m+\tilde{z})}^m  \notag \\
	&= 1 + \sum_{m=1}^\infty  \frac{(2\beta \eta k m)^m}{m!} \br{  1 +\frac{\tilde{z}}{k m}}^m  \notag \\
	&\le  1 + e^{\tilde{z}/k} \sum_{m=1}^\infty \brr{\frac{2\beta \eta k m}{(m/e)} }^m \notag \\
	&\le   1 + e^{\tilde{z}/k}    \notag \\
	&\le 2e^{\tilde{z}/k} \for \beta= \beta_0= \frac{1}{4 e \eta k},
\end{align}
where we use $\norm{\ad^m_{O_1} (O_2)} \le (2\norm{O_1})^m\norm{O_2}$ for arbitrary operators $O_1$ and $O_2$.
By applying the above upper bound to Inequality~\eqref{target_quantity_ineq}, we obtain the desired inequality~\eqref{AKL_type_upp}. 

In the following, we set $E^\ast=0$ without loss of generality.
Under condition~\eqref{AKL_type_upp}, we reconstruct $\ket{\Omega}$ from $\ket{\Omega_{\le \tilde{z}}} $ by 
\begin{align}
	\label{Energy_filtering}
	F_m(H) \ket{\Omega_{\le \tilde{z}}} ,
\end{align}  
where $F_m(H)$ is an $m$th-order polynomial with respect to the Hamiltonian. 
By choosing $F_m(x)$ as
\begin{align}
	F_m(x) = \frac{ T_m \brr{\frac{x -\Delta}{E_m} -1}}{ T_m \brr{\frac{-\Delta}{E_m} -1}},
\end{align}  
where $E_m$, which depends on $m$, is appropriately chosen. 
Then, we can derive from Ref.~\cite[Ineqs. (B11) and (B12)]{Kuwahara_2017}
\begin{align}
	&F_m(0) = 1,\notag \\
	&|F_m(x)| \le e^{-2m /\xi_m} ,\quad \xi_m= \br{ 1+ \frac{2E_m}{\Delta} }^{1/2} \for \Delta \le x \le 2E_m + \Delta,\notag \\
	&|F_m(x)| \le 2^m \br{\frac{x -E_m -\Delta}{E_m} }^{m}e^{-2m /\xi_m} \for x \ge 2E_m + \Delta.
\end{align} 

Then, we decompose 
\begin{align}
	\label{F_m_H_apply}
	F_m(H) \ket{\Omega_{\le \tilde{z}}} 
	&= 
	F_m(H) \br{ \Pi^{H}_{\le 2E_m + \Delta} \ket{\Omega_{\le \tilde{z}}}  + \Pi^{H}_{> 2E_m + \Delta} \ket{\Omega_{\le \tilde{z}}}} \notag \\
	&= \langle \Omega \ket{\Omega_{\le \tilde{z}}} \ket{\Omega} + F_m(H)  \Pi^{H}_{[\Delta, 2E_m + \Delta]} \ket{\Omega_{\le \tilde{z}}} 
	+F_m(H) \Pi^{H}_{> 2E_m + \Delta} \ket{\Omega_{\le \tilde{z}}} ,
\end{align}  
which also yields for an arbitrary $y>0$
\begin{align}
	\norm{ F_m(H) \ket{\Omega_{\le \tilde{z}}} - \langle \Omega \ket{\Omega_{\le \tilde{z}}} \ket{\Omega}}
	&\le  e^{-2m /\xi_m} + \sum_{j=1}^\infty \abs{F_m(2E_m + \Delta + y j)} \cdot \norm{\Pi^{H}_{(2E_m + \Delta + y (j-1) , 2E_m + \Delta + y j]}  
		\ket{\Omega_{\le \tilde{z}}} } \notag \\
	&\le e^{-2m /\xi_m} + \sum_{j=1}^\infty 2^m \br{\frac{E_m+yj}{E_m} }^{m}e^{-2m /\xi_m} 
	\cdot C_{\tilde{z}} e^{-\beta_0 \brr{2E_m + \Delta + y (j-1)}}  \notag \\
	&\le e^{-2m /\xi_m} \brrr{ 1 + C_{\tilde{z}} e^{-\beta_0 \brr{2E_m + \Delta} + m \log(2)  } \sum_{j=1}^\infty e^{y j m/E_m -\beta_0 y (j-1)} }.
\end{align}  
By choosing 
\begin{align}
	E_m = \frac{2m}{\beta_0} ,
\end{align} 
we have 
\begin{align}
	& C_{\tilde{z}} e^{-\beta_0 \brr{2E_m + \Delta} + m \log(2)  } \sum_{j=1}^\infty e^{y j m/E_m -\beta_0 y (j-1)} \notag \\
	&\le C_{\tilde{z}}  \sum_{j=1}^\infty e^{ \beta _0 y j/2 -\beta_0 y (j-1)} = C_{\tilde{z}} \cdot \frac{e^{ \beta _0 y}}{e^{ \beta _0 y /2} -1}   .
\end{align}  
By choosing $y$ such that $e^{ \beta _0 y /2}=2$, we derive 
\begin{align}
	\norm{ F_m(H) \ket{\Omega_{\le \tilde{z}}} - \langle \Omega \ket{\Omega_{\le \tilde{z}}} \ket{\Omega}}
	&\le  e^{-2m /\xi_m} \br{ 1 + 4 C_{\tilde{z}}}. 
\end{align}  

By using the above inequality, we also obtain 
\begin{align}
	\norm{\frac{ F_m(H) \ket{\Omega_{\le \tilde{z}}}}{\langle \Omega \ket{\Omega_{\le \tilde{z}}} } - \ket{\Omega}}
	&\le \frac{1}{ \langle \Omega \ket{\Omega_{\le \tilde{z}}} } e^{-2m /\xi_m} \br{ 1 + 4 C_{\tilde{z}}}  .
\end{align}  
Finally, from the equation of 
\begin{align}
	\Pi_{> \tilde{z} + km} F_m(H) \ket{\Omega_{\le \tilde{z}}} = 0, 
\end{align}  
we obtain the inequality of  
\begin{align}
	\norm{\Pi_{> \tilde{z} + km}  \ket{\Omega}}
	&\le \frac{1 + 4 C_{\tilde{z}}}{\langle \Omega \ket{\Omega_{\le \tilde{z}}}}  e^{-2m /\xi_m} = 
	\frac{1 + 4 C_{\tilde{z}}}{\bra{\Omega} \Pi_{\le \tilde{z}} \ket{\Omega}}  e^{-2m /\xi_m}   . 
\end{align}  
By using $\norm{\Pi_{> \tilde{z} + km}  \ket{\Omega}} \ge \norm{\Pi_{> z}  \ket{\Omega}}$, we reduce the above inequality to the main inequality~\eqref{prob_tight_binding_new/main}.
Note that $z\ge \tilde{z} + km$ from $m=\left \lfloor \frac{z-\tilde{z}}{k} \right \rfloor$. 
This completes the proof. $\square$

\subsection{Method 3: $\mR\mE$ decomposition}

Next, we employ the following convenient decomposition ($\mR\mE$ decomposition) for the Hamiltonian.

Importantly, the $\mR\mE$ decomposition does not define a new Hamiltonian; rather, it provides an alternative \emph{re-decomposition} of the same $k$-local Hamiltonian $H$.
Starting from a given $k$-local expansion $H=\sum_{Z:|Z|\le k} h_Z$, we re-expand each term $h_Z$ into a sum of operators supported on subsets of $Z$, thereby obtaining another $k$-local decomposition [see also Eq.~\eqref{decompostion_H_E_R} below]. 

A key advantage of this re-decomposition is that the resulting terms can be chosen to be ``centered'' with respect to the local reference state $\ket{0}_i$ (a product reference $\ket{0}^{\otimes n}$): for each decomposed term and each site (or block) in its support, the partial expectation in $\ket{0}$ vanishes [see Eq.~\eqref{decompostion_H_E_R}]. Consequently, when restricting to the low-excitation subspace consisting of states close to mean-field product state $\ket{0}^{\otimes n}$, the action of multi-site interaction terms necessarily involves local excitations, making it significantly easier to control operator norms on that subspace (in particular, projected norms admit bounds that scale linearly in the excitation number cutoff).
This point is crucial in satisfying the primary assumption~\eqref{Assump_breve_H_lez} in Proposition~\ref{prob_tight_binding_new}, which will be discussed in the subsequent section (Sec.~\ref{sec:Bound on the Hamiltonian norm in the z-flip subspace}).

For an arbitrary operator $\Phi$, we define $\mR_{i}$ and $\mE_i$ ($i\in \Lambda$)
\begin{align}
	\label{def_superop_R_0_i}
	\mathcal{R}_{i} (\Phi) = \Phi  - \bra{0} \Phi  \ket{0}_i = \Phi  -\mathcal{E}_{i} (\Phi), \quad 
	\mathcal{E}_{i} (\Phi)
	:= \bra{0} \Phi  \ket{0}_i  ,
\end{align}
where $\bra{0_i} \Phi  \ket{0_i}$ means $\bra{0_i} \Phi  \ket{0_i} \otimes \hat{1}_i$; that is, $\mathcal{E}_{i} (\Phi)$ is understood as an operator on the full Hilbert space by tensoring the identity on site $i$. 
Here, the super-operator $\mR_L$ increases the operator norm by at most $2$ times: 
\begin{align}
	\norm{\mathcal{R}_{i} (\Phi)} \le 2 \norm{\Phi} . 
\end{align}
For a two-body interaction $h_{i,i'} = h_i\otimes h_{i'}$ ($i\in L$ and $i'\in L'$), we can obtain 
\begin{align}
	\mathcal{R}_{i} \mathcal{R}_{i'} ( h_i\otimes h_{i'}) =\mathcal{R}_{i} (h_i)\otimes\mathcal{R}_{i'} (h_{i'})= \Bigl(h_i-\bra{0}h_i\ket{0}_i\Bigr)\otimes
	\Bigl(h_{i'}-\bra{0}h_{i'}\ket{0}_{i'}\Bigr). 
\end{align}
Note that $\mathcal{R}_{i} \mathcal{R}_{i'} ( h_i\otimes h_{i'}) $ is still $2$-local since $\bra{0}h_i\ket{0}_i$ and $\bra{0}h_{i'}\ket{0}_{i'}$ are constants. 

Using them, we decompose
\begin{align}
	\label{h_Z_Z_p_decomp}
	h_Z &=\brr{ \prod_{i\in Z } \br{ \mathcal{R}_{i}+\mathcal{E}_{i} } }(h_Z)
	= \sum_{p=1}^{\mN_Z}  \acute{h}_{p, Z_p}, 
\end{align}
where each of $\{Z_p\}_p$ is included in $Z$. 
In the decomposition~\eqref{h_Z_Z_p_decomp}, we define
\begin{align}
	\label{acute_h_p_Z_p}
	\acute{h}_{p, Z_p} := \prod_{i\in Z_p}   \mathcal{R}_{i}   \prod_{i\in Z \setminus Z_p}   \mathcal{E}_{i} (h_Z) 
	=   \prod_{i\in Z_p}   \mathcal{R}_{i} \br{  \bra{0} h_Z \ket{0}_{Z \setminus Z_p}} .
\end{align}

For example, for a three-body interaction $h_{i,i',i''} = h_i\otimes h_{i'}\otimes h_{i''}$ ($i\in L$, $i'\in L'$ and $i''\in L''$), we can obtain 
\begin{align}
	h_i\otimes h_{i'} \otimes h_{i''}  =& \br{ \mathcal{R}_{i}+\mathcal{E}_{i} }  \otimes
	\br{ \mathcal{R}_{i'} + \mathcal{E}_{i'} } 
	\br{ \mathcal{R}_{i''} + \mathcal{E}_{i''} } \br{h_i\otimes h_{i'} \otimes h_{i''}}\notag \\
	=& \mathcal{R}_{i} (h_i)\otimes\mathcal{R}_{i'} (h_{i'})\otimes\mathcal{R}_{i''} (h_{i''}) \notag \\
	&+ \mathcal{E}_{i} (h_i)\otimes\mathcal{R}_{i'} (h_{i'})\otimes\mathcal{R}_{i''} (h_{i''})
	+ \mathcal{R}_{i} (h_i)\otimes\mathcal{E}_{i'} (h_{i'})\otimes\mathcal{R}_{i''} (h_{i''})
	+ \mathcal{R}_{i} (h_i)\otimes\mathcal{R}_{i'} (h_{i'})\otimes\mathcal{E}_{i''} (h_{i''}) \notag \\
	&+ \mathcal{E}_{i} (h_i)\otimes\mathcal{E}_{i'} (h_{i'})\otimes\mathcal{R}_{i''} (h_{i''})
	+ \mathcal{R}_{i} (h_i)\otimes\mathcal{E}_{i'} (h_{i'})\otimes\mathcal{E}_{i''} (h_{i''})
	+ \mathcal{E}_{i} (h_i)\otimes\mathcal{R}_{i'} (h_{i'})\otimes\mathcal{E}_{i''} (h_{i''}) \notag \\
	&+ \mathcal{E}_{i} (h_i)\otimes\mathcal{E}_{i'} (h_{i'})\otimes\mathcal{E}_{i''} (h_{i''}) ,
\end{align}
where $L,L',L''$ are included in the set $\{L_j\}_{j=1}^{\bar{n}}$.
Note that $\mathcal{E}_{i} (h_i)$ is a constant, and hence $\mathcal{E}_{i} (h_i)\otimes \mathcal{R}_{i'} (h_{i'})\otimes\mathcal{R}_{i''} (h_{i''})$ becomes a two-body interaction. 

For the decomposed terms $\{\acute{h}_{p, Z_p}\}_{p=1}^{\mN_Z}$, we can derive 
\begin{align}
	\label{property:tilde_h_p_Z_p}
	&\mN_Z = 2^{|Z|} \le 2^k, \quad \norm{\acute{h}_{p, Z_p}}\le 2^k \norm{h_Z} ,     \notag \\
	&\bra{0} \acute{h}_{p, Z_p} \ket{0}_i = 0 \quad \forall i \in Z_p.
\end{align}
Moreover, using the expression~\eqref{acute_h_p_Z_p}, we decompose the original Hamiltonian into 
\begin{align}
	\label{decompostion_H_E_R}
	\acute{H}= \sum_{Z: |Z| \le k}\sum_{p=1}^{\mN_Z}  \acute{h}_{p, Z_p} .
\end{align}
Then, it satisfies 
\begin{align}
	\label{new_int_strength_upp}
	\sum_{Z: |Z| \le k}\sum_{p: Z_p\supseteq \{i,i'\}} \norm{ \acute{h}_{p, Z_p} } \le   
	\sum_{Z: |Z| \le k}\sum_{p: Z_p\supseteq \{i,i'\}}  2^k \norm{h_Z} 
	\le \sum_{Z: Z \supseteq \{i,i'\}}\sum_{p=1}^{\mN_Z}  2^k \norm{h_Z} 
	\le \frac{2^{2k} g_1}{n} ,
\end{align}
where we use inequality~\eqref{g_0_extensive_2}.

In a similar way, we derive the extensiveness of $\acute{H}$ as follows: 
\begin{align}
	\label{new_int_strength_extensive}
	\sum_{Z: |Z| \le k}\sum_{p: Z_p\ni i } \norm{ \acute{h}_{p, Z_p} } \le   
	\sum_{Z: |Z| \le k}\sum_{p: Z_p\ni i }  2^k \norm{h_Z}  \le \sum_{Z: Z \ni i }\sum_{p=1}^{\mN_Z}  2^k \norm{h_Z} \le 2^{2k} \bar{g} . 
\end{align}

\subsubsection{Properties of the $\mR\mE$ decomposition} \label{sec:prop:RE_decomp}

As an important remark, the $\mR\mE$ decomposition itself does not change the Hamiltonian: 
\begin{align}
	\acute{H}= \sum_{Z: |Z| \le k}\sum_{p=1}^{\mN_Z}  \acute{h}_{p, Z_p} =  \sum_{Z: |Z| \le k} h_{Z} = H .
\end{align}
However, when we focus on a subset-subset interaction, the decomposition yields a difference. 
For example, interactions between $L$ and $L'$ can be denoted by 
\begin{align}
	H_{L,L'} := \sum_{Z: |Z| \le k, Z\cap L\neq \emptyset, \ Z\cap L' \neq \emptyset } h_Z
\end{align}
in the original form, whereas we have 
\begin{align}
	\label{acute_H_L_L'}
	\acute{H}_{L,L'}:=  \sum_{Z: |Z| \le k}\sum_{p: Z_p \cap L\neq \emptyset, \ Z_p \cap L' \neq \emptyset }\acute{h}_{p, Z_p} 
\end{align}
under the $\mR\mE$ decomposition. 
They are not equivalent in general. If we consider $H=h_{i,i'}$ ($i\in L, i' \in L'$), we have 
\begin{align}
	H_{L,L'} = h_{i,i'} ,
\end{align}
while 
\begin{align}
	\acute{H}_{L,L'} = \mathcal{R}_{i} \mathcal{R}_{i'} (h_{i,i'}) ,
\end{align}
where the local interactions $\mathcal{E}_{i} \mathcal{R}_{i'} (h_{i,i'})$ and $\mathcal{R}_{i} \mathcal{E}_{i'} (h_{i,i'})$, as well as the constant term $\mathcal{E}_{i} \mathcal{E}_{i'} (h_{i,i'})$, are removed. 
In general, one can formally write 
\begin{align}
	\label{Decompose_RE_vs_original}
	H_{L,L'} = \acute{H}_{L,L'} + \bra{0} H_{L,L'} \ket{0}_{L'}+ \bra{0} H_{L,L'} \ket{0}_{L} - \br{ \bra{0} \otimes \bra{0} H_{L,L'}  \ket{0} \otimes \ket{0}}_{L,L'}
\end{align}
by adding the local and constant terms.

\subsection{Interaction strength under the $\mR\mE$ decomposition}

Based on the $\mR\mE$ decomposition, we consider the hopping amplitude in Eq.~\eqref{def:breve_Hami}.
To apply Proposition~\ref{prob_tight_binding_new}, we need to derive the property~\eqref{Assump_breve_H_lez} for the norm of 
$H_{\le z} = \Pi_{\le z} H \Pi_{\le z}$. 

For this purpose, we decompose the total Hamiltonian into 
\begin{align}
	\label{acute_H_definition}
	H= \sum_{i\in \Lambda} \acute{H}_i + \acute{H}^{(\ge 2)} + \check{E}_0 ,
\end{align}
where $\check{E}_0=\bra{0}H\ket{0}_\Lambda$ is the $0$-local (i.e., constant term),
$\{\acute{H}_i\}_{i\in \Lambda}$ are the on-site potentials, and $\acute{H}^{(\ge 2)}$ consists of more-than-two local terms: 
\begin{align}
	\acute{H}^{(\ge 2)} :=  \sum_{Z: |Z| \le k}\sum_{p: |Z_p|\ge 2} \acute{h}_{p, Z_p} .
\end{align}
We also note that 
\begin{align}
	\acute{H}_i  :=  \sum_{Z: |Z| \le k}\sum_{p: Z_p = \{i\}} \acute{h}_{p, Z_p}  , \quad \check{E}_0 = \sum_{Z: |Z| \leq k} \sum_{p: Z_{p} = \emptyset} \acute{h}_{p,Z_{p}}.
\end{align}

Here, our purpose is to prove 
\begin{align}
	\norm{ \Pi_{\le z} \acute{H}^{(\ge 2)} \Pi_{\le z} }= \orderof{z} ,
\end{align}
which will be proven in the following proposition: 
\begin{prop}\label{Prop:renormalized_interaction}
	The norm of the interactions $\acute{H}^{(\ge 2)}$ is upper bounded under the projection $ \Pi_{\le z} $ as follows:
	\begin{align}
		\label{Prop/main_ineq:renormalized_interaction}
		\norm{ \Pi_{\le z}  \acute{H}^{(\ge 2)}\Pi_{\le z}  } 
		\le 2^{2k+2} g_1 z  .
	\end{align}
\end{prop}

\textit{Proof of Proposition~\ref{Prop:renormalized_interaction}.}
We consider an arbitrary quantum state $\ket{\psi}$ such that 
\begin{align}
	\label{Eq.invariance_projection}
	\Pi_{\le z} \ket{\psi} = \ket{\psi} 
\end{align}
and estimate an upper bound for 
\begin{align}
	\label{Starting_inequality_amplitude}
	| \bra{\psi} \acute{H}^{(\ge 2)} \ket{\psi} | 
	&=
	\abs{ \bra{\psi} \sum_{Z: |Z| \le k}\sum_{p: |Z_p|\ge 2}  
		\acute{h}_{p, Z_p}\ket{\psi} } \notag \\
	&\le \sum_{i\in \Lambda}\sum_{i'\in i^\co}  \sum_{Z: |Z| \le k}\sum_{p: Z_p \supseteq \{i,i'\}} 
	\abs{ \bra{\psi}\acute{h}_{p, Z_p} \ket{\psi} } ,
\end{align}
where, in the last inequality, we use $|Z_p|\ge 2$ to ensure the existence of $\{i,i'\}$ ($i\neq i'$) such that $Z_p \supseteq \{i,i'\}$.

Then, from Eq.~\eqref{property:tilde_h_p_Z_p}, we obtain 
\begin{align}
	\bra{\psi} P_i  \acute{h}_{p, Z_p}  P_i \ket{\psi} =\bra{\psi} P_{i'}  \acute{h}_{p, Z_p}  P_{i'} \ket{\psi} = 0   ,
\end{align}
where we define $P_i=(\ket{0}\bra{0})_i$ and $P_{i'}=(\ket{0}\bra{0})_{i'}$.

Then, for $Q_i= 1- P_i$ and $Z_p \supseteq \{i,i'\}$, we obtain 
\begin{align}
	&\bra{\psi} \acute{h}_{p, Z_p}   \ket{\psi}   \notag \\  
	&=  \bra{\psi} (P_i + Q_i)(P_{i'}+Q_{i'}) \acute{h}_{p, Z_p}   (P_i + Q_i)(P_{i'}+Q_{i'}) \ket{\psi}   \notag \\
	&=   \bra{\psi} P_i (P_{i'}+Q_{i'}) \acute{h}_{p, Z_p}   Q_i (P_{i'}+Q_{i'}) \ket{\psi}  
	+   \bra{\psi} Q_i (P_{i'}+Q_{i'}) \acute{h}_{p, Z_p}   (P_{i'}+Q_{i'}) \ket{\psi}   \notag \\ 
	&=  \bra{\psi} P_i P_{i'} \acute{h}_{p, Z_p}   Q_i Q_{i'} \ket{\psi}  
	+  \bra{\psi} P_i Q_{i'} \acute{h}_{p, Z_p}   Q_i  \ket{\psi}   +   \bra{\psi} Q_i P_{i'} \acute{h}_{p, Z_p}   Q_{i'} \ket{\psi} + \bra{\psi} Q_i Q_{i'} \acute{h}_{p, Z_p}  \ket{\psi},   
\end{align}
which yields 
\begin{align}
	\label{inequality_sum_Z_Z_p_ave}
	& \sum_{Z: |Z| \le k}\sum_{p: Z_p \supseteq \{i,i'\}} 
	\abs{ \bra{\psi}\acute{h}_{p, Z_p} \ket{\psi} } \notag \\
	&\le\sum_{Z: |Z| \le k}\sum_{p: Z_p \supseteq \{i,i'\}} \norm{  \acute{h}_{p, Z_p} } \br{2 \sqrt{\bra{\psi}Q_i Q_{i'} \ket{\psi}} + 2\sqrt{\bra{\psi}Q_i\ket{\psi}\bra{\psi}Q_{i'} \ket{\psi}}}.
\end{align}
By combining inequalities~\eqref{Starting_inequality_amplitude} and \eqref{inequality_sum_Z_Z_p_ave}, we derive
\begin{align}
	\label{Proof_inequality_reduce_1}
	&\norm{ \Pi_{\le z}  \acute{H}^{(\ge 2)}  \Pi_{\le z} } \notag\\
	&\le 
	\sup_{\ket{\psi}: \Pi_{\le z} \ket{\psi} = \ket{\psi}  }\sum_{i\in \Lambda}\sum_{i'\in i^\co} 
	\br{2 \sqrt{\bra{\psi}Q_i Q_{i'} \ket{\psi}} + 2\sqrt{\bra{\psi}Q_i\ket{\psi}\bra{\psi}Q_{i'} \ket{\psi}}} \sum_{Z: |Z| \le k}\sum_{p: Z_p \supseteq \{i,i'\}}  \norm{ \acute{h}_{p, Z_p} } \notag \\
	&\le \frac{2^{2k+1} g_1}{n} \sup_{\ket{\psi}: \Pi_{\le z} \ket{\psi} = \ket{\psi}  }\sum_{i\in \Lambda}\sum_{i'\in i^\co}  
	\br{\sqrt{\bra{\psi}Q_i Q_{i'} \ket{\psi}} + \sqrt{\bra{\psi}Q_i\ket{\psi}\bra{\psi}Q_{i'} \ket{\psi}}},
\end{align}
where we used inequality~\eqref{new_int_strength_upp}.  

Using the Cauchy-Schwarz inequality, we obtain 
\begin{align}
	\label{Q_i_Q_i'_sum}
	\sum_{i\in \Lambda}\sum_{i'\in i^\co}  \sqrt{\bra{\psi}Q_i Q_{i'} \ket{\psi}} 
	&\le 
	\br{ \sum_{i\in \Lambda}\sum_{i'\in \Lambda} 1}^{1/2}  \br{ \sum_{i\in \Lambda}\sum_{i'\in \Lambda} \bra{\psi}Q_i Q_{i'} \ket{\psi}}^{1/2}  \notag \\
	&= |\Lambda|   \br{\bra{\psi}  \Pi_{\le z}  \sum_{i\in \Lambda} Q_i \sum_{i'\in \Lambda}Q_{i'}  \Pi_{\le z} \ket{\psi}}^{1/2}   \notag \\
	&\le  |\Lambda| \norm{ \Pi_{\le z}  \sum_{i\in \Lambda} Q_i \Pi_{\le z}  }^{1/2} \cdot \norm{ \Pi_{\le z}  \sum_{i'\in \Lambda} Q_{i'} \Pi_{\le z}  }^{1/2} \notag \\
	& =  n z  ,
\end{align}
where we use Eq.~\eqref{Eq.invariance_projection} from the first line to the second line. 
The same inequality is applied to $\sqrt{\bra{\psi}Q_i\ket{\psi}\bra{\psi}Q_{i'} \ket{\psi}}$.
Therefore, by combining the above estimation with Inequality~\eqref{Proof_inequality_reduce_1}, we arrive at the desired inequality~\eqref{Prop/main_ineq:renormalized_interaction}. 
This completes the proof. $\square$

In a similar way, we also prove the following lemma:
\begin{lemma} \label{lem:norm_conditional_phi0}
	For the interaction between the site $i$ and the subset $i^\co$, which is denoted by $\acute{H}_{i,i^\co}$ as in Eq.~\eqref{acute_H_L_L'}, the conditional norm 
	$
	\norm{ \bra{\phi_0} \acute{H}_{i,i^\co} \ket{\phi_0}_{i^\co}} 
	$
	under the state $\ket{\phi_0}_{i^\co}$ 
	is bounded from above by
	\begin{align}
		\label{lem:norm_conditional_phi0_main_/ineq}
		\norm{ \bra{\phi_0} \acute{H}_{i,i^\co} \ket{\phi_0}_{i^\co}} \le  \frac{2^{2k+3} g_1}{\lambda_0} \br{ \frac{\delta_{\rm Rob}}{\Delta}}^{1/2} .
	\end{align}
	Here, $\ket{\phi_0}_{i^\co}$ is defined by Schmidt decomposition with the largest coefficient $\lambda_{0}$ as has been defined in Eq.~\eqref{Schmidt_decomposition}.  
\end{lemma}

\textit{Proof of Lemma~\ref{lem:norm_conditional_phi0}.} 
The proof is similar to Proposition~\ref{Prop:renormalized_interaction}.
We consider an arbitrary quantum state $\ket{\psi}$ such that 
\begin{align}
	\br{\ket{\phi_0} \bra{\phi_0}}_{i^\co} \ket{\psi} = \ket{\psi}  ,
\end{align} 
where $ \ket{\psi}$ becomes a product state between $i$ and $i^\co$. 
Then, we have $\norm{ \bra{\phi_0} \acute{H}_{i,i^\co} \ket{\phi_0}_{i^\co}} = \sup_{\psi} | \bra{\psi} \acute{H}_{i,i^\co} \ket{\psi} | $. 
We derive a similar inequality to~\eqref{Starting_inequality_amplitude}:
\begin{align}
	| \bra{\psi} \acute{H}_{i,i^\co} \ket{\psi} | 
	\le \sum_{i'\in i^\co}  \sum_{Z: |Z| \le k}\sum_{p: Z_p \supseteq \{i,i'\}} 
	\abs{ \bra{\psi}\acute{h}_{p, Z_p} \ket{\psi} } . 
\end{align}

Moreover, we can derive 
\begin{align}
	\label{Proof_inequality_reduce_1_rederive}
	\norm{ \bra{\phi_0} \acute{H}_{i,i^\co} \ket{\phi_0}_{i^\co}} 
	&\le \frac{2^{2k+1} g_1}{n} \sup_{\ket{\psi}: \br{\ket{\phi_0} \bra{\phi_0}}_{i^\co} \ket{\psi} = \ket{\psi}  }\sum_{i'\in i^\co}  
	\br{\sqrt{\bra{\psi}Q_i Q_{i'} \ket{\psi}} + \sqrt{\bra{\psi}Q_i\ket{\psi}\bra{\psi}Q_{i'} \ket{\psi}}} \notag \\
	&\le \frac{2^{2k+2} g_1}{n} \sum_{i'\in i^\co}
	\sqrt{\bra{\phi_0} Q_{i'} \ket{\phi_0}_{i^\co} }  ,
\end{align}
which corresponds to inequality~\eqref{Proof_inequality_reduce_1}. 
Finally, we consider $\bra{\phi_0} Q_{i'} \ket{\phi_0}_{i^\co}$.
By using $P_i \ket{\Omega}=\lambda_0 \ket{0}_i \otimes \ket{\phi_0}_{i^\co}$, we have 
\begin{align}
	\label{Proof_inequality_reduce_1_rederive_Qi_part}
	\bra{\phi_0} Q_{i'} \ket{\phi_0}_{i^\co} &=\frac{1}{\lambda_0^2} \bra{\Omega} P_i  Q_{i'} P_i \ket{\Omega}  =\frac{1}{\lambda_0^2} \bra{\Omega} P_i  Q_{i'} \ket{\Omega} = \frac{1}{\lambda_0^2} \br{\bra{\Omega} Q_{i'} \ket{\Omega}  - \bra{\Omega} Q_i  Q_{i'} \ket{\Omega}  } \notag \\
	& \le \frac{1}{\lambda_0^2}  \norm{Q_{i'} \ket{\Omega}}^2   
	\le  \frac{2\delta_{\rm Rob}}{\lambda_0^2 \Delta},
\end{align}
where we use $P_i=1-Q_i$ and inequality~\eqref{Q_i_proj_norm}. 

Therefore, by combining inequalities~\eqref{Proof_inequality_reduce_1_rederive} and \eqref{Proof_inequality_reduce_1_rederive_Qi_part}, we obtain
\begin{align}
	\norm{ \bra{\phi_0} \acute{H}_{i,i^\co} \ket{\phi_0}_{i^\co}} 
	\le 2^{2k+3} g_1\br{ \frac{\delta_{\rm Rob}}{\lambda_0^2 \Delta}}^{1/2} ,
\end{align}
which gives the main inequality~\eqref{lem:norm_conditional_phi0_main_/ineq}.
This completes the proof. $\square$

\subsection{Method 4: Bound on the Hamiltonian norm in the $z$-flip subspace}
\label{sec:Bound on the Hamiltonian norm in the z-flip subspace}

By Proposition~\ref{Prop:renormalized_interaction}, under the $\mR\mE$ decomposition, the higher order terms $\acute{H}^{(\ge 2)}$ can be efficiently controlled as 
\begin{align}
	\label{higher_order_norm}
	\norm{ \Pi_{\le z}  \acute{H}^{(\ge 2)} \Pi_{\le z} } \le 2^{2k+2} g_1 z  . 
\end{align}
Therefore, the remaining task is to control the norm of 
\begin{align}
	\norm{\sum_{i\in \Lambda} \Pi_{\le z} \acute{H}_i \Pi_{\le z} } ,\quad 
	\acute{H}_i :=   \sum_{Z: |Z| \le k}\sum_{p: Z_p=\{i\}} \acute{h}_{p, Z_p} . 
\end{align}

For this purpose, we need to prove that the mean-field state $\ket{0}_i$ is an approximate eigenstate of $\acute{H}_i$ as shown in the following proposition (see Sec.~\ref{Proof_of_Prop_mean_field} for the proof). 
\begin{prop} \label{prop:mean_field_local_Ham_re}
	For an arbitrary on-site potential $\acute{H}_i$ with $i\in \Lambda$, the norm of $Q_i \acute{H}_i P_i$ is upper-bounded as follows:
	\begin{align}
		\norm{Q_i \acute{H}_i P_i}  \le \delta_0, \quad 
		\delta_0:= \frac{\bar{g}}{\lambda_0}  \br{ \frac{2\delta_{\rm Rob}}{\Delta}}^{1/2} 
		\br{2^{2k+5/2} + 1 + \frac{3}{\lambda_0}} ,
		\label{H_i_off_diagonal_norm_bound}
	\end{align} 
	where $P_i$ and $Q_i$ were defined in Eq.~\eqref{P_i_Q_i_definitiion/mean_field}. 
\end{prop}

By using the proposition with inequality~\eqref{higher_order_norm}, we derive the upper bound for $\norm{ \Pi_{\le z} \br{H -\check{E}_0}\Pi_{\le z} } $ in the following ways.
We begin with
\begin{align}
	\label{Hopping_proof_1}
	\norm{ \Pi_{\le z} \br{H -\check{E}_0}\Pi_{\le z} }  
	&\le 
	\norm{ \Pi_{\le z}  \acute{H}^{(\ge 2)} \Pi_{\le z} }+ \norm{ \sum_{i\in \Lambda} \Pi_{\le z}  \acute{H}_i \Pi_{\le z} }  \notag \\
	&\le 2^{2k+2} g_1 z +\norm{ \sum_{i\in \Lambda}  \Pi_{\le z} P_i \acute{H}_i Q_i \Pi_{\le z} + \Pi_{\le z} Q_i \acute{H}_i P_i \Pi_{\le z} + \Pi_{\le z} Q_i \acute{H}_i Q_i \Pi_{\le z} } ,
\end{align}
where we use $P_i \acute{H}_i P_i=0$. 

For the summation of $\norm{ \sum_{i\in \Lambda}  \Pi_{\le z} Q_i \acute{H}_i Q_i \Pi_{\le z} }$, we immediately obtain 
\begin{align}
	\label{Hopping_proof_2}
	\norm{\sum_{i\in \Lambda}  \Pi_{\le z} Q_i \acute{H}_i Q_i \Pi_{\le z} } 
	&=\sup_{\psi} \abs{ \bra{\psi}\sum_{i\in \Lambda}  \Pi_{\le z} Q_i \acute{H}_i Q_i \Pi_{\le z} \ket{\psi}}  \notag \\
	&\le\sup_{\psi}\sum_{i\in \Lambda}   \abs{ \bra{\psi}\Pi_{\le z} Q_i \acute{H}_i Q_i \Pi_{\le z} \ket{\psi}}\notag \\
	&\le  \sup_{\psi}\sum_{i\in \Lambda}   \norm{\acute{H}_i} \cdot \bra{\psi}\Pi_{\le z} Q_i \Pi_{\le z} \ket{\psi} \notag \\
	&\le2^{2k} \bar{g}  \sup_{\psi} \bra{\psi}\Pi_{\le z} \sum_{i\in \Lambda} Q_i \Pi_{\le z} \ket{\psi} \le 2^{2k} \bar{g} z ,
\end{align}
where we use  inequality~\eqref{new_int_strength_extensive} to obtain $\norm{\acute{H}_i} \le 2^{2k} \bar{g}$. 

For the summation of $\norm{ \sum_{i\in \Lambda}  \Pi_{\le z} Q_i \acute{H}_i P_i \Pi_{\le z} }$, we use 
\begin{align}
	\label{Hopping_proof_3}
	\norm{\sum_{i\in \Lambda}  \Pi_{\le z} Q_i \acute{H}_i P_i \Pi_{\le z} }
	&\le\sup_{\psi,\psi': \norm{\ket{\psi}}=\norm{\ket{\psi'}}=1}\sum_{i\in \Lambda}   \abs{ \bra{\psi}\Pi_{\le z} Q_i \acute{H}_i P_i \Pi_{\le z} \ket{\psi'}}\notag \\
	&\le \sup_{\psi: \norm{\ket{\psi}}=1} \sum_{i\in \Lambda}   \norm{ \bra{\psi}\Pi_{\le z} Q_i} \cdot \norm{Q_i \acute{H}_i P_i } \notag \\
	&\le \delta_0 \sup_{\psi: \norm{\ket{\psi}}=1} \sqrt{ \sum_{i\in \Lambda} 1} \sqrt{ \sum_{i\in \Lambda} \norm{ \bra{\psi}\Pi_{\le z} Q_i}^2} \notag \\
	&\le \delta_0 \sqrt{nz}.
\end{align}
By combining the upper bounds~\eqref{Hopping_proof_2} and \eqref{Hopping_proof_3} with inequality~\eqref{Hopping_proof_1}, we derive 
\begin{align}
	\label{Hopping_proof_fin }
	\norm{ \Pi_{\le z} \br{H - \check{E}_0}\Pi_{\le z} }  
	&\le 2^{2k+2} g_1 z + 2^{2k} \bar{g} z  + 2 \delta_0 \sqrt{nz}  \notag \\
	&\le 2^{2k+3} \bar{g} z + 2 \delta_0 \sqrt{nz} ,
\end{align}
where we use $g_1 \le \bar{g}$.

\subsection{Determining the parameters in Proposition~\ref{prob_tight_binding_new} and completing the proof of Theorem~\ref{Them:prob_dist_z}}

From the assumption~\eqref{assump_for_n_ge_n_0}, we have 
\begin{align}
	\frac{2\delta_{\rm Rob}}{\Delta}  = 
	\frac{1}{\Delta}   \br{\frac{12 \gamma g_1 d^2}{\Delta}}^2 \frac{4\bar{g}}{n}   \le \frac{1}{2}.
\end{align}
Then, we have $\lambda_0^2 \ge 1/2$ ($\lambda_0\ge \sqrt{1/2}$) from inequality~\eqref{mean/field_error_bo8nd}, and hence we upper-bound $\delta_0$ in Eq.~\eqref{H_i_off_diagonal_norm_bound} by
\begin{align}
	\delta_0 &=\frac{\bar{g}}{\lambda_0}  \br{ \frac{2\delta_{\rm Rob}}{\Delta}}^{1/2} 
	\br{2^{2k+5/2} + 1 + \frac{3}{\lambda_0}}  \notag \\
	&\le 2 \sqrt{2} \bar{g} \brr{  \br{\frac{12 \gamma g_1 d^2}{\Delta}}^2 \frac{\bar{g}}{n\Delta}}^{1/2} 
	\br{2^{2k+5/2} + 1 + 3 \cdot 2^{1/2}}   \notag \\
	&\le  2^{2k+8} \gamma d^2 \bar{g} \br{ \frac{\bar{g}}{\Delta}}^{3/2}  n^{-1/2},
\end{align}
where we use $g_1\le \bar{g}$ and $24 \sqrt{2}\br{2^{2k+5/2} + 1 + 3 \cdot 2^{1/2}}  \le 2^{2k+8}$ for $k\ge 1$.
By applying the above inequality to inequality~\eqref{Hopping_proof_fin }, we have 
\begin{align}
	\norm{ \Pi_{\le z} \br{H -\check{E}_0}\Pi_{\le z} }  
	&\le 2^{2k+3} \bar{g} z + 2 \delta_0 \sqrt{nz}  \notag \\
	&\le 2^{2k+3} \bar{g} z + 2^{2k+9} \gamma d^2  \br{ \frac{\bar{g}}{\Delta}}^{3/2}  \cdot  \bar{g} z^{1/2}  \notag \\
	&\le 2^{2k+3} \bar{g} z \brr{1 +  z^{-1/2}\cdot 2^6  \gamma d^2  \br{ \frac{\bar{g}}{\Delta}}^{3/2}  }  \notag \\
	&=: \frac{\eta}{2} z \brr{1 +   z^{-1/2} \cdot \tilde{z}^{1/2}} , 
\end{align}
with
\begin{align}
	&\eta = 2^{2k+4} \bar{g} ,\quad  \tilde{z}:= 2^{12} \gamma^2 d^4  \br{\bar{g}/\Delta}^{3} .
\end{align}
Therefore, for $z\ge \tilde{z}$, we have 
\begin{align}
	\label{Pi_le_z_H-check_E_0}
	\norm{ \Pi_{\le z} \br{H - \check{E}_0}\Pi_{\le z} }  \le  \eta z.
\end{align}

Moreover, the above definition of $\tilde{z}$ ensures $\tilde{z} \ge z_0$ with $z_0$ defined in inequality~\eqref{definitiion_z_0}, and hence 
\begin{align}
	\bra{\Omega} \Pi_{\le \tilde{z}} \ket{\Omega} \ge  \norm{ \Pi_{\le z_0} \ket{\Omega} }^2\ge \frac{1}{2}.
\end{align}
Therefore, by applying Proposition~\ref{prob_tight_binding_new} with the above choice of $\tilde{z}$,
we prove the main inequality~\eqref{Them:prob_dist_z_main_ineq} in Theorem~\ref{Them:prob_dist_z}. 
$\square$

\subsection{Proof of Proposition~\ref{prop:mean_field_local_Ham_re}}
\label{Proof_of_Prop_mean_field}

Using $P_{i} \ket{\Omega} = \lambda_{0} \ket{0}_{i} \otimes \ket{\phi_{0}}_{i^{c}}$, we begin with the following relation: 
\begin{align}
	\label{proof_mean_field_prop_start}
	\norm{ \br{ \ket{\phi_0}\bra{\phi_0}}_{i^\co}H  P_i \ket{\gs}}
	&= \norm{ \br{ \ket{\phi_0}\bra{\phi_0}}_{i^\co} H \lambda_0 \ket{0}_i\otimes \ket{\phi_0}_{i^\co} } \notag \\
	&=\lambda_0\norm{\bra{\phi_0} H  \ket{\phi_0}_{i^\co } \ket{0}_i} .
\end{align}
Then, from $H=  \acute{H}_{i} +   \acute{H}_{i^\co} + \acute{H}_{i,i^\co}+ \check{E}_0$~\footnote{For readers' convenience, we show the explicit expressions: 
	\begin{align}
		&H = \sum_{Z : |Z| \leq k} h_{Z} = \sum_{Z: |Z| \leq k} \sum_{p: Z_{p} \subset Z} \acute{h}_{p,Z_{p}},  \quad 
		\acute{H}_{i} = \sum_{Z: |Z| \leq k} \sum_{p: Z_{p} = \{i\}} \acute{h}_{p,Z_{p}}, \quad 
		\acute{H}_{i^{c}} = \sum_{Z: |Z| \leq k} \sum_{\substack{p: Z_{p} \subset i^{c} \\ Z_{p} \neq \emptyset}} \acute{h}_{p,Z_{p}}, \notag \\
		&\acute{H}_{i,i^{c}} = \sum_{Z: |Z| \leq k} \sum_{\substack{p: Z_{p} \ni i  \\ Z_{p} \cap i^{c} \neq \emptyset}} \acute{h}_{p,Z_{p}},  \quad 
		\check{E}_0 = \sum_{Z: |Z| \leq k} \sum_{p: Z_{p} = \emptyset} \acute{h}_{p,Z_{p}}.
\end{align}}, we derive  
\begin{align}
	\label{check_h'_i_mE_i}
	\bra{\phi_0} H \ket{\phi_0}_{i^\co } 
	&=\acute{H}_{i} + \bra{\phi_0}   \acute{H}_{i^\co} \ket{\phi_0}_{i^\co } +  \bra{\phi_0} \acute{H}_{i,i^\co}  \ket{\phi_0}_{i^\co }  +  \check{E}_0 
	=: \check{h}_i - \check{\mE}_i,
\end{align}
where we use $\bra{\phi_0} \acute{H}_{i}  \ket{\phi_0}_{i^\co } = \acute{H}_{i}$ and define $\check{h}_i$ and $\check{\mE}_i$ as follows: 
\begin{align}
	\check{h}_i:=\acute{H}_{i} + \bra{\phi_0} \acute{H}_{i,i^\co}  \ket{\phi_0}_{i^\co },\quad 
	\check{\mE}_i:=- \bra{\phi_0}   \acute{H}_{i^\co} \ket{\phi_0}_{i^\co }  -\check{E}_0   .
\end{align}

Then, by combining~Eqs.~\eqref{proof_mean_field_prop_start} and \eqref{check_h'_i_mE_i}, we obtain 
\begin{align}
	\bra{\phi_0} H  \ket{\phi_0}_{i^\co } \ket{0}_i= (\check{h}_{i} - \check{\mE}_{i}) \ket{0}_{i}  \quad 
	&\Rightarrow \quad \norm{ \check{h}_i \ket{0}_i - \check{\mE}_i \ket{0}_i } = \frac{\norm{ \br{ \ket{\phi_0}\bra{\phi_0}}_{i^\co}H  P_i \ket{\gs}}}{\lambda_0} \notag \\
	&\Rightarrow \quad \norm{ Q_i \check{h}_i  P_i} \le \frac{\norm{ \br{ \ket{\phi_0}\bra{\phi_0}}_{i^\co}H  P_i \ket{\gs}}}{\lambda_0} , \label{closeness_h_check'}
\end{align}
where we use $\norm{ Q_i \br{ \check{h}_i \ket{0}_i - \check{\mE}_i \ket{0}_i}} =\norm{ Q_i \check{h}_i \ket{0}_i } = \norm{ Q_i \check{h}_i P_i }$.
This yields 
\begin{align}
	\norm{Q_i \acute{H}_i P_i} \le \norm{ \acute{H}_i - \check{h}_i} +  \frac{\norm{ \br{ \ket{\phi_0}\bra{\phi_0}}_{i^\co}H  P_i \ket{\gs}}}{\lambda_0}
	= \norm{\bra{\phi_0} \acute{H}_{i,i^\co}  \ket{\phi_0}_{i^\co }} +   \frac{\norm{ \br{ \ket{\phi_0}\bra{\phi_0}}_{i^\co}H  P_i \ket{\gs}}}{\lambda_0}. 
	\label{closeness_h_check'//2}
\end{align} 
Thus, our task is to derive upper bounds for $ \norm{\bra{\phi_0} \acute{H}_{i,i^\co}  \ket{\phi_0}_{i^\co }}$ and $\norm{ \br{ \ket{\phi_0}\bra{\phi_0}}_{i^\co}H  P_i \ket{\gs}}$.

First, Lemma~\ref{lem:norm_conditional_phi0} gives the upper bound of 
\begin{align}
	\label{lem:norm_conditional_phi0_apply_mean}
	&\norm{\bra{\phi_0} \acute{H}_{i,i^\co}  \ket{\phi_0}_{i^\co }}  \le   \frac{2^{2k+3} g_1}{\lambda_0} \br{ \frac{ \delta_{\rm Rob}}{\Delta}}^{1/2} .
\end{align} 
Second, to upper-bound $\norm{ \br{ \ket{\phi_0}\bra{\phi_0}}_{i^\co}H  P_i \ket{\gs}}$, we first decompose 
\begin{align}
	\label{start_proof_lem:prop:mean_field_lo}
	\norm{ \br{ \ket{\phi_0}\bra{\phi_0}}_{i^\co}H  P_i \ket{\gs}}  
	&=\norm{ \br{ \ket{\phi_0}\bra{\phi_0}}_{i^\co} H  Q_i \ket{\gs}} \notag \\
	&\le \norm{ \br{ \ket{\phi_0}\bra{\phi_0}}_{i^\co} \widehat{H}_i  Q_i \ket{\gs}} + \norm{ \br{ \ket{\phi_0}\bra{\phi_0}}_{i^\co} H_{i^\co} Q_i \ket{\gs}}  \notag \\
	&\le \br{  \norm{  \widehat{H}_i } + \norm{  H_{i^\co} \ket{\phi_0}_{i^\co} } } \norm{Q_i \ket{\gs}}   ,
\end{align}
where we use $H \ket{\Omega} = 0$ in the first equation. 
From inequality~\eqref{g_ext_full_Ham}, we have $\norm{  \widehat{H}_i } \le \bar{g}$.
Moreover, by using $\norm{\ket{\psi}_{i^\co}} = \norm{\ket{0}_i \otimes \ket{\psi}_{i^\co}}$ for an arbitrary unnormalized state $\ket{\psi}_{i^\co}$, we obtain
\begin{align}
	\norm{  H_{i^\co} \ket{\phi_0}_{i^\co} } 
	&=  \norm{  H_{i^\co} \ket{0}_i \ket{\phi_0}_{i^\co} }  = \frac{1}{\lambda_0} \norm{  H_{i^\co} P_i \ket{\Omega}}  = \frac{1}{\lambda_0} \norm{  \br{H - \widehat{H}_{i}} P_i \ket{\Omega}}   \notag \\
	&\le \frac{1}{\lambda_0} \br{\norm{\widehat{H}_{i}} +  \norm{H P_i \ket{\Omega}}}  \notag \\
	&\le  \frac{3}{\lambda_0} \norm{\widehat{H}_{i}}  
	\le \frac{3\bar{g}}{\lambda_0}  ,
\end{align}
where we use $\norm{H P_i \ket{\Omega}} = \norm{[H, P_i] \ket{\Omega}}= \norm{[\widehat{H}_{i}, P_i] \ket{\Omega}} \le 2 \norm{\widehat{H}_i}$.   
Therefore, by applying the above inequality to \eqref{start_proof_lem:prop:mean_field_lo} with $\norm{Q_i \ket{\Omega} }^2 \le 2\delta_{\rm Rob}/\Delta$ in Eq.~\eqref{Q_i_proj_norm}, we have 
\begin{align}
	\norm{ \br{ \ket{\phi_0}\bra{\phi_0}}_{i^\co}H  P_i \ket{\gs}}  \le  \bar{g}   \br{1+ \frac{3}{\lambda_0}} \br{\frac{2\delta_{\rm Rob}}{\Delta}}^{1/2} . \label{lem:prop:mean_field_local_Ham_1_re/main}
\end{align}

By applying inequalities~\eqref{lem:norm_conditional_phi0_apply_mean} and \eqref{lem:prop:mean_field_local_Ham_1_re/main} to inequality~\eqref{closeness_h_check'//2}, we prove the upper bound of 
\begin{align}
	\label{lem:norm_conditional_phi0_apply_mean_final}
	\norm{Q_i \acute{H}_i P_i} \le  \frac{\bar{g}}{\lambda_0}  \br{ \frac{2\delta_{\rm Rob}}{\Delta}}^{1/2} 
	\br{2^{2k+5/2} + 1 + \frac{3}{\lambda_0}} .
\end{align} 
This completes the proof. $\square$

\section{Uniform Hamiltonians} \label{Sec:Uniform Hamiltonians}

\subsection{Constant Entanglement bound}

In this section, we specialize to the case of uniform all-to-all (permutation-invariant) Hamiltonians, which is defined as follows:
\begin{definition}[Uniform Hamiltonian]
	Let $H$ be a $k$-local Hamiltonian acting on $\mathcal{H} = (\mathbb{C}^{d})^{\otimes n}$ with $n > d$. 
	We say that the Hamiltonian is uniform all-to-all, meaning that it is invariant under arbitrary relabelings of the sites.
	Equivalently,
	\begin{align}
		[H,P_{\pi}] = 0 \qquad \forall \pi \in S_{n},
	\end{align}
	where $P_{\pi}$ denotes the unitary operator on $\mathcal{H}$ defined by
	\begin{align}
		P_{\pi} \br{\ket{v_{1}} \otimes \ket{v_{2}} \otimes \cdots \otimes \ket{v_{n}}} = \ket{v_{\pi^{-1}(1)}} \otimes \ket{v_{\pi^{-1}(2)}} \otimes  \cdots \otimes \ket{v_{\pi^{-1}(n)}} 
	\end{align}
	for all $\ket{v_{i}} \in \mathbb{C}^{d}$. 
\end{definition}

We prove that if a Hamiltonian is permutation invariant and admits a unique ground state separated by a constant spectral gap, then the bipartite entanglement entropy of that state is bounded by a constant independent of the system size.
In the following, we consider the R\'enyi entanglement $S^{(\alpha)}_{A,B}(\ket{\gs})$ as 
\begin{align}
	\label{Def:entanglement_entropy_Renyi}
	S^{(\alpha)}_{A,B}(\ket{\gs}) := \frac{1}{1-\alpha} \log \brr{ \tr\br{  \rho^\alpha_{A} }},
\end{align}
where $\rho_{A} := \tr_{B}\left( \ket{\gs} \bra{\gs} \right)$, and the entanglement entropy in Eq.~\eqref{Def:entanglement_entropy} corresponds to the case of $\alpha=1$. 
Then, we prove the following theorem:
\begin{theorem} \label{Thm:Constant_Entanglement_Uniform}
	Assume that $H$ is uniform and has a unique ground state $\ket{\gs}$ separated from the rest of the spectrum by a constant spectral gap $\Delta = \Omega(1)$.
	Then, for every bipartition $\Lambda = A \cup B$, the bipartite R\'enyi entanglement of the ground state satisfies
	\begin{align}
		\label{Thm:Constant_Entanglement_Uniform_main_ineq}
		S^{(\alpha)}_{A,B}(\ket{\gs}) \leq  \frac{1}{1-\alpha} \log\br{ \mathfrak{C}_\alpha} \for 0 < \alpha <1,
	\end{align}
	where $\mathfrak{C}_\alpha$ is positive constant depending only on $d,k,\bar{g}$ and $\Delta$ (through the constants $C_{1},C_{2}$ from Theorem~\ref{Them:prob_dist_z}), and is independent of the system size $n$.
	More explicitly, 
	$\mathfrak{C}_\alpha$ is given by
	\begin{align}
		\label{mathcal_C_alpha}
		\mathfrak{C}_\alpha &:= (\tilde{z}+1)^{d-1}  + 4^{d-1} C^{2\alpha}_{1} \brr{ 2 + \frac{(2d)!}{(2 \alpha C_{2})^{2d}}} .
	\end{align}
\end{theorem}

{\bf Remark.} By using the fact that the R\'enyi entanglement is monotonically decreasing with $\alpha$, we obtain the upper bound of the entanglement entropy $S_{A,B}(\ket{\gs})$ by $S^{(\alpha)}_{A,B}(\ket{\gs})$ with $\alpha<1$. 
Then, by choosing $\alpha=1/2$ for example, we derive 
\begin{align}
	S_{A,B}(\ket{\gs}) \leq 2 \log\br{ \mathfrak{C}_{1/2}} = 2\log \brrr{(\tilde{z}+1)^{d-1}  + 4^{d-1} C_1 \brr{ 2 + \frac{(2d)!}{C_{2}^{2d}}} } .
\end{align}

\subsection{Proof of Theorem~\ref{Thm:Constant_Entanglement_Uniform}}

\subsubsection{Preliminaries}
We begin by identifying the symmetry sector of the unique ground state.
\begin{lemma} \label{symmetry_sector_uniform_gs}
	Under the assumptions of Theorem~\ref{Thm:Constant_Entanglement_Uniform}, the unique ground state $\ket{\gs}$ belongs to the totally symmetric subspace
	\begin{align}
		\mathrm{Sym}^{n}(\mathbb{C}^{d}) := \brrr{\ket{\psi} \in \mathcal{H}\,:\, P_{\pi} \ket{\psi} = \ket{\psi} \quad \forall \pi \in S_{n}}.
	\end{align}
	Moreover,
	\begin{align}
		\mathrm{dim}(\mathrm{Sym}^{n}(\mathbb{C}^{d})) = \binom{n + d - 1}{d-1}.
	\end{align}
\end{lemma}
\begin{proof}
	As in the preceding discussion, we shift $H$ by a constant so that the ground-state energy is zero. 
	Let $\mathcal{G} := \ker(H)$ denote the ground eigenspace.
	Since $[H,P_{\pi}] = 0$ for all $\pi \in S_{n}$, the space $\mathcal{G}$ is invariant under permutations:
	\begin{align}
		P_{\pi} \mathcal{G}\subseteq \mathcal{G}.
	\end{align}
	By uniqueness of the ground state, $\mathrm{dim}(\mathcal{G}) = 1$, so $\mathcal{G}= \mathrm{span}\{\ket{\gs}\}$.
	Hence for each $\pi \in S_{n}$ there exists a scalar $\chi(\pi) \in \mathbb{C}$ such that
	\begin{align}
		P_{\pi} \ket{\gs} = \chi(\pi) \ket{\gs}.
	\end{align}
	Since $P_{\pi}$ is unitary, $|\chi(\pi)| = 1$, and $\chi: S_{n} \rightarrow \mathrm{U}(1)$ defines a one-dimensional character of $S_{n}$.
	
	For $n \geq 2$, the only one-dimensional irreducible representations of $S_{n}$ are the trivial character and the sign character~\cite{Sagan2001,Serre1977}.
	If $\ket{\gs}$ is transformed according to the sign character, it would lie in the antisymmetric subspace $\wedge^{n}(\mathbb{C}^{d})$, whose dimension equals $\binom{d}{n}$.
	Since $n > d$, this dimension vanishes, so the antisymmetric case is impossible.
	Therefore, the trivial character must hold:
	\begin{align}
		P_{\pi} \ket{\gs} = \ket{\gs} \qquad \forall \pi \in S_{n},
	\end{align}
	which shows $\ket{\gs} \in \mathrm{Sym}^{n}(\mathbb{C}^{d})$.
	
	Finally, $\mathrm{Sym}^{n}(\mathbb{C}^{d})$ is spanned by occupation-number configurations $(n_{1},\cdots,n_{d})$ satisfying $n_{1},\cdots,n_{d} \in \mathbb{Z}_{\geq 0}$ and $n_{1} + \cdots n_{d} = n$.
	The number of such configurations equals $\binom{n+d-1}{d-1}$~\cite{gilmore1972geometry,harrow2013church}.
\end{proof}

\medskip

We next exploit the permutation symmetry to control the entanglement generated by the low-deviation sector.
In particular, we show that projecting the ground state onto configurations with at most $z$ local deviations yields a state whose Schmidt rank---and hence entanglement entropy---is only polynomial in $z$, uniformly over all bipartitions.

\begin{corol} \label{Entropy_bound_low_deviations}
	In the uniform all-to-all setting described above, consider the unnormalized projected state 
	\begin{align}
		\ket{\gs_{\leq z}} :=\Pi_{\leq z} \ket{\gs},\label{proj_z_gs}
	\end{align}
	where $\Pi_{\leq z}$ is the projector of the local-deviation number operator $M = \sum_{i \in \Lambda} Q_{i}$ onto the sector with at most $z$ deviations.
	Then for any bipartition $\Lambda = A \cup B$, we have
	\begin{align}
		\mathrm{SR}_{A|B}(\ket{\gs_{\leq z}}) \le \binom{z + d-1}{d-1} ,
	\end{align} 
	where $\mathrm{SR}_{A|B}(\cdots)$ is the Schmidt rank across the bipartition $A|B$ with $A \sqcup B=\Lambda$.
\end{corol}
\begin{proof}
	By Lemma~\ref{symmetry_sector_uniform_gs}, the ground state $\ket{\gs}$ is fully symmetric.
	Since $M = \sum_{i} Q_{i}$ is permutation invariant, its spectral projector $\Pi_{\leq z}$ is also permutation invariant.
	Hence $\Pi_{\leq z} \ket{\gs}$ is fully symmetric, and therefore $\ket{\gs_{\leq z}} \in \mathrm{Sym}^{n}(\mathbb{C}^{d})$.
	
	Fix an arbitrary bipartition $\Lambda = A \cup B$.
	It suffices to upper bound the Schmidt rank of $\ket{\gs_{\leq z}}$ across $A |B$.
	Because $\ket{\gs_{\leq z}}$ is supported on the sector with at most $z$ local deviations from the dominant local Schmidt vector $\ket{0}$, the only degrees of freedom that can appear in the reduced state on $A$ are the symmetric patterns specifying how many sites occupy each of the $(d-1)$ non-dominant local states $\ket{1},\cdots,\ket{d-1}$, with total count at most $z$.
	For each $t \in \{0,1,\cdots,z\}$, the number of such patterns with total non-dominant weight $t$ equals the number of $(d-1)$-tuples of nonnegative integers summing to $t$, namely $\binom{t + d - 2}{d-2}$.
	Therefore, the reduced state on $A$ is supported on a subspace of dimension at most
	\begin{align}
		\sum_{t = 0}^{z} \binom{t + d - 2}{d - 2} = \binom{z + d - 1}{d-1},
	\end{align}
	uniformly over the choice of $A$ and $n$. This completes the proof. 
\end{proof}

\subsubsection{Proof of Theorem~\ref{Thm:Constant_Entanglement_Uniform}}
We now combine the symmetry reduction (Lemma~\ref{symmetry_sector_uniform_gs}), the low-deviation entanglement entropy bound (Corollary~\ref{Entropy_bound_low_deviations}), and the stretched-exponential concentration estimate (Theorem~\ref{Them:prob_dist_z}) to prove that the ground-state entanglement entropy is bounded by a constant independent of the system size.

Fix an arbitrary bipartition $\Lambda = A \cup B$ and write the Schmidt decomposition of the nondegenerate ground state
\begin{align}
	\ket{\gs} = \sum_{j \geq 1} \mu_{j} \ket{\psi_{A,j}} \otimes \ket{\psi_{B,j}}, \qquad \mu_{1} \geq \mu_{2} \geq \cdots \geq 0, \qquad \sum_{j \geq 1} \mu_{j}^{2} = 1.
\end{align}
We will upper bound the R\'enyi entanglement in Eq.~\eqref{Def:entanglement_entropy_Renyi}, i.e., 
\begin{align}
	\label{target_renyi_ent}
	S^{(\alpha)}_{A,B}(\ket{\gs}) = \frac{1}{1-\alpha} \log\br{ \sum_{j \geq 1} \mu_{j}^{2\alpha}}.
\end{align}

We first bound the distance between $\ket{\gs}$ and $\ket{\gs_{\leq z}}$ defined in Eq.~\eqref{proj_z_gs}. Using $\ket{\gs} = \Pi_{\leq z} \ket{\gs} + \Pi_{>z}\ket{\gs}$, we obtain
\begin{align}
	\norm{\ket{\gs} - \ket{\gs_{\leq z}}} \leq  \norm{\Pi_{> z} \ket{\gs}} \leq  C_{1} e^{-C_{2} \sqrt{z}},  \label{eq:gs_norm_difference}
\end{align}
where we use Theorem~\ref{Them:prob_dist_z} in the second inequality. 
Moreover, by Corollary~\ref{Entropy_bound_low_deviations}, the projected state $\ket{\gs_{\leq z}}$ has uniformly bounded Schmidt rank across the bipartition $A|B$:
\begin{align}
	\mathrm{SR}_{A|B}(\ket{\gs_{\leq z}}) \leq \binom{z + d - 1}{d-1} = \prod_{s=1}^{d-1}\frac{z + d - s}{d-s} \le (z+1)^{d-1} =: D_z. \label{eq:SR_bound_Dz}
\end{align}

We now convert the approximation error into a bound on the tail of Schmidt coefficients.
By the Eckart-Young theorem, for any unnormalized $\ket{\phi}$ of Schmidt rank at most $D$,
\begin{align}
	\sum_{j > D} \mu_{j}^{2} \leq \norm{\ket{\gs} - \ket{\phi}}^{2}.
\end{align}
Applying this to $\ket{\phi} = \ket{\gs_{\leq z}}$ (whose Schmidt rank is at most $D(z)$ by Eq.~\eqref{eq:SR_bound_Dz}) yields, 
\begin{align}
	\sum_{j \in (D_z,D_{z+1}]} \mu_{j}^{2} \le  \norm{\ket{\gs} -  \ket{\gs_{\leq z}}}^2 \le  C^2_{1} e^{-2 C_{2} \sqrt{z}} \for z\ge \tilde{z} ,
\end{align}
where $\tilde z$ was defined in Eq.~\eqref{Them:prob_dist_z_basic_para}. 
From the above inequality, we also obtain 
\begin{align}
	\sum_{j \in (D_z,D_{z+1}]} \mu_{j}^{2\alpha} \le D_{z+1}  C^{2\alpha}_{1} e^{-2 \alpha C_{2} \sqrt{z}} =   C^{2\alpha}_{1} (z+2)^{d-1} e^{-\br{ z/\Phi}^{1/2}} , \quad 
	\Phi:= \frac{1}{(2 \alpha C_{2})^2} .
\end{align}

Therefore, we obtain 
\begin{align}
	\sum_{j = 1}^\infty \mu_{j}^{2\alpha} 
	&\le \sum_{j: j\le D_{\tilde{z}}}  \mu_{j}^{2\alpha}  + \sum_{z=\tilde{z}}^\infty \sum_{j \in (D_z,D_{z+1}]}\mu_{j}^{2\alpha} \notag \\
	&\le D_{\tilde{z}} +C^{2\alpha}_{1} \sum_{z=0}^\infty (z+2)^{d-1} e^{-\br{ z/\Phi}^{1/2}}  \notag\\
	&\le D_{\tilde{z}} + (2^{d-1}+3^{d-1}) C^{2\alpha}_{1} + C^{2\alpha}_{1} \sum_{z=2}^\infty 4^{d-1} (z-1)^{d-1}   e^{-\br{ z/\Phi}^{1/2}}   \notag\\
	&\le D_{\tilde{z}} + (2^{d-1}+3^{d-1}) C^{2\alpha}_{1} + 4^{d-1} C^{2\alpha}_{1} \Phi^{d}  \cdot 2 \cdot (2d-1)! \notag \\
	&\le (\tilde{z}+1)^{d-1}  + 4^{d-1} C^{2\alpha}_{1} \brr{ 2 + \frac{(2d)!}{(2 \alpha C_{2})^{2d}}} = \mathfrak{C}_\alpha,
\end{align}
where we use $\sum_{z=1}^\infty (z-1)^{s}   e^{-\br{z/\Phi}^{1/\kappa}} \le \Phi^{s+1} \kappa \Gamma[\kappa(s+1)]$ with $\Gamma$ the gamma function~\cite[Supplementary Lemma~1 therein]{kuwahara2022optimal}, and $\mathfrak{C}_\alpha$ was defined in Eq.~\eqref{mathcal_C_alpha}. 
By applying the above upper bound to Eq.~\eqref{target_renyi_ent}, we obtain the desired inequality~\eqref{Thm:Constant_Entanglement_Uniform_main_ineq}.
This completes the proof. $\square$

\subsection{Exact polynomial time algorithm}

The symmetry reduction established above has an immediate algorithmic consequence.
Since the unique ground state lies in the fully symmetric sector, the search for the ground state can be restricted to a dramatically smaller Hilbert space.
This yields the following exact and efficient simulation result.
\begin{corol}\label{corol:poly_time_simulation}
	Under the same assumptions as in Theorem~\ref{Thm:Constant_Entanglement_Uniform}, the ground-state energy and a corresponding ground state of $H$ can be computed exactly in time
	\begin{align}
		\mathcal{O}(n^{3(d-1)}),
	\end{align}
	for fixed local dimension $d$.	
\end{corol}
\begin{proof}
	By Lemma~\ref{symmetry_sector_uniform_gs}, the unique ground state lies in totally symmetric subspace $\mathrm{Sym}^{n}(\mathbb{C}^{d})$.
	Since $[H,P_{\pi}] = 0$, this subspace is invariant under $H$, so restricting $H$ to $\mathrm{Sym}^{n}(\mathbb{C}^{d})$ does not change the ground-state energy or eigenvector. 
	Let $D(n)$ denote the dimension of the symmetric subspace $\mathrm{Sym}^{n}(\mathbb{C}^{d})$, namely $\binom{n + d - 1}{d - 1}$.
	Exact diagonalization of a $D(n) \times D(n)$ matrix requires time $\mathcal{O}(D(n)^{3})$ using standard dense-matrix algorithms.
	Since $D(n) = \Theta(n^{d-1})$, this yields an overall runtime $\mathcal{O}(n^{3(d-1)})$.
	The resulting eigenvalue and eigenvector coincide with the exact ground-state energy and ground state of $H$.
\end{proof}

\subsection{Poly-logarithmic time algorithm to compute the ground state}

In the uniform (permutation-invariant) setting, the concentration bound of Theorem~\ref{Them:prob_dist_z} implies that, for every target state-vector accuracy $\epsilon > 0$, the true ground state $\ket{\gs}$ can be approximated within error $\epsilon$ by its low-deviation component $\Pi_{\leq z} \ket{\gs}$, provided that $z = \Theta(\log^{2}(1 / \epsilon))$; moreover, this truncated sector admits a compact symmetric basis.
In this subsection, we describe an algorithm that computes an $\mathcal{O}(\epsilon)$-approximate ground-state energy and a corresponding approximate ground-state vector within the truncated symmetric subspace.
Under the standing nondegeneracy and gap assumptions, the state-vector error is of order $\sqrt{\epsilon / \Delta}$.
The total runtime is
\begin{align}
	\orderof{\mathrm{polylog}(1 / \epsilon) \cdot \log(n/ \epsilon)},
\end{align}
under the real-algebraic modeling assumptions.

\subsubsection{Setup}
Let $H$ be a permutation-invariant $k$-local Hamiltonian acting on $\mathcal{H}= (\mathbb{C}^{d})^{\otimes n}$.
We fix once and for all a computational basis $\{\ket{\phi_a}\}_{a=0}^{d-1}$ of $\mathbb{C}^{d}$,  and use the induced tensor-product basis on $\mathcal{H}$.

We choose a local one-body reference state
\begin{align}
	\ket{e_{0}(u)} \in \mathbb{C}^{d},
\end{align}
which depends continuously on a parameter $u$.
The vector $\ket{e_{0}(u)}$ plays the role of a dominant local component: we measure deviations of the many-body configurations relative to the product state $\ket{e_{0}(u)}^{\otimes n}$.
Different choices of $u$ correspond to different local directions, and the optimal choice of $u$ will be determined variationally. 
This uniform choice across all sites is justified by permutation invariance.
Since $H$ is permutation-invariant and has a nondegenerate ground state $\ket{\gs}$, the reduced density matrix on any single site $\rho_{i} := \tr_{i^{\co}}(\ket{\gs} \bra{\gs})$ is identical for all $i$.
Consequently, all sites share the same spectral decomposition, and in particular the same leading eigenvector.
Therefore, it is consistent to select a single reference state $\ket{e_{0}(u)}$ (intended to approximate this dominant local component) and use it uniformly on every site.

To give a concrete description of the variational parameter $u$, we represent it as a choice of a local orthonormal basis of $\mathbb{C}^{d}$.
Equivalently, we associate to $u$ a unitary matrix $U(u) \in \mathrm{U}(d)$, and define
\begin{align}
	\ket{e_{a}(u)} := U(u) \ket{\phi_a}, \qquad a = 0,1,\cdots,d-1,
\end{align}
where $\ket{\phi_a}_{a = 0}^{d-1}$ is the fixed computational basis.
In particular, the reference direction introduced above satisfies $\ket{e_{0}(u)} = U(u) \ket{0}$.
We parameterize $U(u)$ by its real and imaginary matrix elements.
Writing $U_{ab} = p_{ab} + i q_{ab}$ with real variable $p_{ab}, q_{ab}$, we collect them into $x_{u} = \{p_{ab},q_{ab}\}_{a,b = 0}^{d-1} \in \mathbb{R}^{2d^{2}}$ subject to the polynomial unitarity constraints
\begin{align}
	\label{constrain_s_u}
	U(u)^{\dagger} U(u) = \mathds{1}_{d}.
\end{align}
These are quadratic polynomial equations in $x_{u}$.
Since $d$ is treated as constant, the total number of real variables $x_{u}$ to describe $u$ is $\mathcal{O}(d^{2})$.

Given $\ket{e_{0}(u)}$, define the local deviation projector $Q_{i}(u) := 1 - \ket{e_{0}(u)}\bra{e_{0}(u)}_{i}$ and the global deviation number operator $M(u) = \sum_{i = 1}^{n} Q_{i}(u)$.
The operator $M(u)$ counts the number of sites that are not in the reference state $\ket{e_{0}(u)}$.
For an integer $z \geq 0$, let $\Pi_{\leq z}(u)$ denote the spectral projector of $M(u)$ onto eigenvalues $\leq z$.
This projector restricts the Hilbert space to configurations with at most $z$ local deviations (or ``flips'') relative to $\ket{e_{0}(u)}^{\otimes n}$.

Because $H$ is permutation-invariant, the ground state lies in the totally symmetric subspace $\mathrm{Sym}^{n}(\mathbb{C}^{d})$ by Lemma~\ref{symmetry_sector_uniform_gs}. 
The truncated space $\Pi_{\leq z}(u) \mathrm{Sym}^{n}(\mathbb{C}^{d})$ admits a natural occupation-number basis indexed by
\begin{align}
	\mathbf{n} = (n_{0},n_{1},\cdots,n_{d-1}) \in \mathbb{Z}_{\geq 0}^{d}, \qquad \sum_{a = 0}^{d-1} n_{a} = n, \qquad \sum_{a = 1}^{d-1} n_{a} \leq z,
\end{align}
where $n_{a}$ denotes the number of sites occupying $\ket{e_{a}(u)}$.
For each such multi-index $\mathbf{n}$, the corresponding normalized symmetric state is given by
\begin{align}
	\ket{\mathbf{n};u} = \sqrt{\frac{1}{n! \cdot n_{0}! n_{1}! \cdots n_{d-1}!}} \sum_{ \pi \in S_{n}} P_{\pi} \br{\bigotimes_{a = 0}^{d-1} \ket{e_{a}(u)}^{\otimes n_{a}}},
\end{align}
where $P_{\pi}$ denotes the permutation operator on the $n$ sites.
The collection of all such states forms an orthonormal basis of $\Pi_{\leq z} (u) \mathrm{Sym}^{n}(\mathbb{C}^{d})$.
Since the number of choices of $(n_{1},\cdots,n_{d-1})$ with total weight at most $z$ equals $\binom{z+ d- 1}{d-1}$, the dimension of this truncated symmetric sector is
\begin{align}
	m(z) := \binom{z + d - 1}{d-1} = \Theta(z^{d-1}).
\end{align}

We define the truncated Hamiltonian
\begin{align}
	H^{(z)}(u) := \Pi_{\leq z}(u) H \Pi_{\leq z}(u)
\end{align}
viewed as an $m(z) \times m(z)$ Hermitian matrix in the symmetric occupation-number basis. 
Write the $k$-local Hamiltonian as $H = \sum_{Z : |Z| \leq k} h_{Z}$, where each $h_{Z}$ acts nontrivially only on the sites in $Z$.
We assume that all matrix entries of each local term $h_{Z}$ in the computational basis are specified with finite bit complexity (e.g., rationals or algebraic numbers, with polynomially bounded bit-length in the input size).
For occupation-number states $\ket{\mathbf{n};u}$, we have
\begin{align}
	\bra{\mathbf{n};u} H \ket{\mathbf{n}';u} = \sum_{Z: |Z| \leq k} \bra{\mathbf{n}; u} h_{Z} \ket{\mathbf{n}';u}.
\end{align}
Each symmetric basis state $\ket{\mathbf{n};u}$ is obtained by symmetrizing tensor products of the local vectors $\{\ket{e_{a}(u)}\}_{a = 0}^{d-1}$.
Expanding each symmetric occupation-number state $\ket{\mathbf{n};u}$ in the fixed computational basis, its coefficients are polynomial functions of the real variables $x_{u}$ parameterizing the local basis.
Since each local term $h_{Z}$ acts on at most $k$ sites and its matrix elements in the computational basis are fixed numbers of finite bit complexity, the quantity $\bra{\mathbf{n};u} h_{Z} \ket{\mathbf{n}';u}$ is a finite sum of products of at most $2k$ local basis amplitudes, each of which is linear in the real variables $x_{u}$.
Hence, each matrix element of $H^{(z)}(u)$ is a polynomial in $x_{u}$ of degree at most $2k$.
These polynomials are evaluated on the feasible set of real variables $x_{u}$ satisfying the polynomial unitarity constraints introduced above.
Although the symmetric states $\ket{\mathbf{n};u}$ are defined through permutation sums, these sums are never expanded explicitly in computations of matrix elements. Because each term $h_{Z}$ acts on at most $k$ sites, matrix elements between symmetric occupation-number states can be evaluated combinatorially using multinomial counting identities. In particular, the evaluation reduces to sums over local amplitudes of the vectors $\{\ket{e_{a}(u)}\}_{a = 0}^{d-1}$ and can be performed without enumerating permutations.

\subsubsection{Algorithm: $\poly(z) \cdot \log(n/ \epsilon)$-time optimization over $u$}

We now describe a worst-case algorithmic route to compute an $\epsilon$-approximate ground-state energy and an explicit approximate ground-state vector within the truncated symmetric sector, in total time $\poly(z) \cdot \log(n / \epsilon)$ (and polynomial dependence on the input bit-length).

Fix a truncation level $z$.
For each parameter $u$ (equivalently $x_{u} \in \mathbb{R}^{2d^{2}}$) satisfying the polynomial unitarity constraints, define the truncated variational optimum
\begin{align}
	E_{z}^{\star} := \min_{u} \lambda_{\min}\br{H^{(z)}(u)}.
\end{align}
Here, for a Hermitian matrix $X$, we write $\lambda_{\min}(X)$ for its smallest eigenvalue, i.e.,
\begin{align}
	\lambda_{\min}(X) := \min_{\norm{\ket{v}} = 1} \bra{v} X \ket{v}.
\end{align}
Equivalently, $\lambda_{\min}(X)$ is the unique value such that $X - \lambda_{\min}(X) \mathds{1}\succeq 0$ and $\det(X - \lambda_{\min}(X) \mathds{1}) = 0$.
Our objective is to approximate $E_{z}^{\star}$ within additive error $\pm \epsilon$ and to output a corresponding approximate eigenvector supported on $\Pi_{\leq z}(u) \in \mathrm{Sym}^{n}(\mathbb{C}^{d})$.

By the setup above, each matrix element of $H^{(z)}(u)$ in the symmetric occupation-number basis is a polynomial in the real variables $x_{u}$ of degree at most $2k$.
Moreover, each entry can be computed without expanding permutations, by evaluating multinomial coefficients and local-amplitude products arising from the $k$-local terms.
Consequently, there is a uniform procedure that, given a basis-index pair $(\mathbf{n},\mathbf{n}')$ and a point $x_{u}$, evaluates $\bra{\mathbf{n};u} H \ket{\mathbf{n}';u}$ using $\poly(m(z)) = \poly(z)$ arithmetic operations and polynomially many-bit operations on the input coefficients.

Introduce an auxiliary real variable $\lambda \in \mathbb{R}$ and define the characteristic polynomial
\begin{align}
	P(x_{u},\lambda) := \det \br{H^{(z)}(u) - \lambda \mathds{1}}.
\end{align}
For each feasible $x_{u}$, the equality $P(x_{u},\lambda) = 0$ is equivalent to ``$\lambda$ is an eigenvalue of $H^{(z)}(x_{u})$.''
Hence, the global optimum can be written as the constrained polynomial optimization
\begin{align}
	E_{z}^{\star} = \min_{x_{u},\lambda} \brrr{\lambda \,:\, \mathcal{C}(x_{u}) \, \wedge \, P(x_{u},\lambda) = 0}, \label{eq:Ez_star_polynomial_form}
\end{align}
where $\mathcal{C}(x_{u})$ denotes the polynomial feasibility constraints encoding $U(u)^{\dagger} U(u) = \mathds{1}_{d}$ and any gauge-fixing constraints (to avoid redundancies).
This avoids introducing a positive semidefinite constraint $H^{(z)}(u) - \lambda \mathds{1} \succeq 0$ (and the attendant principal-minor explosion), replacing it with one polynomial equality.

Although the determinant formally contains $m(z)!$ terms when expanded, we never expand it explicitly.
Instead, we compute the characteristic polynomial $P(x_{u},\lambda) = \det(H^{(z)}(u) - \lambda \mathds{1})$ using a division-free algorithm that operates over an arbitrary commutative ring.
Standard examples include the Berkowitz algorithm~\cite{berkowitz1984computing}, which computes the characteristic polynomial in $\mathcal{O}(m(z)^{4})$ ring operations without division, and the Bareiss fraction-free elimination scheme~\cite{bareiss1968sylvester}, which performs Gaussian elimination while avoiding intermediate fractions and is well suited for integer or polynomial rings.
In our setting, the underlying ring is the polynomial ring in the real variables $(x_{u},\lambda)$.
Since $d$ and $k$ are treated as constants, the number of real variables in $x_{u}$ is constant, and the entries of $H^{(z)}(u)$ have degree at most $2k$ in $x_{u}$.
Consequently, the characteristic polynomial has total degree $\mathcal{O}(m(z)) = \mathcal{O}(z^{d-1})$ in the variables $(x_{u},\lambda)$.

Let $r$ denote the number of real variables in $(x_{u},\lambda)$.
In the present parameterization, we have $x_{u} \in \mathbb{R}^{2d^{2}}$ and $\lambda \in \mathbb{R}$, so $r = 2d^{2} + 1$, which is constant since $d$ is fixed.
The number of monomials of total degree at most $D$ in $r$ variables is therefore bounded by
\begin{align}
	\binom{D+r}{r} = \mathcal{O}(D^{r}).
\end{align}
Since $r$ is a constant, this quantity grows polynomially in $D$.
Consequently, because the characteristic polynomial $P(x_{u},\lambda)$ has total degree $\mathcal{O}(m(z))$, its dense representation has size polynomial in $m(z)$.
Each ring operation performed by the Berkowitz or Bareiss algorithm corresponds to additions and multiplications of such polynomials, which can be carried out in time polynomial in their representation size.
Because these algorithms require $\poly(m(z))$ ring operations, the overall bit-complexity of constructing $P$ is bounded by $\poly(m(z)) = \poly(z)$, with constants depending only on $d$ and $k$ and with polynomial dependence on the input bit-length.

Assume coarse numerical bounds $\mathcal{L} \leq E_{z}^{\star} \leq \mathcal{U}$ are available, e.g., via Gershgorin bounds or simple operator-norm bounds on $H^{(z)}(u)$ over the feasible set~\cite{horn2012matrix}. Here, we can set $\mathcal{U}- \mathcal{L} = \orderof{n}$. 
To approximate the optimum value $E_{z}^{\star}$, it suffices to solve the following threshold decision problem: given a candidate value $t$, determine whether there exists a parameter choice $u$ such that $\lambda_{\min}(H^{(z)}(u)) \leq t$.
If such a parameter exists, then $E_{z}^{\star} \leq t$, whereas otherwise $E_{z}^{\star} > t$.
This allows one to recover $E_{z}^{\star}$ to additive accuracy $\epsilon$ via binary search over the interval $[\mathcal{L},\mathcal{U}]$.
For any threshold $t \in [\mathcal{L},\mathcal{U}]$, we therefore consider the decision problem:
\begin{align}
	\mathsf{Decide}(t): \quad \exists x_{u} \in \mathbb{R}^{2d^{2}}\quad \exists \lambda \in \mathbb{R}\quad \text{s.t.} \quad \br{\mathcal{C}(x_{u}) \, \wedge \, \lambda \leq t \, \wedge \, P(x_{u}, \lambda) = 0}. \label{eq:Decide_t}
\end{align}
All constraints appearing in $\mathsf{Decide}(t)$ are polynomial equalities and inequalities in the real variables $(x_{u},\lambda)$.
Hence, the problem is an instance of the existential theory of the reals (ETR), i.e., the feasibility of a semialgebraic set.

Crucially, the total number of real variables in Eq.~\eqref{eq:Decide_t} is $r = 2d^{2} + 1$, which is a constant when $d$ is a constant.
Standard real-algebraic decision procedures (quantifier elimination, critical point methods, roadmap methods) solve Eq.~\eqref{eq:Decide_t} in time~\cite{renegar1992computational,basu2006algorithms}
\begin{align}
	(sD)^{\mathcal{O}(r)} \cdot \poly(\text{input bit-length}),
\end{align}
where $s$ is the number of polynomial constraints and $D$ is the maximum degree.
In our setting, $r = \mathcal{O}(d^2)$, $s = \orderof{d^2}$ from Eq.~\eqref{constrain_s_u}, and $D = \orderof{m(z)} = \poly(z)$, hence $\mathsf{Decide}(t)$ can be solved in $\poly(z)$ time up to polynomial factors in the input bit-length.

We now approximate $E_{z}^{\star}$ by performing a binary search over the interval $[\mathcal{L},\mathcal{U}]$ using the decision procedure $\mathsf{Decide}(t)$. 
Starting from the initial interval $[\mathcal{L},\mathcal{U}]$, at each iteration we set
\begin{align}
	t = \frac{\mathcal{L} + \mathcal{U}}{2},
\end{align}
and evaluate $\mathsf{Decide}(t)$.
If the answer is \textsc{Yes}, then there exists $u$ such that $\lambda_{\min}(H^{(z)}(u)) \leq t$, implying $E_{z}^{\star} \leq t$, and we update the upper bound $\mathcal{U} \leftarrow t$.
Otherwise, $E_{z}^{\star} > t$ and we update the lower bound $\mathcal{L} \leftarrow t$.
Thus, each iteration halves the length of the interval containing $E_{z}^{\star}$.
After $K$ iterations, the interval length becomes $(\mathcal{U} - \mathcal{L})/2^{K}$. 
Therefore, choosing $K = \mathcal{O}\br{\log \frac{\mathcal{U} - \mathcal{L}}{\epsilon}}$ ensures that the interval width is at most $\epsilon$.
Since each call to $\mathsf{Decide}(t)$ can be solved in time $\poly(z)$, the total runtime of the procedure is
\begin{align}
	\poly(z) \cdot \log \br{\frac{\mathcal{U} - \mathcal{L}}{\epsilon}}.
\end{align}
Let $\widehat{E}_{z}$ denote the midpoint of the final interval. Then $\abs{\widehat{E}_{z} - E_{z}^{\star}} \leq \epsilon$.

To output a state vector, we also need a corresponding parameter choice $\widehat{u}$ (or $\widehat{x}_{u}$) and an eigenvector of $H^{(z)}(u)$.
Many ETR decision procedures can be implemented to return a feasible point (a witness) when $\mathsf{Decide}(t)$ is true. 
Using the final threshold $t = \widehat{E}_{z} + \epsilon$ (for instance), extract a feasible $(\widehat{x}_{u},\widehat{\lambda})$ satisfying Eq.~\eqref{eq:Decide_t} with $\widehat{\lambda} \leq \widehat{E}_{z} + \epsilon$.
Then form the numeric matrix $H^{(z)}(\widehat{u})$ by evaluating its entries at $\widehat{x}_{u}$.
Finally, compute the minimum-eigenpair of the $m(z) \times m(z)$ Hermitian matrix $H^{(z)}(u)$ by standard numerical linear algebra (dense diagonalization in time $\mathcal{O}(m(z)^{3}) = \poly(z)$ for fixed $d$).
Let $\ket{\varphi_{z}}$ be the resulting normalized eigenvector in the truncated symmetric basis; then
\begin{align}
	\abs{\bra{\varphi_{z}} H \ket{\varphi_{z}} - E_{z}^{\star}} = \mathcal{O}(\epsilon).
\end{align}

Finally, we aim to prove that $\bra{\varphi_{z}} H \ket{\varphi_{z}}$ and $\ket{\varphi_{z}}$ approximate the ground energy and the ground state, respectively.
We first prove
\begin{align}
	\abs{\bra{\varphi_{z}} H \ket{\varphi_{z}} - \bra{\gs} H \ket{\gs} } = \mathcal{O}(\epsilon).
\end{align}
For this purpose, we define 
\begin{align}
	\ket{\gs_{\le z}} := \frac{\Pi_{\leq z} \ket{\gs}}{\norm{\Pi_{\leq z} \ket{\gs}}} ,
\end{align}
where $\Pi_{\leq z}$ was defined in \ref{sec:Main statement} for the exact mean-field basis. 
Then, the concentration bound (Theorem~\ref{Them:prob_dist_z}) ensures that the true ground state satisfies
\begin{align}
	\norm{\ket{\gs} - \ket{\gs_{\le z}}} \le c_1 e^{-c'_1 z^{1/2}},
\end{align}
which yields 
\begin{align}
	E_{z}^{\star} -  \bra{\gs} H \ket{\gs} \le \bra{\gs_{\le z}} H \ket{\gs_{\le z}}- \bra{\gs} H \ket{\gs} &= \sum_{z_1=z}^n  \bra{\gs_{\le z_1}} H \ket{\gs_{\le z_1}}- \bra{\gs_{\le z_1+1}} H \ket{\gs_{\le z_1+1}} \notag \\
	&\le \sum_{z_1=z}^n \norm{\Pi_{\leq z_1+1} (H - \check{E}_0) \Pi_{\leq z_1+1}} \cdot \norm{\ket{\gs_{\le z_1}}- \ket{\gs_{\le {z_1+1}}}} \notag \\
	& \le c'_1e^{-c'_2 z^{1/2}} ,
\end{align} 
where we use inequality~\eqref{Pi_le_z_H-check_E_0} to upper-bound $\norm{\Pi_{\leq z_1+1}(H - \check{E}_0) \Pi_{\leq z_1+1}}$. 

From the above inequalities, once we choose $z = \Theta(\log^{2}(1 / \epsilon))$, we can ensure 
\begin{align}
	\abs{\bra{\varphi_{z}} H \ket{\varphi_{z}} - \bra{\gs} H \ket{\gs} } \le 
	\abs{E_{z}^{\star} -  \bra{\gs} H \ket{\gs} } + \abs{\bra{\varphi_{z}} H \ket{\varphi_{z}} -E_{z}^{\star} } =  \mathcal{O}(\epsilon) .
\end{align}
As long as the Hamiltonian is gapped, we also obtain 
\begin{align}
	\abs{\bra{\varphi_{z}} H \ket{\varphi_{z}} - \bra{\gs} H \ket{\gs} } \ge \br{ 1- |\langle \varphi_{z} \ket{\gs}|^2} \Delta  ,
\end{align}
which gives 
\begin{align}
	\norm{ \ket{\varphi_{z}} - \ket{\gs}}=  \frac{\mathcal{O}(\epsilon^{1/2})}{\Delta^{1/2} }
\end{align}
under an appropriate phase choice as $|\langle \varphi_{z} \ket{\gs}|= \langle \varphi_{z} \ket{\gs}$.

Therefore, for $z = \Theta(\log^{2}(1 / \epsilon))$, the truncated computation yields an $\mathcal{O}(\epsilon)$-accurate ground-state energy and a corresponding state satisfying $\norm{\ket{\varphi_{z}} - \ket{\gs}} = \mathcal{O}\br{\sqrt{\epsilon / \Delta}}$. 
The total runtime is
\begin{align}
	\orderof{\poly(z) \cdot \log \br{n/\epsilon} }= \orderof{ \polylog(1 / \epsilon) \cdot \log(n/\epsilon) },
\end{align}
for fixed $d,k$ and $z = \Theta(\log^{2}(1 / \epsilon))$. 

\begin{algorithm}[h]
	\caption{Polylogarithmic-time approximation of the ground state in the symmetric sector}
	\label{Polylog_ETR_Algorithm}
	\SetKwInOut{Input}{input}
	\SetKwInOut{Output}{output}
	\Input{Permutation-invariant $k$-local Hamiltonian $H$, target accuracy $\epsilon$}
	\Output{Approximate energy $\widehat{E}_z$ and approximate ground-state vector $\ket{\varphi_z}$}
	
	\SetKwBlock{Begin}{begin}{end}
	\Begin{
		Introduce optimization variables $x_u \in \mathbb{R}^{2d^2}$ parameterizing a unitary $U(u)$ that defines a local basis $\{\ket{e_a(u)}\}_{a=0}^{d-1}$\;
		
		Set truncation level $z \leftarrow \Theta(\log^2(1/\epsilon))$\;
		
		Let $\Pi_{\le z}(u)$ denote the projector onto configurations with at most $z$ deviations from $\ket{e_0(u)}^{\otimes n}$\;
		
		For each $x_u$, work in the truncated symmetric occupation-number basis $\{\ket{\mathbf{n};u}\}$ indexed by $\mathbf{n}=(n_0,\dots,n_{d-1})$ (counts of sites in $\ket{e_a(u)}$), with $\sum_{a=1}^{d-1} n_a \le z$\;
		
		Compute the truncated Hamiltonian $H^{(z)}(u) \leftarrow \Pi_{\le z}(u)\, H\, \Pi_{\le z}(u)$\;
		
		Construct the characteristic polynomial $P(x_u,\lambda) \leftarrow \det(H^{(z)}(u)-\lambda \mathds{1})$ (e.g.\ Berkowitz algorithm~\cite{berkowitz1984computing})\;
		
		Obtain coarse bounds $\mathcal{L},\mathcal{U}$ on $E_z^\star$\;
		
		\While{$\mathcal{U}-\mathcal{L} > \epsilon$}{
			
			$t \leftarrow (\mathcal{L}+\mathcal{U})/2$\;
			
			Solve the ETR feasibility problem
			
			\Indp
			$\exists x_u,\lambda:\ \mathcal{C}(x_u)\wedge \lambda \le t \wedge P(x_u,\lambda)=0$\;
			\Indm
			
			\uIf{feasible}{
				$\mathcal{U} \leftarrow t$\;
			}
			\Else{
				$\mathcal{L} \leftarrow t$\;
			}
			
		}
		
		$\widehat{E}_z \leftarrow (\mathcal{L}+\mathcal{U})/2$\;
		
		Solve the ETR feasibility problem for $t = \widehat{E}_z+\epsilon$ and extract a feasible witness $(\widehat{x}_u,\widehat{\lambda})$\;

		Form the numeric matrix $H^{(z)}(\widehat{u})$\;
		
		Compute the minimum eigenpair of $H^{(z)}(\widehat{u})$ (dense diagonalization in time $\mathcal{O}(m(z)^3)$)\;
		
		Let $\ket{\varphi_z}$ be the resulting normalized eigenvector\;
		
		\KwRet $(\widehat{E}_z,\ket{\varphi_z})$\;
	}	
\end{algorithm}

\subsubsection{Numerical Implementation}

We now describe a numerical implementation of the truncated variational construction introduced in Algorithm~\ref{Polylog_ETR_Algorithm}.
The numerical calculation is based on the same finite-dimensional variational principle as the formal algorithm: one minimizes the lowest eigenvalue of the Hamiltonian restricted to the low-deviation symmetric sector while optimizing over the local reference basis.

For numerical purposes, we specialize to the qubit permutation-invariant setting, where the local reference basis can be parameterized by two Bloch-sphere angles.
Rather than using a symbolic ETR solver, we solve the corresponding threshold problem by a deterministic branch-and-bound procedure with a finite decision tolerance over these two parameters.
Thus, the formal and numerical procedures address the same truncated variational problem, but use different realizations of the threshold oracle.

For the numerical demonstration, we consider the permutation-invariant Lipkin--Meshkov--Glick (LMG) Hamiltonian
\begin{align}
	H_{\mathrm{LMG}} = - \frac{1}{n} \sum_{1 \leq i < j \leq n} \br{\sigma_{i}^{x} \sigma_{j}^{x} + \gamma \sigma_{i}^{y} \sigma_{j}^{y}} - h \sum_{i = 1}^{n} \sigma_{i}^{z} \label{All_LMG}
\end{align}
where $\sigma_{i}^{x}, \sigma_{i}^{y}, \sigma_{i}^{z}$ denote the Pauli matrices at site $i$ and $n$ denotes the size of the system.
Using the collective spin operators $J_{\alpha} = \sum_{i = 1}^{n} \sigma_{i}^{\alpha} / 2$ for $\alpha = x,y,z$, the Hamiltonian restricted to the symmetric Dicke sector can be written as
\begin{align}
	H_{\mathrm{LMG}} = - \frac{2}{n} \br{J_{x}^{2} + \gamma J_{y}^{2}} - 2h J_{z} + \frac{1 + \gamma }{2} \mathds{1}.
\end{align}
For qubits, this symmetric sector has dimension $n + 1$.

Since $d = 2$, the local reference direction is parameterized by a point on the Bloch sphere.
We write
\begin{align}
	U(\theta,\phi) = \begin{pmatrix}
		\cos(\theta / 2) & - e^{- i \phi} \sin(\theta / 2) \\
		e^{i \phi} \sin(\theta / 2) & \cos(\theta / 2)
	\end{pmatrix}, \qquad 0 \leq \theta \leq \pi, \quad 0 \leq \phi \leq 2 \pi. \label{eq:local_basis_theta_phi}
\end{align}
The columns of $U(\theta,\phi)$ define a local orthonormal basis.
We denote the first column by $\ket{e_{0}(\theta,\phi)}$, which plays the role of the local reference state, and the second column by $\ket{e_{1}(\theta,\phi)}$, which is the orthogonal single-flip direction.
In the qubit symmetric sector, a basis vector is then specified only by the number of sites occupying $\ket{e_{1}(\theta,\phi)}$.
We denote by $\ket{r;\theta,\phi}$ the normalized symmetric Dicke state with $r$ sites in $\ket{e_{1}(\theta,\phi)}$ and $n-r$ sites in $\ket{e_{0}(\theta,\phi)}$.
Therefore, the sector with at most $z$ local deviations is spanned by $\brrr{\ket{r; \theta,\phi} \,:\, r = 0,1,\cdots, z}$, and has dimension $z+1$.
For each pair $(\theta,\phi)$, we construct the truncated matrix
\begin{align}
	H^{(z)}(\theta,\phi) = \brr{ \bra{r; \theta,\phi} H_{\mathrm{LMG}} \ket{r';\theta,\phi} }_{0 \leq r,r' \leq z}. \label{eq:numerical_truncated_matrix} 
\end{align}
The numerical objective function is then
\begin{align}
	f_{z}(\theta,\phi) := \lambda_{\min}\br{H^{(z)}(\theta,\phi)}, \qquad E_{z}^{\star} = \min_{\theta,\phi} f_{z}(\theta,\phi). \label{eq:numerical_objective_fz}
\end{align}
This is the same finite-dimensional truncated optimization problem as in Algorithm~\ref{Polylog_ETR_Algorithm}, specialized to the qubit permutation-invariant case. 
The general local-basis parameter $u$ is reduced to the two Bloch-sphere angles $(\theta,\phi)$.

We next explain how the threshold step is implemented numerically.
As in Algorithm~\ref{Polylog_ETR_Algorithm}, the outer layer of the procedure is organized as a binary search over the energy threshold $t$.
The numerical modification lies in the realization of the threshold oracle.
In the formal algorithm, the condition $E_{z}^{\star} \leq t$ is formulated as the real-algebraic threshold problem $\mathsf{Decide}(t)$ in Eq.~\eqref{eq:Decide_t}.
In the present numerical implementation, we consider the corresponding finite-dimensional threshold problem
\begin{align}
	\mathrm{Decide}_{z}(t): \qquad \exists \, (\theta,\phi) \quad \text{s.t.} \quad f_{z}(\theta,\phi) \leq t \label{eq:numerical_threshold_problem}
\end{align}
and solve it by a deterministic branch-and-bound search over the compact domain $[0, \pi] \times [0,2 \pi]$.
At each step of the outer binary search, this procedure serves as a finite-tolerance numerical version of the formal threshold oracle.

It recursively partitions the parameter domain into rectangular boxes.
For a box $B$ with center $c_{B} = (\theta_{B}, \phi_{B})$ and Euclidean radius $r_{B}$, we evaluate $f_{z}(c_{B})$ and use a conservative Lipschitz lower bound 
\begin{align}
	\inf_{(\theta,\phi) \in B} f_{z}(\theta,\phi) \geq f_{z}(c_{B}) - L_{\mathrm{Lip}} r_{B}. \label{eq:BB_Lipschitz_lower_bound}
\end{align}
For the LMG Hamiltonian in Eq.~\eqref{All_LMG}, we use the global bound
\begin{align}
	L_{\mathrm{Lip}} = \sqrt{2} \brr{\frac{4J^{2}}{n} (1 + |\gamma|) + 2|h| J}, \qquad J = \frac{n}{2} \label{LMG_global_Lipschitz_constant}
\end{align}
which follows directly from the collective-spin representation and is uniform over the Bloch-sphere parameters and the truncation level $z$.
Thus, $f_{z}(c_{B})$ provides an explicit candidate energy, while Eq.~\eqref{eq:BB_Lipschitz_lower_bound} gives a lower bound on the minimum over the box.

Boxes whose lower bounds are already above the best candidate energy found so far are discarded, while the remaining boxes are subdivided further.
For each trial threshold $t$, the branch-and-bound routine returns a lower estimate $L_{\mathrm{BB}}$ and an upper estimate $U_{\mathrm{BB}}$ for $\min_{\theta,\phi} f_{z}(\theta,\phi)$.
The upper estimate $U_{\mathrm{BB}}$ is the best value found at an explicit parameter point, whereas the lower estimate $L_{\mathrm{BB}}$ is obtained from the Lipschitz-based lower bounds over the remaining active boxes.
Thus, $L_{\mathrm{BB}} \leq E_{z}^{\star} \leq U_{\mathrm{BB}}$.
If $U_{\mathrm{BB}} \leq t$, then an explicit witness $(\theta,\phi)$ with $f_{z}(\theta,\phi) \leq t$ has been found, and the threshold query is answered affirmatively.
If $L_{\mathrm{BB}} > t$, then no point in the parameter domain can reach energy below $t$, and the threshold query is answered negatively.
If $t$ lies within the unresolved interval $[L_{\mathrm{BB}}, U_{\mathrm{BB}}]$, the query is treated using the prescribed internal decision tolerance.

We denote the outer energy tolerance by $\epsilon_{\mathrm{energy}}$.
For each fixed $z$, the binary search over $t$ is continued until the reported binary-search interval has width at most $\epsilon_{\mathrm{energy}}$.
We denote the internal decision tolerance of the branch-and-bound procedure by $\epsilon_{\mathrm{decision}}$ and set $\epsilon_{\mathrm{decision}} = \epsilon_{\mathrm{energy}} / 4$.
For computational practicality, each threshold query is limited to at most $30000$ adaptive branch-and-bound box-splitting iterations.
Importantly, $\epsilon_{\mathrm{energy}}$ is not the final state-approximation parameter that determines the choice of $z$ in Algorithm~\ref{Polylog_ETR_Algorithm}.
Rather, it controls the numerical resolution of the fixed-$z$ optimization.
In the computations below, we set $\epsilon_{\mathrm{energy}} = 10^{-8}$.

\begin{figure}[t]
	\begin{center}
		\includegraphics[width=0.5\textwidth]{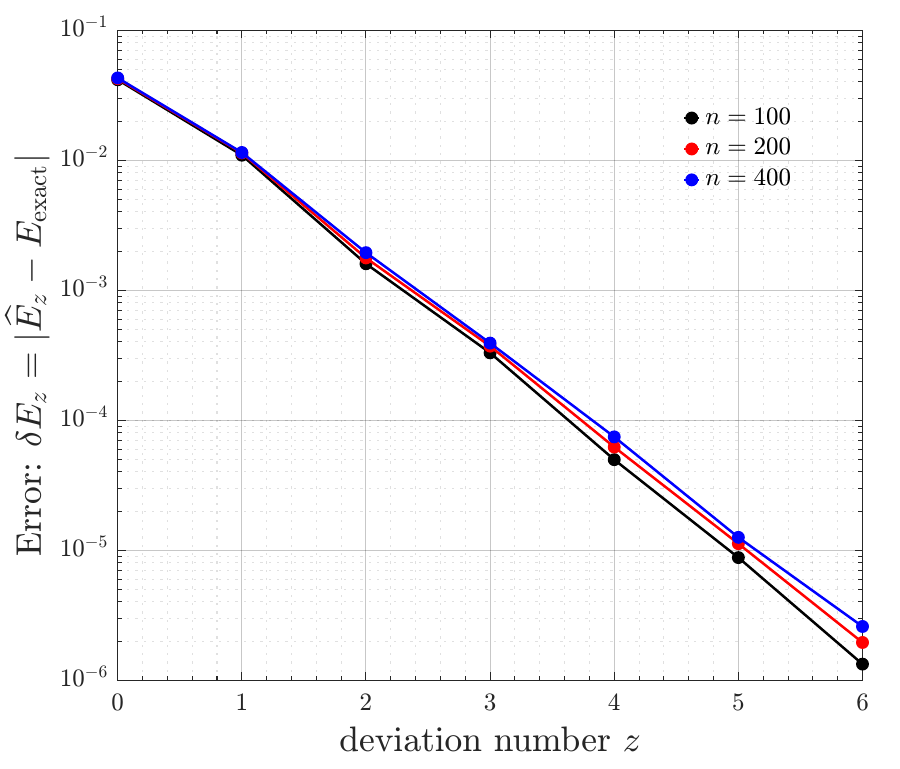}
	\end{center}
	\caption{Numerical implementation of the truncated variational construction for the LMG Hamiltonian in Eq.~\eqref{All_LMG}.
		For each $n \in \{100,200,400\}$ and $z = 0,1,\cdots,6$ we compute the numerical truncated energy $\widehat{E}_{z}$ by optimizing the local reference basis and plot the absolute error $\delta E_{z} = \abs{\widehat{E}_{z} - E_{\mathrm{exact}}}$.
		The exact benchmark $E_{\mathrm{exact}}$ is obtained by diagonalizing the full symmetric Dicke sector.
	}
	\label{fig:LMG_truncated_algorithm_error}
\end{figure}

To quantify the convergence with respect to the truncation level, we compare the computed truncated energy $\widehat{E}_{z}$ with the exact ground-state energy in the full symmetric Dicke sector.
Here $\widehat{E}_{z}$ denotes the witness energy obtained from the optimized local basis returned by the numerical branch-and-bound/binary-search procedure for the fixed-$z$ optimization problem, while $E_{z}^{\star}$ denotes the exact optimum of that truncated problem.
The exact benchmark $E_{\mathrm{exact}}$ is obtained by diagonalizing the $(n+1)$-dimensional symmetric-sector Hamiltonian.
This diagonalization is used to assess the accuracy of the truncated optimization, and not as part of the truncated variational search.

In the numerical data shown in Fig.~\ref{fig:LMG_truncated_algorithm_error}, we use $\gamma = 0.5$ and $h = 1.5$ in Eq.~\eqref{All_LMG}, which corresponds to the gapped paramagnetic regime of the LMG model.
We compute the truncated energies for $n \in \{100,200,400\}$ and $z = 0,1,\cdots,6$.
For each system size $n$, we plot the absolute energy error $\delta E_{z} = \abs{\widehat{E}_{z} - E_{\mathrm{exact}}}$ as a function of the truncation level $z$.
The resulting $z$-sweep illustrates the central mechanism behind the truncated variational algorithm: after optimizing the local reference direction, a small number of local deviations already gives a rapidly improving approximation to the exact ground-state energy.

We finally clarify how this plotted error should be interpreted.
In Algorithm~\ref{Polylog_ETR_Algorithm}, a target accuracy $\epsilon$ is specified first, and the low-deviation concentration bound determines the required truncation scale, with $z = \Theta(\log^{2}(1 / \epsilon))$, so that the truncated subspace approximates the true ground state to the desired accuracy. 
In the present numerical demonstration, we vary $z$ explicitly in order to display the convergence of the truncated energy as the low-deviation sector is enlarged.
For a fixed value of $z$, the computed value $\widehat{E}_{z}$ is the witness energy returned by the numerical branch-and-bound/binary-search procedure.
Accordingly, the plotted difference 
\begin{align}
	\delta E_{z} = \abs{\widehat{E}_{z} - E_{\mathrm{exact}}}
\end{align}
contains both the finite-truncation error and the residual error of the fixed-$z$ numerical optimization. 
The purpose of the calculation is to illustrate the convergence of the low-deviation variational space as the truncation level $z$ is increased; the numerical results are not used in proving the analytical error bounds or the formal complexity result.

\vspace{1em}
\bigskip

\section{Numerical Investigation of Entanglement Entropy}

To support our analytical results, we perform numerical simulations of the entanglement entropy for two all-to-all interacting systems: the Lipkin-Meshkov-Glick (LMG) model and the bilinear fermion system.

\subsection{Lipkin-Meshkov-Glick model}

We examine  the LMG model introduced in Eq.~\eqref{All_LMG}, which satisfies the bounded local-energy assumption due to the Kac normalization.
Previous studies of the ferromagnetic anisotropic LMG model, typically for $0 \leq \gamma < 1$, have established a second-order quantum critical point at $h_{c} = 1$, separating a symmetric phase for $h > 1$ from a symmetry-broken phase for $0 < h < 1$.
For a bipartition into blocks of size $L$ and $n - L$ with $L,n \gg 1$ and finite $L / n$, the ground-state bipartite entanglement entropy remains bounded away from criticality, tends to zero as $h \rightarrow \infty$, and displays logarithmic critical behavior at $h = 1$~\cite{PhysRevA.71.064101,dusuel2005continuous,vidal2007entanglement}.

\begin{figure}[h]
	\begin{center}
		\includegraphics[width=\textwidth]{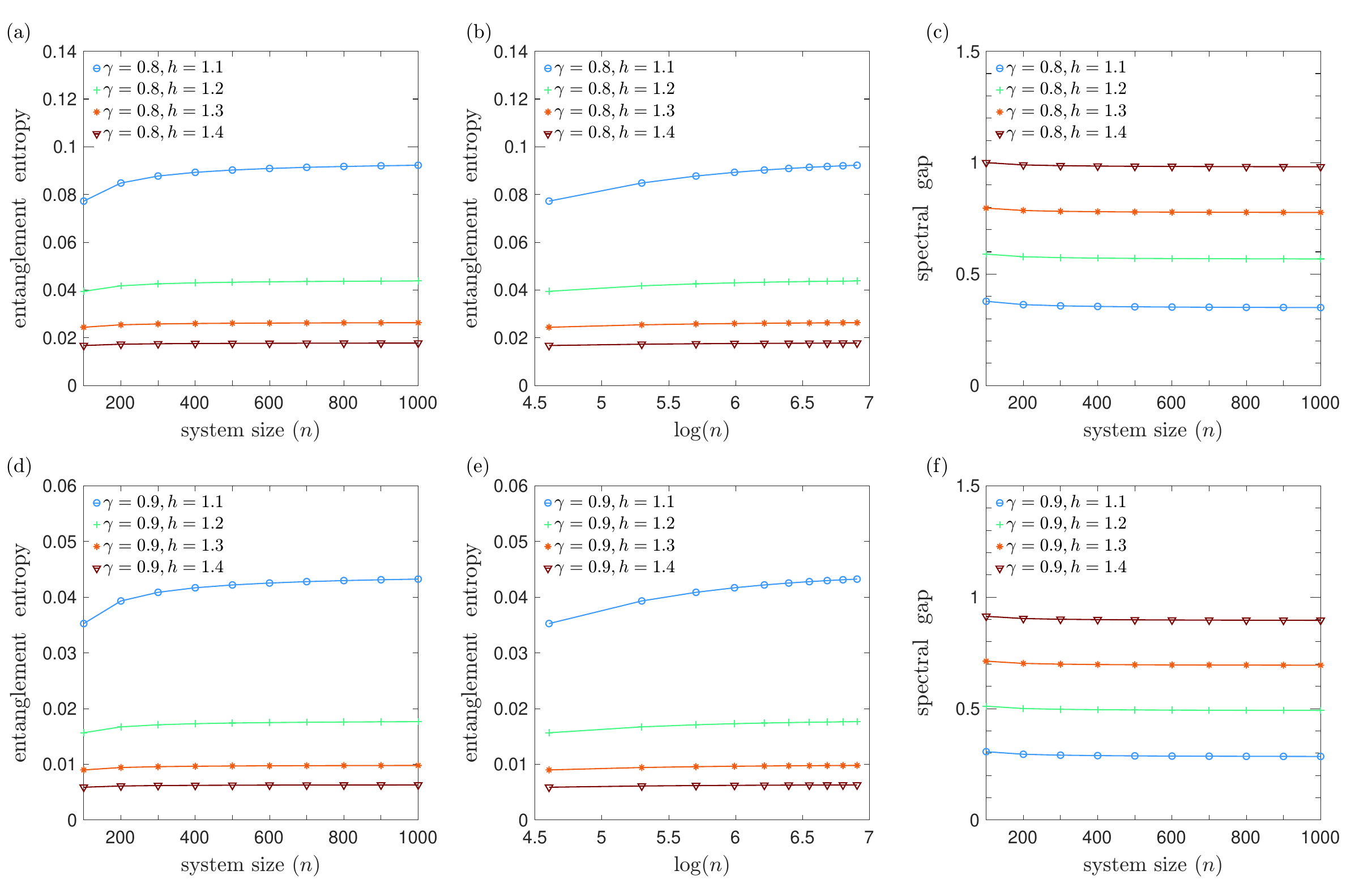}
	\end{center}
	\caption{The figures present the entanglement entropy and energy gaps for the Lipkin-Meshkov-Glick model in Eq.~\eqref{All_LMG} at $\gamma = 0.8$ and $\gamma = 0.9$ across various values of $h$. Panels (a-c) focus on $\gamma = 0.8$, detailing the entropy as a function of $n$, its logarithmic scaling, and the energy gap. Panels (d-f) display analogous measurements for $\gamma = 0.9$, showcasing consistent observations and trends.
	}
	\label{fig:FigLMG}
\end{figure}

By exploiting the permutation symmetry of the LMG model, we restrict the problem to the symmetric Hilbert space spanned by Dicke states, whose dimension scales linearly with system size, and perform exact diagonalization to compute the excitation gap and the entanglement entropy between the two halves of the system.
In line with earlier studies, the numerical data recover the expected features of the model and are in agreement with the analytical predictions of Theorem~\ref{Thm:Constant_Entanglement_Uniform}.
To address the gapped regime, we focus on the case where $h > 1$ and $h > \gamma$~\cite{dusuel2004finite,dusuel2005continuous}.
In the thermodynamic limit, the gap in this regime is $2 \sqrt{\br{h-1} \br{h-\gamma}}$.
Figures~\ref{fig:FigLMG}(a, b, c) present results for $\gamma = 0.8$ at various values of $h$, with Figure~\ref{fig:FigLMG}(a) displaying the entanglement entropy as a function of $n$, Figure~\ref{fig:FigLMG}(b) showing these data with a logarithmic scale on the $x$-axis, and Figure~\ref{fig:FigLMG}(c) illustrating the energy gap. Figures~\ref{fig:FigLMG}(d, e, f) for $\gamma = 0.9$ at various values of $h$ present analogous results, displaying the entanglement entropy in (d), its logarithmic representation in (e), and the energy gap in (f).
As $n$ increases, the gap tends to saturate, and we observe that the gap converges to a value of $2 \sqrt{\br{h-1} \br{h-\gamma}}$, while the entanglement entropy increases more slowly than $\log(n)$ and eventually saturates.

\subsection{All-to-all bilinear fermions}

\begin{figure}[h]
	\begin{center}
		\includegraphics[width=\textwidth]{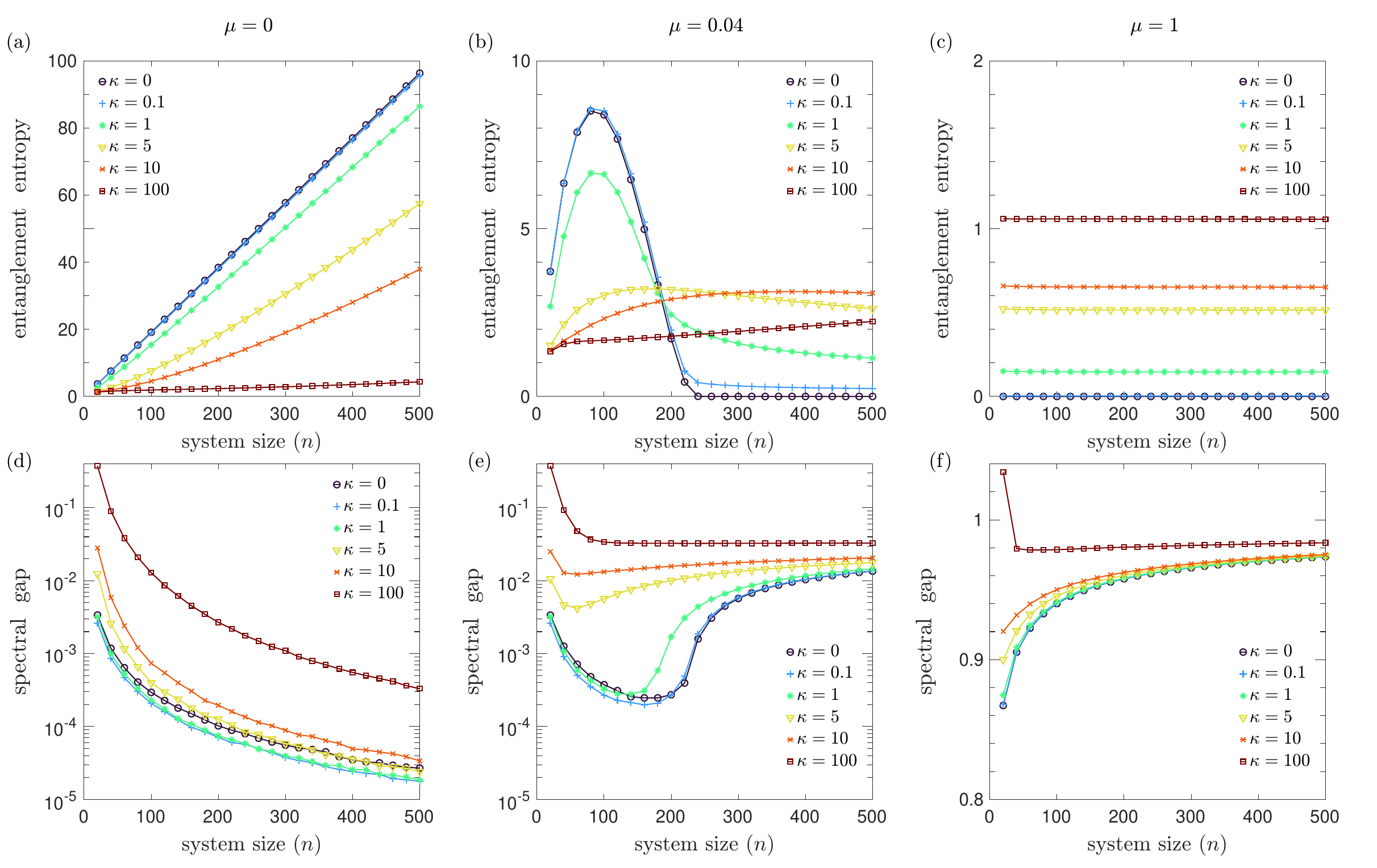}
	\end{center}
	\caption{Entanglement entropy and energy gap for the all-to-all bilinear fermion model in Eq.~\eqref{All_BdG}. Panels (a)-(c) depict the entanglement entropy between two halves of the system as a function of system size $n$ for $\mu = 0$, $0.04$, and $1$, respectively. Each graph shows the results for six distinct values of $\kappa$, ranging from 0 to 100. Panels (d)-(f) depict the energy gap for $\mu = 0$, $0.04$, and $1$, respectively, where the gap is defined as the difference between the energy of the first excited state and that of the ground state. All results are averaged over 1000 samples, with the values of $t_{i,j}$ in Eq.~\eqref{All_BdG} randomly chosen between 0 and 1.
	}
	\label{fig:FigBdG}
\end{figure}

We consider an all-to-all bilinear fermion model, incorporating both pair creation and annihilation terms together with an on-site term, given by
\begin{align}
	H = \sum_{1 \leq i < j \leq n} - \frac{t_{ij}}{n} (c_{i}^{\dagger} c_{j} + c_{j}^{\dagger} c_{i}) + \sum_{1 \leq i < j \leq n} \frac{\kappa}{n} (c_{i}^{\dagger} c_{j}^{\dagger} + c_{j} c_{i}) - \mu \sum_{i = 1}^{n} c_{i}^{\dagger} c_{i}. \label{All_BdG}
\end{align}
Here, $c_{i}$ and $c_{i}^{\dagger}$ denote the annihilation and creation operators for a spinless fermion at site $i$. 
The first term in Eq.~\eqref{All_BdG} describes the all-to-all hopping between sites, while the second term accounts for pair creation and annihilation. 
The final term corresponds to the on-site potential.
Exploiting the bilinear structure of the Hamiltonian, we reduce the problem to a Bogoliubov--de Gennes Hamiltonian of linear size and compute the spectral gap and the entanglement entropy between two halves using standard techniques for Gaussian fermionic states, including the correlation-matrix method~\cite{peschel2003calculation}.
These simulations were performed to provide numerical support for our analytical results.
We evaluate the spectral gap and the entanglement entropy as functions of the system size $n$ and the parameters $\mu$ and $\kappa$.
We take $t_{i,j}$ to be independent random variables uniformly distributed on $[0,1]$, and average the results over 1000 samples.
Figure~\ref{fig:FigBdG} presents the results for three different values of $\mu$. 
For $\mu = 0$, the gap decreases as the system size increases. 
The entanglement entropy exhibits linear growth when $\kappa = 0$, and although it continues to increase for $\kappa \neq 0$, the growth rate decreases as $\kappa$ increases.
For $\mu = 0.04$, the gap approaches $\Omega(1)$, and the entanglement entropy begins to saturate as the system size increases.
For the larger value $\mu = 1$, the $\Omega(1)$ gap is more clearly developed, and the saturation of the entanglement entropy becomes more apparent.

\newpage

\end{document}